%% file: main.tex
\documentclass[10pt]{article}
\pdfoutput=1
\usepackage{jheplikemod}
\usepackage{ifpdf,tocloft,rotating,textcomp,mathrsfs,amssymb,amsmath,amsthm,mathtools,float,caption,tikz-cd,subcaption,xcolor,epsfig,dcolumn,upgreek,setspace,enumitem,array,multirow,bigdelim,arydshln,appendix,xparse,microtype,bm,braket,dsfont,nccmath,scalerel}
\usetikzlibrary{cd}
\usetikzlibrary{shapes.geometric,arrows,svg.path}
\usepackage[export]{adjustbox}
\usepackage[utf8]{inputenc}
\definecolor{lapis}{rgb}{0.0.0470,0.2941,0.5568}

\newcommand{\lab}[1]{\bigg\langle{#1\bigg\rangle}}
\newcommand{\ab}[1]{\langle{#1\rangle}}

\newcommand{\z}{\overline{z}}
\newcommand{\Q}{\overline{Q}}
\newcommand{\jj}{\overline{j}}
\newcommand{\J}{\overline{J}}
\newcommand{\K}{\overline{K}}
\newcommand{\dfunction}{\delta^{2}(z-z_{i},\z-\z_{i})}
\newcommand{\el}{\overline{L}}
\newcommand{\h}{\overline{h}}
\newcommand{\rom}[1]{\mathrm{#1}}
\newcommand{\mref}[1]{(\ref{#1})}
\usepackage{hyperref}
\hypersetup{colorlinks=true,urlcolor=lapis,linkcolor=lapis,citecolor=lapis,pdfauthor={Nikhil Kalyanapuram}, pdftitle={},	pdfdisplaydoctitle=true, pdfstartview=FitH}
\newtheorem*{acknowledgements}{Acknowledgements}
\newtheorem*{outline}{Outline}
\allowdisplaybreaks
\graphicspath{{figures/}}
\renewcommand{\thefigure}{\thesection.\arabic{figure}}

\title{Infrared {\normalfont\emph{and}} Holographic Aspects {\normalfont\emph{of the}} $S$-Matrix {\normalfont\emph{in}} Gauge Theory {\normalfont\emph{and}} Gravity}
\author{\vspace{-0.5cm}Nikhil Kalyanapuram}
\affiliation{Department of Physics and Institute for Gravitation and the Cosmos, Pennsylvania State University, University Park PA 16802, USA}
\emailAdd{nkalyanapuram@psu.edu}
\abstract{Soft theorems in gauge theory and gravity encode the universal properties of scattering amplitudes as the zero frequency limit of one or more external states is approached. When the participating particles are treated in the massless limit, the soft theorems are known to depend only on the directions of the states on null infinity. Leveraging this fact, we develop dual two-dimensional descriptions of soft theorems, recasting them as Ward identities of such dual models on the celestial sphere. This is done by first postulating putative holographic representations of the hard scattering amplitudes and dressing the asymptotic operators with appropriate two-dimensional analogues of the Wilson line. The soft theorems are then recovered by inserting currents that generate the Ward identities of the dual models. In addition to providing naturally holographic representations of the soft theorems, we see that it becomes possible to develop presentations of the asymptotic symmetries associated to these theorems entirely in terms of the dual two-dimensional fields. In the course of carrying out this analysis for soft theorems at leading order and beyond, we find that the two-dimensional dual description of soft theorems may be directly inferred by drawing analogies with the existing framework of asymptotic symmetry charges. We find that the soft charges generating the asymptotic symmetries can be brought into correspondence with dual two-dimensional currents, while the two-dimensional Wilson loops can be related to the hard parts of the conserved charges. Consequently, we rewrite the triality of asymptotic symmetries, soft theorems and conserved charges directly in the language of a class of two-dimensional theories.}

\begin{document}
\maketitle
\setcounter{page}{1}
\clearpage

\input{sec1intro}

\clearpage
\begin{outline}\normalfont
In Section \ref{sec:2}, we present a class of two-dimensional models on $\mathbb{CP}^{1}$ that can be used to compute two quantities of interest - the soft $S$-matrix and the soft factor due to infrared radiation, keeping the focus mainly on QED and gravity. Section \ref{sec:2.1} reviews the soft $S$-matrix in QED and gravity along with setting up the specific basis in which we work. In Section \ref{sec:2.2} the soft $S$-matrices in QED and gravity are recast as correlation functions of operators constructed out of manifestly two-dimensional fields. The soft theorems due to real emission are reviewed in Section \ref{sec:2.3}, which is followed by Section \ref{sec:2.4}, in which the soft theorems are shown to arise out of Ward identities in the two-dimensional models. The extension to Yang-Mills theory is presented in Section \ref{sec:2.5}.

In Section \ref{sec:3} we relate the two-dimensional models obtained to the asymptotic symmetries are known to provide the leading order soft theorems in gauge theory and gravity. We express the generators of large gauge transformations in terms of the dual field for the soft photon in Section \ref{sec:3.1}, which is followed by an overview of the generators of superrotations and supertranslations on null infinity in Section \ref{eq:3.2}. We then realize the generators of supertranslations in terms of the dual field of the soft graviton in Section \ref{sec:3.3}.

Section \ref{sec:4} extends the results of the preceding two sections to soft theorems beyond leading order, specifically to the subleading soft photon theorem and the subleading and subsubleading soft graviton theorems. The hierarchy of soft theorems in the celestial basis is first recalled in Section \ref{sec:4.1}. The subleading soft photon theorem, the subleading soft graviton theorem and the subsubleading soft graviton theorem and the symmetry generators thereof are treated in turn in Sections \ref{sec:4.2}, \ref{sec:4.3} and \ref{sec:4.4} respectively. Some comments on the double copy structures that show up are made in Section \ref{sec:4.5}.

In Section \ref{sec:5} we discuss and survey in detail four directions of future research beyond the ideas presented in this work.

In Appendix \ref{app}, we provide a detailed review of the conventional approach to soft theorems in terms of conserved charges due to asymptotic symmetries. We set up the essential language of Ward identities at null infinity in Appendix \ref{app:a1}. In Appendix \ref{app:a2} the relation of the leading soft photon theorem to the generators of large gauge transformations is detailed. The equivalence of supertranslation Ward identities to the leading soft graviton theorem is explained in Appendix \ref{app:a3}. We explain how the subleading soft photon theorem is derived from a Ward identity in Appendix \ref{app:a4}. Finally, in Appendix \ref{app:a5} we review the Ward identities due to the subleading and subsubleading soft graviton theorems.
\end{outline}
\vfill
\begin{acknowledgements}\normalfont
I would like to thank Nima Arkani-Hamed, Shamik Banerjee, Jacob Bourjaily, Eduardo Casali, Claude Duhr, Sudip Ghosh, Raghav Govind Jha, Alok Laddha, Alfredo Guevara, Lorenzo Magnea, Andrew McLeod, Akavoor Manu, Yasha Neiman, Monica Pate, Radu Roiban, Suraj Shankar, Vasudev Shyam and Cristian Vergu for discussions on various topics related to the present work. Research in the Bourjaily group is supported in part by an ERC Starting Grant (No. 757978) and a grant from the Villum Fonden (No. 15369).
\end{acknowledgements}
\clearpage
\input{sec2correlators}
\clearpage
\input{sec3asymptotic}
\clearpage
\input{sec4beyond}

\clearpage
\input{sec5conclusion}
\clearpage
\appendix
\titleformat{name=\section}[display]
{\normalfont}
{\footnotesize\textsc{Appendix \thesection}}
{0pt}
{\Large\bfseries}
[\vspace{-10pt}\color{lapis}\rule{\textwidth}{0.6pt}]
\input{sec6appendix}
\clearpage
\addcontentsline{toc}{section}{References}
\bibliographystyle{JHEP}
\bibliography{main}
\end{document}

%% file: sec1intro.tex
\section{Introduction}\label{sec:1}
The boundary structure of spacetime has for several decades provided a rich avenue of research. Indeed, this has been the case not only in the study of gravity, but in the study of quantum fields in curved backgrounds, specifically those backgrounds which are asymptotically flat. The study of these spacetimes and the specifics of massless fields propagating on them goes back a long way, to seminal work of Penrose et al. \cite{Penrose:1962sjs,Penrose:1962ij,Penrose:1964ge,Penrose:1965am,Penrose:1980yx, Newman:1968uj}. In particular, it turns out that the properties of massless fields on such backgrounds are especially interesting in the asymptotic regime, or in other words, when studied on the conformal boundaries\footnote{We use the term conformal boundary to mean the null boundary of asymptotically flat spacetimes. The conformal boundary is built out of the end points of propagating null geodesics, and is labelled by a choice of null coordinate and sphere variables $(z,\z)$.} of asymptotically flat backgrounds.

A result of particular interest of the study of such systems is a hierarchy of conserved quantities in QED and linearized gravity, studied by Newman and Penrose \cite{Newman:1968uj} on the conformal boundary. In the case of QED, the first of these conservation laws corresponds to the conservation of electric charge, with the higher conserved charges encode multipole charges. In the case of gravity, the hierarchy of conservation laws starts with the conservation of mass on the conformal boundary.

Independent of the study of such conservation laws was the realization that the group of symmetries of an asymptotically flat spacetime was far richer than the expected Poincar\'e group. It was observed by several authors \cite{Bondi:1962px,PhysRev.128.2851} that the group of symmetries preserving the structure of the metric on the conformal boundary actually corresponded to a class of transformations that translated the null boundary in a local fashion. Working in retarded coordinates $(u,z,\z)$, these so called \emph{supertranslations} act as

\begin{equation}
    u \longrightarrow u + f(z,\z),
\end{equation}
where $f$ is an arbitrary function on the Riemann surface $\mathbb{CP}^{1}$. This group of symmetries, now known as the Bondi-Metzner-Sachs (BMS) group turns out to be the true symmetry group of asymptotically flat spacetimes.

These results are interesting in their own right, shedding quite a bit of light on how gravity fundamentally alters the global structure of even those spacetimes that are flat at infinity. However, at the time they were first discovered, it wasn't clear that they were directly related to any results already known in the context of quantum field theory. Intuitively however one might expect that such asymptotic phenomena would have implications for the infrared properties of QFTs with massless properties, which arise due to long-distance interactions mediated by massless quanta. 

It turned out that this was indeed the case; in a number of papers (among others) by several authors \cite{Strominger:2013jfa,Strominger:2013lka,He:2014laa,He:2014cra,Campiglia:2014yka,Strominger:2014pwa,Kapec:2014zla,Pasterski:2015tva,He:2015zea,Campiglia:2015qka,Campiglia:2015kxa,Campiglia:2016jdj,Campiglia:2016hvg,Campiglia:2016efb,Kapec:2016jld,He:2017fsb,Kapec:2017tkm,Laddha:2017vfh,Kapec:2017gsg,Kapec:2017tkm,H:2018ktv,Campiglia:2018dyi,He:2019jjk,He:2019pll,Campiglia:2019wxe,He:2019ywq,Barnich:2019vzx}, it was understood that there was an intimate web of relations between the triad of conserved quantities, asymptotic symmetries and the infrared dynamics of gauge theory and gravity. Specifically, it was shown that the conserved quantities controlled asymptotic symmetries on the conformal boundary; in the case of QED they gave rise to large gauge symmetries, while they generated the BMS group in the case of gravity. Moreover, the conserved charges naturally implied Ward identities at the level of the $S$-matrix, which appropriately regarded ended up being equivalent the soft theorems in QED \cite{PhysRev.52.54,PhysRev.76.790,Yennie:1961ad,Jauch:1976ava,Weinberg:1964ew,Weinberg:1965nx} and gravity \cite{Weinberg:1964ew,Weinberg:1965nx,DeWitt:1967uc} already well-understood by field theorists\footnote{This web of relations has been treated fairly extensively, but the literature is somewhat scattered. We provide a detailed review of these ideas in Appendix \ref{app}.}. This triality has generated much interest for another reason however, namely the possibility that a detailed study of it might guide the effort towards a concrete realization of flat space holography. 

In its most precise form, an explicit expression of flat space holography at the level of the $S$-matrix for massless particles would amount to prescribing an intrinsically two-dimensional representation thereof. More precisely, each external state involved in the scattering process would have to be mapped to an operator $O_{\Delta_{i}}(z_{i},\z_{i})$\footnote{The $\Delta_{i}$ are used in generic fashion, to indicate parameters in the dual models such as conformal weights (in the event that the dual two-dimensional theory is a CFT).} with the operator product expansions between these operators defined so that one has

\begin{equation}
    \bra{\rom{out}}\mathcal{S}\ket{\rom{in}} = \ab{O_{\Delta_{1}}(z_{1},\z_{1})\dots O_{\Delta_{n}}(z_{n},\z_{n})},
\end{equation}
which amounts to a fundamentally two-dimensional definition of the originally four-dimensional $S$-matrix. Experience suggests that this is likely a herculean task, as exact dynamical dualities between four-dimensional and two-dimensional theories are generically difficult to realize\footnote{We remark that formal two-dimensional observables can be defined for gauge theories and gravity by carefully treating the corresponding fields at null infinity. In the specific case of gravity, such an analysis seems to indicate that gravity is somehow naturally holographic (see \cite{Laddha:2020kvp} and references therein for further details).}. Accordingly, insofar as the issue of flat space holography can be dealt with, a more productive approach perhaps is to focus on specific dynamical sectors of the theories in question, rather than the entire models. It is in this context that the recent results on soft theorems are encouraging. Indeed, the following is a natural question - is it possible to recover the infrared behaviour of gauge theory and gravity as direct dynamical consequences of two-dimensional models, rather than by having recourse to their full field theory presentations? Recently in two articles \cite{Kalyanapuram:2020epb,Kalyanapuram:2021bvf} by the present author, progress was made in answering this question, and the current work constitutes a systematic and comprehensive extension of the results therein. Throughout the work, our emphasis will be on studying these models for QED and gravity, making more often parenthetical remarks about non-Abelian gauge theories.

Starting with the case of quantum electrodynamics, we recall the famous result due to Yennie, Frautschi and Suura \cite{Yennie:1961ad}, who observed that in the eikonal approximation soft divergences due to the exchange of low energy quanta between external legs exponentiate. When all the external particles are massless, the result of this exponentiation is that the bare scattering amplitude $\mathcal{M}^{\rom{QED}}_{0}$ is adjusted according to the relation

\begin{equation}
    \mathcal{M}^{\rom{QED}} = \mathcal{A}_{n,\rom{soft}}^{\rom{QED}}\mathcal{M}^{\rom{QED}}_{0},
\end{equation}
where the function $\mathcal{A}_{n,\rom{soft}}^{\rom{QED}}$ in dimensional regularization ($d = 4+\epsilon$) is expanded as

\begin{equation}
    \ln\left(\mathcal{A}^{\rom{QED}}_{n,\rom{soft}}\right) = -\frac{1}{8\pi^{2}\epsilon}\sum_{i\neq j}e_{i}e_{j}\ln|z_{i}-z_{j}|^{2},
\end{equation}
where the $e_{i}$ are charges of the external states and the variables $(z_{i},\z_{i})$, labelling directions on $\mathbb{CP}^{1}$, are due to the following expansion of massless momenta

\begin{equation}
    p_{i} = \omega_{i}(1+z_{i}\z_{i},z_{i}+\z_{i},-i(z_{i}-\z_{i}),1-z_{i}\z_{i}).
\end{equation}
In performing eikonal exponentiation, this soft factor is computed by evaluating an infinite set of Feynman diagrams and carrying out a resummation. The form of the logarithm however suggests that it should arise out of a specific two-dimensional quantum field theory, namely that of the Coulomb gas. Specifically, it is sufficient to notice that given an action of the form

\begin{equation}\label{eq:1.6}
    \mathcal{I} = \int dz\wedge d\z \left(D_{z}\phi(z,\z)D_{\z}\phi(z,\z)\right),
\end{equation}
the correlation function between the conformal primary operators of such a theory has precisely the form required to obtain the soft exponent. It only requires one to correctly specify the conformal weights to obtain the desired result. We see as a consequence of this observation that soft exponentiation in QED can be rewritten as a dynamical consequence of free bosons on the celestial sphere. 

A similar phenomenon of soft exponentiation is known to hold in the case of gravity, and is exact to all-loop order in the eikonal approximation \cite{Weinberg:1965nx,DeWitt:1967uc,Naculich:2011ry}. It turns out that the right model to describe this exponentiation is a generalization of the Coulomb gas model of the general form

\begin{equation}\label{eq:1.7}
    \mathcal{I} = \int dz\wedge d\z \left(D^{2}_{z}\sigma(z,\z)D^{2}_{\z}\sigma(z,\z)\right),
\end{equation}
which is the action controlling fields that obey the biharmonic equation on the celestial sphere\footnote{It is known that such actions appear in the description of a class of topological defects in two dimensions known as crystal dislocations and disclinations.}. 

The dynamical origins of these actions end up being closely related to the Ward identities that yield soft theorems in QED and gravity. Specifically, it is known that these Ward identities hold at every point on $\mathbb{CP}^{1}$, which in analytic terms means that the corresponding conserved charges are labelled by an arbitrary function $f(z,\z)$ on $\mathbb{CP}^{1}$. In the case of QED, this conserved charge (generating large gauge transformations) remains invariant under the shift

\begin{equation}\label{eq:1.8}
    f(z,\z) \longrightarrow f(z,\z) + c
\end{equation}
for any constant $c$ while the corresponding global symmetry in the case of gravity is a four parameter shift

\begin{equation}\label{eq:1.9}
    f(z,\z) \longrightarrow f(z,\z) + c_{1} + c_{2}z + c_{3}\z + c_{4}z\z.
\end{equation}
Indeed, the shift \mref{eq:1.8} is precisely the global symmetry enjoyed by the action in \mref{eq:1.6} while the shift \mref{eq:1.9} is the global transformation under which the action \mref{eq:1.7} stays invariant. This tells us that these asymptotic actions correspond to the behaviour of smearing functions defining the Ward identities when they are allowed to become dynamical. We discuss this correspondence in detail in Appendix \ref{app}. 

These global symmetries however provide another upshot, namely in deriving the \emph{soft theorems} as Ward identities. Soft theorems, fixed entirely by Lorentz
invariance and unitarity, tell us how scattering amplitudes are modified by the \emph{radiation} of real soft particles. In the case of our two-dimensional models, they appear more naturally; the global symmetry of \mref{eq:1.6} yields the following current\footnote{We have listed the currents that lead to the positive helicity soft theorems; the negative helicity counterparts are obtained by the replacement $D_{z}\rightarrow D_{\z}$}

\begin{equation}
    j_{\rom{QED}}(z) = D_{z}\phi(z,\z),
\end{equation}
while from those of \mref{eq:1.7} we find

\begin{equation}
    j_{\rom{grav}}(z,\z) = D_{z}^{2}\sigma(z,\z). 
\end{equation}
Inserting these currents into the correlation functions that compute the soft $S$-matrix supplies precisely the soft theorems of the respective four-dimensional duals, namely QED and gravity. In other words, the soft $S$-matrix is a dynamical output of our two-dimensional models, while the soft theorems are realized as Ward identities. We discuss these dual models and the extensions required to understand multiple soft emissions in Section \ref{sec:2}. We also comment briefly on how these models must be refined to account for infrared behaviour in Yang-Mills in the same section.

The fact that the soft $S$-matrices and soft theorems at leading order in QED and gravity are recovered so naturally in terms of two-dimensional models leads to a natural question - is it possible to derive once again the triality between asymptotic symmetries, soft theorems and conserved quantities using the language of these models? This question is studied in Section \ref{sec:3}, where we extract the generators of asymptotic symmetries corresponding to the soft theorems in terms of the currents $j_{\rom{QED}}$, $j_{\rom{grav}}$ and their conjugate counterparts on $\mathbb{CP}^{1}$.

On the conformal boundary, large gauge transformations are known to act multiplicatively, where their infinitesimal actions on boundary operators (denoted generically by $\mathcal{O}$) take the form

\begin{equation}
    [j_{\rom{QED},p},\mathcal{O}(z_{0},\z_{0})] = ie z^{p}_{0}\mathcal{O}(z_{0},\z_{0})),
\end{equation}
where $e$ is the charge of the operator $\mathcal{O}$ (the action due to antiholomorphic generators are analogous). These generators may be recast in terms of two-dimensional fields by making use of the fact that they are in fact the coefficients of the mode expansion of the soft current. Specifically, we find that

\begin{equation}
    j_{\rom{QED},p} = \frac{1}{2\pi i}\oint_{c_{z_{0}}}z^{p}D_{z}\phi(z,\z)dz,
\end{equation}
and

\begin{equation}
    \overline{j}_{\rom{QED},p} = \frac{1}{2\pi i}\oint_{c_{z_{0}}}\z^{p}D_{\z}\phi(z,\z)d\z,
\end{equation}
where $c_{z_{0}}$ is a small counterclockwise circle around the point $z_{0}$. This realizes large gauge transformations as modes of the soft currents, which in turn are Noether currents of the dual two-dimensional model, completing the original triality directly on the celestial sphere. We also derive the generators of large gauge transformations for Yang-Mills theory in terms of two-dimensional fields and verify that they obey the expected commutation relations.

The same analysis is repeated for the case of gravity and the BMS group, although it is more involved. The BMS group has a specific presentation on the celestial sphere in terms of a collection of generators denoted by $P_{p,q}$ ($p$ and $q$ are integers), which have the effect of translating the retarded coordinate according to

\begin{equation}
    P_{p,q}: u \longrightarrow u + z^{p+1}\z^{p+1}.
\end{equation}
It is upon specialising to a specific subalgebra generated by the elements $\lbrace{P_{-1,q},P_{0,q},P_{p,-1},P_{p,0}\rbrace}$\footnote{These generators form a closed algebra when combined with the generators of Lorentz rotations.} that the leading soft graviton is obtained as a Ward identity.  Indeed, we see that it is precisely this subset of generators that can be recast as modes over specific combinations of derivatives of the two-dimensional field $\sigma(z,\z)$. Specifically, we will demonstrate that we have in terms of the soft graviton current $j_{\rom{grav}}(z,\z)$ the following

\begin{equation}
    P_{p,-1} = \frac{1}{2\pi i}\oint_{c_{z_{0}}}z^{p}\left(D_{\z}j_{\rom{grav}}(z,\z)\right)dz,
\end{equation}
and
\begin{equation}
    P_{p,0} = \frac{1}{2\pi i}\oint_{c_{z_{0}}}z^{p}\left(j_{\rom{grav}}(z,\z) - \z D_{\z}j_{\rom{grav}}(z,\z)\right)dz,
\end{equation}
such that $P_{-1,q}$ and $P_{0,q}$ are defined by making the replacements $D_{z}\rightarrow D_{\z}$ and $\z\rightarrow z$.

It's possible to leverage the form of the two-dimensional models to make concrete statements about soft theorems beyond leading order as well. To do so, we exploit two distinct phenomena already observed at leading order. First, deriving the soft theorems from the dual models proceeds according to a split already observed in the formulation in terms of Ward identities. Schematically, given a conserved charge $Q$, the commutation of which with the $S$-matrix is known to imply a soft theorem, it is always possible to decompose $Q$ into two parts, referred to as soft and hard pieces respectively. The soft part actually creates the infrared particle, while the hard piece acts on the hard external states and supplies the soft factor. This decomposition is mapped into the dual models by identifying the soft part of the charge with the Noether current of the model. The hard part is related to the exponential operators whose correlator is the soft $S$-matrix. 

Second, the soft factors in QED and gravity obey specific identities which immediately suggest generalization. In the case of QED, the reader will note that we have

\begin{equation}\label{eq:1.18}
    D_{\z}\frac{1}{z-z'} = \pi\dfunction,
\end{equation}
while the analogous identity in the case of gravity takes the form

\begin{equation}
    D^{2}_{\z}\frac{(\z-\z')}{z-z'} = \pi\dfunction.
\end{equation}
In each case, the soft factor is essentially a Green's function - of the holomorphic differential operator $D_{\z}$ in QED and of the second degree operator $D_{\z}^{2}$ in gravity. A natural generalization of these transforms is furnished by the function defined as

\begin{equation}
    \Delta_{N}(z,\z) = \frac{1}{(N-1)!}\frac{(\z-\z')^{N-1}}{z-z'}.
\end{equation}
Indeed, it can be established by induction that $\Delta_{N}$ is the Green's function of $D_{z}^{N}$. The case of $N=1$ is just \mref{eq:1.18}; assuming the case for $N-1$ we have

\begin{equation}
    D_{\z}\Delta_{N}(z,\z) = \frac{1}{(N-2)!}\frac{(\z-\z')^{N-2}}{z-z'} + 0,
\end{equation}
where the latter zero arises due to the delta function appearing when $D_{\z}$ acts on the denominator. The right side is the the Green's function of $D_{\z}^{N-1}$, completing the induction.

This observation comes in handy when we recall that it is precisely expressions of this form that one encounters in dealing with soft theorems in QED and gravity beyond leading order. Accordingly, in Section \ref{sec:4}, we consider the following higher order soft theorems in turn by making use of the preceding identities - the subleading soft theorem in QED, the subleading soft theorem in gravity and the subsubleading soft theorem in gravity. More precisely, we find that the higher order soft theorems are defined in terms of dual higher derivative theories, generalizing the dual models used at leading order\footnote{A similar idea was suggested in the discussion section of \cite{Pasterski:2021dqe}.}

For the subleading soft theorem in QED, the highest power of $N$ required is $2$. It is known due to studies of the Ward identity corresponding to this theorem that the conserved charges are labelled by vector fields on the celestial sphere. Informed by the duality between the smearing function and fields in the leading order theorems, we suggest that it is the following action that acts as the dual two-dimensional description of the subleading soft theorem in QED

\begin{equation}
    \mathcal{I}^{(1)}_{\rom{QED}} = -\int\left(D^{2}_{z}V^{z}_{1}(z,\z)D^{2}_{\z}V^{z}_{2}(z,\z) + D^{2}_{z}V^{\z}_{1}(z,\z)D^{2}_{\z}V^{\z}_{2}(z,\z)\right)dz\wedge d\z,
\end{equation}
where $(V^{A}_{1},V^{A}_{2})$ are a pair of vector fields on the sphere. The following Noether currents of this model, given by 

\begin{equation}
    j^{V}(z,\z) = D_{z}^{2}V^{z}_{1}(z,\z),
\end{equation}
and its conjugate obtained by replacing $D_{z}\rightarrow D_{\z}$ and $z\rightarrow \z$ play the role of soft currents. For the right choice of dressing operators\footnote{We provide the full expressions for these dressing operators in Section \ref{sec:4}.}, we find that inserting this current into a correlation function of such operators yields precisely the soft photon theorem at subleading order.

Finding the two-dimensional model for the soft graviton theorem at subleading order is now a matter of generalizing the approach used in the case of QED. We first observe that the highest value of $N$ required is $3$, in accordance with which we expect a theory containing kinetic terms of sixth order. Furthermore, we have once again the fact that asymptotic charges encoding the subleading soft graviton theorem are labelled by vector fields. Putting these together, we find that the action given by

\begin{equation}
    \mathcal{I}^{(1)}_{\rom{grav}} = \int \left(D^{3}_{z}Y^{z}_{1}(z,\z)D^{3}_{\z}Y^{z}_{2}(z,\z)+ D^{3}_{z}Y^{\z}_{1}(z,\z)D^{3}_{\z}Y^{\z}_{2}(z,\z)\right)dz\wedge d\z
\end{equation}
turns out to be the correct choice. Indeed, this time, the positive helicity soft charge is prescribed by the Noether current

\begin{equation}
    j^{Y}(z,\z) = D^{3}_{z}Y^{z}_{1}(z,\z),
\end{equation}
with the negative helicity case obtained the usual way. A properly defined dressing operator is shown to correctly yield the subleading soft graviton theorem upon insertion of the soft current.

The subsubleading soft graviton theorem is studied using two-dimensional models as a final case. Here, after noting that $N$ takes the maximum value of $3$, one expects a kinetic term of eighh order. This time however, making note of the fact that the smearing operator in this case is known to be a second rank, symmetric tensor on the sphere, we consider the action given by 

\begin{equation}
    \mathcal{I}^{(2)}_{\rom{grav}} = \int \left(D^{4}_{z}X^{zz}_{1}(z,\z)D^{4}_{\z}X^{zz}_{2}(z,\z) + D^{4}_{z}X^{\z\z}_{1}(z,\z)D^{4}_{\z}X^{\z\z}_{2}(z,\z)\right)dz\wedge d\z,
\end{equation}
where $(X^{AB}_{1},X^{AB}_{2})$ this time denotes a pair of second rank symmetric tensors (with off diagonal elements set to zero). Keeping track of the global symmetries of this model results in the use of the following positive helicity soft current

\begin{equation}
    j^{X}(z,\z) = D^{4}_{z}X^{zz}_{1}(z,\z),
\end{equation}
and its negative helicity analogue. When inserted into a correlation function of dressing operators defined in accordance with the corresponding hard charge, the subsubleading soft graviton current is recovered as a direct consequence. 

Just as the soft theorems at leading order are a consequence of conservation laws on null infinity, there are an infinity of conservation laws which may be inferred from the soft theorems are higher order. Now there are several ways of going about this, and a detailed description of these conservation laws in a traditional analysis is given in the appendix. However, in Section \ref{sec:4} we chose to adopt a somewhat different approach along the lines of \cite{Banerjee:2019prz,Banerjee:2020kaa,Banerjee:2020vnt,Banerjee:2020zlg,Banerjee:2021cly}. In these works, the soft theorems (say for positive helicity radiation) were expanded in powers of $\z$, and generators of asymptotic symmetries were extracted as modes of the corresponding coefficients.

To repeat this analysis, we perform the following decomposition of the soft operators for the higher order soft theorems. For the subleading theorem in QED for example, we have

\begin{equation}
    j^{V}(z,\z) = J^{V}(z) + \z K^{V}(z).
\end{equation}
Deriving this decomposition ultimately amounts to finding general solutions of the equations obeyed by the vector fields $Y^{A}_{a}$ on the celestial sphere, and isolating out terms proportional to differing powers of $\z$. For example, in the preceding case, the term proportional to $\z$ is obtained by noting that a further derivative along $\z$ is enough to isolate it. We have accordingly the definitions

\begin{equation}
    K^{V}(z) = D_{\z}j^{V}(z,\z),
\end{equation}
and

\begin{equation}
    J^{V}(z) =j^{V}(z,\z) - \z D_{\z}j^{V}(z,\z).
\end{equation}
This process can be carried out iteratively for the more complicated cases that we encounter in gravity. Indeed, for the subleading soft theorem in gravity we find that

\begin{equation}
    j^{Y}(z,\z) = L^{Y}(z) + 2\z J^{Y}(z) + \z^{2}K^{Y}(z),
\end{equation}
while for the soft current of the subsubleading theorem in gravity the decomposition takes the form

\begin{equation}
    j^{X}(z,\z) = M^{X}(z) + \z L^{X}(z) + \z^{2}K^{X}(z) + \z^{3}J^{X}(z).
\end{equation}
In addition to identifying the various holomorphic contributions to the soft theorems at higher order by making these decompositions, further contact with \cite{Banerjee:2019prz,Banerjee:2020kaa,Banerjee:2020vnt,Banerjee:2020zlg,Banerjee:2021cly} is made by computing the modes of these currents. We observe in the course of these calculations one aspect not considered by these latter authors. When the soft currents are broken up into holomorphic pieces, the decomposition process leads to a proliferation of a large number of contact terms. Generically, when working in the non-collinear limit, these contact terms vanish. Additionally, they do not affect the mode operators as they do not contribute to the corresponding contour integrals. However, in the collinear limit, they do damage holomorphicity. Accordingly, all statements made in relation to the existing literature about the modes should be understood keeping this caveat in mind.

In the course of deriving the dual models on the celestial sphere for soft theorems, there is a hint of underlying double copy structures relating the models in QED to those in gravity. In the simplest case of the leading order soft theorems, the correspondence is most precise; the action and currents for the leading order theorem in gravity are obtained by simply performing a squaring of all derivative operators present in the relevant expressions in QED. In light of the fact that the literature on the double copy is rather vast, and that the main thrust of this work is on soft theorems, we restrict our comments on the possible interpretation of these double copy structures to the short subsection \ref{sec:4.5} at the end of Section \ref{sec:4}.

In the main body of this work, unless stated otherwise, it is to be understood that we are working in the so-called regulated Mellin basis (of which we provide a brief review in Section \ref{sec:2}). When expressed in this basis, scattering amplitudes in massless quantum field theories are more naturally seen to be valued on the conformal boundary, and it is accordingly easier to make statements about null infinity in a more self contained fashion. 

In the appendix, where we provide an extensive review and summary of the relations between soft theorems and Ward identities that are known at the time of writing, we have chosen to make use instead of the momentum basis. Wherever a relation needs to be drawn between quantities expressed in two different basis, we have indicated that a basis change must be performed.

%% file: sec2correlators.tex
\section{Soft Divergences from Two-Dimensional Correlators}\label{sec:2}
In this section, we study the leading order soft theorems in gauge theory and gravity, first maintaining a focus on QED on the gauge theory side and considering the subtleties that come with Yang-Mills theory next. Due to the specific divergent structure of the leading order soft theorems, we employ factorization theorems to construct two-dimensional models that are conjectured to completely capture the leading order soft dynamics.

We remark that most of this section has been written as a global review of the work done by the present author in \cite{Kalyanapuram:2020epb,Kalyanapuram:2021bvf} for those readers unfamiliar with these papers. As mentioned in the introduction, in the main body of the paper we will be making statements that apply to scattering amplitudes in the so-called Mellin basis, alternatively known as the boost eigenbasis. Accordingly, let us begin with a brief overview of the conventions that we will be using before getting into the details of how we construct two-dimensional models for soft theorems.

\subsection{Infrared Divergences in Gauge Theory and Gravity}\label{sec:2.1}
Let us start with the simplest case of scattering amplitudes in quantum electrodynamics\footnote{Historically, QED provided an especially convenient laboratory to study infrared divergences (see \cite{PhysRev.52.54,Chung:1965zza,Kibble:1968sfb,Kibble:1969ip,Kibble:1969ep,Kibble:1969kd,Kulish:1970ut,Steinmann:1973pla}) due to the relatively simple analytic structure of its $S$-matrix.}. Consider a scattering amplitude in QED involving $n$ particles, which we denote by $\mathcal{A}^{\rom{QED}}_{n}$. Due to the masslessness of the photon, scattering amplitudes in QED are naturally divergent. The two sources of these infinities are virtual exchanges of soft photons between external legs and the radiation of low energy photons given an energy threshold $\Delta E$. In particular, the universality of the virtual contribution is captured by factorization \cite{Collins:1988ig,Collins:1989bt,Collins:1989gx,Feige:2013zla,Feige:2014wja}, due to which we have\footnote{There is more than one way (commanding differing degrees of support based on the audience) to define the soft part of the $S$-matrix. One way, which we used in the introduction, is by making use of the eikonal theorem and exponentiation along the lines of \cite{Yennie:1961ad,Weinberg:1965nx,DeWitt:1967uc,Gatheral:1983cz}. We have chosen to make use of factorization along the lines of \cite{Collins:1988ig,Collins:1989bt,Collins:1989gx,Sen:1982bt,Sterman:2002qn,Dixon:2008gr,Kidonakis:1998nf,Aybat:2006mz,Feige:2013zla,Feige:2014wja} with the eikonal assumption. This way, the soft contribution is isolated from the collinear piece, while giving us room to enjoy the advantages of exponentiation as well.}

\begin{equation}
    \mathcal{A}^{\rom{QED}}_{n} = \mathcal{A}^{\rom{QED}}_{n,\rom{soft}}\times \mathcal{H}^{\rom{QED}}_{n}.
\end{equation}
In this factorized representation, the quantity $\mathcal{A}^{\rom{QED}}_{n,\rom{soft}}$ carries all divergences which arise out of exchanges of soft particles. $\mathcal{H}^{\rom{QED}}_{n}$ is known as the \emph{hard} $S$-matrix, and is infrared finite. It may contain collinear divergences, which are themselves known to factorize as well\footnote{The resolution of a scattering amplitude into soft, collinear and hard sectors can be done systematically using soft-collinear effective theory (SCET). The reader may refer to \cite{Bauer:2000ew,Bauer:2000yr,Bauer:2001ct,Bauer:2001yt,Beneke:2002ph,Beneke:2002ni,Hill:2002vw,Cheung:2009sg,Becher:2011dz,Chiu:2012ir,Becher:2014oda}.}. When soft and collinear divergences are factored out, the remaining hard contribution is finite throughout phase space, and can be defined using asymptotic operators \cite{Hannesdottir:2019opa,Hannesdottir:2019rqq}.

Notably, we have chosen to not present the factorized amplitude including soft radiation due to the following reason. It is well known that soft radiation and soft exchanges interact in a manner that eliminates their infrared divergent contributions. However, in a strict sense the scattering amplitude in the presence of soft radiation is not quite an $n$ particle amplitude, but rather an $n+r$ particle amplitude when $r$ soft particles are radiated\footnote{I would like to thank Lorenzo Magnea for emphasizing this point.}. For the time being, we will employ this representation, and consider the issue of soft radiation later. Accordingly, the scattering amplitude as we have defined is manifestly divergent. The divergence is easily expressed in dimensional regularization; defining $\epsilon = d - 4$, we have

\begin{equation}
    \ln\left(\mathcal{A}^{\rom{QED}}_{n,\rom{soft}}\right) = -\frac{1}{8\pi^{2}\epsilon}\sum_{i\neq j}\eta_{i}\eta_{j}e_{i}e_{j}\ln|z_{i}-z_{j}|^{2},
\end{equation}
where the $e_{i}$ are the electric charges of the external states. Naturally, when one of the external states is a photon, the corresponding $e_{i}$ vanishes. The $z_{i}$ label locations of the massless external states on the celestial sphere, and indeed, this representation provides the first hint that the soft part of the $S$-matrix is perhaps naturally holographic. 

This formalism is readily generalized to the case of gravity. As Weinberg showed in \cite{Weinberg:1964ew,Weinberg:1965nx}, infrared divergences in gravity are universal at leading order. Factorization \cite{Akhoury:2011kq} once again leads to the following form of an $n$ particle amplitude $\mathcal{A}^{\rom{grav}}_{n}$ in gravity

\begin{equation}
    \mathcal{A}^{\rom{grav}}_{n} = \mathcal{A}^{\rom{grav}}_{n,\rom{soft}}\times \mathcal{H}^{\rom{grav}}_{n},
\end{equation}
where the meanings of the various terms are clear by analogy, with the qualification that $\mathcal{H}^{\rom{grav}}_{n}$ contains no collinear divergences\footnote{This can be shown by writing gravity in the soft-collinear expansion \cite{Beneke:2012xa,Okui:2017all,Chakraborty:2019tem}.} at leading order \cite{Weinberg:1965nx,DeWitt:1967uc,Akhoury:2011kq} - which is another way of saying that there are no gravitational jets.

The soft part of the gravitational $S$-matrix also admits of a simple exponential representation, which takes the form

\begin{equation}
    \ln\left(\mathcal{A}^{\rom{grav}}_{n,\rom{soft}}\right) = -\frac{\kappa^{2}}{8\pi^{2}\epsilon}\sum_{i<j}\eta_{i}\eta_{j}\omega_{i}\omega_{j}|z_{i}-z_{j}|^{2}\ln|z_{i}-z_{j}|^{2},
\end{equation}
where $\kappa = \sqrt{8\pi G}$ is the gravitational coupling constant and the $\omega_{i}$ represent the energies of the external states. 

Note that we have made use of the signature variables $\eta_{i}$, which take the values $\pm 1$ for incoming and outgoing states respectively. Alternatively, the some of the charges/energies can be made to assume negative values, but we refrain from using this representation.

The reader will notice simply by inspection that the soft $S$-matrix for gravity is not holographic in the strongest sense of the word. Indeed, it depends manifestly not only on the positions of the external states on the celestial sphere, but also on the energies of those particles. It turns out that there is a particular basis change, known as the Mellin transform, which has the effect of recasting this presentation into one that is more naturally thought of as being holographic. This transformation, first considered in \cite{Pasterski:2016qvg,Pasterski:2017ylz} takes an amplitude $\mathcal{A}_{n}$ in momentum space and transforms it according to

\begin{equation}
    \widetilde{\mathcal{A}}_{n}(\lbrace{\Lambda_{i},z_{i},\z_{i}\rbrace}) = \int_{0}^{\infty} \mathcal{A}_{n}(\lbrace{\omega_{i},z_{i},\z_{i}\rbrace})\prod_{i=1}^{n}\omega_{i}^{\Lambda_{i}}d\omega_{i},
\end{equation}
which eliminates the dependence on the energies $\omega_{i}$ in favour of a dependence on the weights $\Lambda_{i}$. Convergence issues make this transformation best suited for UV complete theories like Yang-Mills. A refinement of the Mellin transform that yields instead a three dimensional holographic presentation was provided in \cite{Banerjee:2018fgd,Banerjee:2018gce} and applied to a wide class of problems in \cite{Banerjee:2019aoy,Banerjee:2019tam,Banerjee:2019prz,Banerjee:2020kaa,Banerjee:2020zlg,Banerjee:2020vnt}. It takes the form

\begin{equation}
    \widetilde{\mathcal{A}}_{n}(\lbrace{\Lambda_{i},u_{i},z_{i},\z_{i}\rbrace}) = \int_{0}^{\infty} \mathcal{A}_{n}(\lbrace{\omega_{i},z_{i},\z_{i}\rbrace})\prod_{i=1}^{n}\omega_{i}^{\Lambda_{i}}e^{i\eta_{i}u_{i}\omega_{i}}d\omega_{i},
\end{equation}
where the $u_{i}$ denote retarded time parameters, which are regulated (after having reinstated the signature variables $\eta_{i}$) according to $u_{i}\rightarrow u_{i}+i\eta_{i}\delta$ for positive and infinitesimal $\delta$. 

To understand how to proceed in describing holographic models for the soft $S$-matrices, we should then first ask what happens to the soft factors in particular when the Mellin transform is applied. For QED, we observe that

\begin{equation}\label{eq:2.7}
\begin{aligned}
   & \int_{0}^{\infty} \mathcal{A}^{\rom{QED}}_{n}(\lbrace{\omega_{i},z_{i},\z_{i}\rbrace})\prod_{i=1}^{n}\omega_{i}^{\Lambda_{i}}e^{i\eta_{i}u_{i}\omega_{i}}d\omega_{i} = \\
   & \mathcal{A}^{\rom{QED}}_{n,\rom{soft}}(\lbrace{z_{i},\z_{i}\rbrace})\times  \int_{0}^{\infty}\mathcal{H}^{\rom{QED}}_{n}(\lbrace{\omega_{i},z_{i},\z_{i}\rbrace})\prod_{i=1}^{n}\omega_{i}^{\Lambda_{i}}e^{i\eta_{i}u_{i}\omega_{i}}d\omega_{i}.
\end{aligned}
\end{equation}
Clearly from the preceding expression, the soft factor factorizes even in the Mellin basis with no qualitative change. This is simply due to the fact that the soft factor has no dependence on the energies of the external states. An essentially similar phenomenon is observed for the case of gravity, with the following factorization taking place

\begin{equation}
\begin{aligned}
   & \int_{0}^{\infty} \mathcal{A}^{\rom{grav}}_{n}(\lbrace{\omega_{i},z_{i},\z_{i}\rbrace})\prod_{i=1}^{n}\omega_{i}^{\Lambda_{i}}e^{i\eta_{i}u_{i}\omega_{i}}d\omega_{i} = \\
   & \mathcal{A}^{\rom{grav}}_{n,\rom{soft}}\left(\lbrace{-i\eta_{i}\partial_{u_{i}},z_{i},\z_{i}\rbrace}\right)\times  \int_{0}^{\infty}\mathcal{H}^{\rom{grav}}_{n}(\lbrace{\omega_{i},z_{i},\z_{i}\rbrace})\prod_{i=1}^{n}\omega_{i}^{\Lambda_{i}}e^{i\eta_{i}u_{i}\omega_{i}}d\omega_{i},
\end{aligned}
\end{equation}
where

\begin{equation}
      \ln\left(\mathcal{A}^{\rom{grav}}_{n,\rom{soft}}\left(\lbrace{-i\eta_{i}\partial_{u_{i}},z_{i},\z_{i}\rbrace}\right) \right) = \frac{\kappa^{2}}{8\pi^{2}\epsilon}\sum_{i<j}|z_{i}-z_{j}|^{2}\ln|z_{i}-z_{j}|^{2}\partial_{u_{i}}\partial_{u_{j}}.
\end{equation}
Accordingly, factorization continues to hold even for the gravitational case, with the qualification that the soft factor is now operator valued. It is clear from the analytic forms of the soft factors in QED and gravity that they are in a literal sense holographic. In the case of QED, the soft $S$-matrix is only dependent on the locations of the external states on $\mathbb{CP}^{1}$. On the other hand, the soft factor in gravity is dependent both on the punctures, but has the effect of translating the hard $S$-matrix along null infinity. This is the expected behaviour as we shall soon see, since the BMS algebra arises as a natural consequence of this phenomenon. 

With all of this information in hand, our task is to recast these soft factors as correlation functions of fields defined intrinsically on the celestial sphere, which we will now consider.

\subsection{Two-Dimensional Models for Soft Factors}\label{sec:2.2}
To prescribe two-dimensional models that capture the soft factors in some intrinsically holographic fashion (a notion that we will make precise very soon), it is best to phrase the entire problem in more familiar language by making reference to the very well understood idea of a \emph{dressed asymptotic state} (see \cite{Chung:1965zza,Kibble:1968sfb,Kibble:1969ip,Kibble:1969ep,Kibble:1969kd,Kulish:1970ut,Anupam:2019oyi} for examples of state dressing in QED and QCD). In the case of QED, let us take the quantity

\begin{equation}
  \widetilde{\mathcal{H}}^{\rom{QED}}_{n}(\lbrace{\Lambda_{i},u_{i},z_{i},\z_{i}\rbrace}) = \int_{0}^{\infty}\mathcal{H}^{\rom{QED}}_{n}(\lbrace{\omega_{i},z_{i},\z_{i}\rbrace})\prod_{i=1}^{n}\omega_{i}^{\Lambda_{i}}e^{i\eta_{i}u_{i}\omega_{i}}d\omega_{i},
\end{equation}
and suppose that there exist asymptotic operators $\mathcal{O}_{\Lambda_{i}}(u_{i},z_{i},\z_{i})$ on null infinity, whose operator product expansions are defined so as to ensure that the following prescription holds

\begin{equation}\label{eq:2.11}
  \widetilde{\mathcal{H}}^{\rom{QED}}_{n}(\lbrace{\Lambda_{i},u_{i},z_{i},\z_{i}\rbrace}) = \lab{\mathcal{O}_{\Lambda_{1}}(u_{1},z_{1},\z_{1})\dots \mathcal{O}_{\Lambda_{n}}(u_{n},z_{n},\z_{n})}.
\end{equation}
We emphasize that this particular prescription is conjectural; finding a genuine representation of this form would require the explicit construction of the corresponding boundary theory. The simplest way to do so would be simply taking boundary values of the bulk fields and reading off the right operator product expansions by matching the needed amplitudes. Preferable of course would be the construction of a dual boundary model. We won't elaborate on this point further, but it would certainly make for an interesting study.

To present the right holographic duals for the soft $S$-matrix, we ask how we can leverage what we already know about the representation of soft theorems in conventional quantum field theory. The simplest way that this is done is by using Wilson lines and using them to dress the Hamiltonian. Accordingly, the right way to proceed is probably to dress the external operators $\mathcal{O}_{\Lambda_{i}}$ by operators which themselves yield the soft $S$-matrix when the expectation value is taken. Now the question to answer is - what is the two-dimensional theory that we need to use to define the Wilson lines?

To answer this, we note the following identity

\begin{equation}\label{eq:2.12}
    D_{z}D_{\z}\ln|z-z'|^{2} = \pi\delta^{2}(z-z',\z-\z').
\end{equation}
The operator $D_{z}D_{\z}$ is simply the Laplacian on $\mathbb{CP}^{1}$ - in other words the kinetic operator for a free boson on the celestial sphere. This tells us that if we consider the following conformal model

\begin{equation}\label{eq:2.13}
    S^{}_{\rom{QED}} = -8\pi \epsilon \int D_{z}\phi^{}(z,\z)D_{\z}\phi^{}(z,\z) dz \wedge d\z,
\end{equation}
on the celestial sphere, the operator product expansion

\begin{equation}
    \ab{\phi^{}(z,\z)\phi^{}(z',\z')} = \frac{1}{8\pi^2\epsilon}\ln|z-z'|^{2}
\end{equation}
can be readily inferred. We have of course used the convention

\begin{equation}
    dz\wedge d\z = \frac{dzd\z}{2i}.
\end{equation}
The theory defined by equation \mref{eq:2.13} is the theory of a single free boson in two dimensions, and is well known to be a conformal field theory called the Coulomb gas model. Originally found by Dotsenko and Fateev \cite{Dotsenko:1984ad,Dotsenko:1984nm}, it has since been applied variously to topological models \cite{Kosterlitz_1973}, string theory \cite{Polyakov:1981rd,Polchinski:1998rq,Polchinski:1998rr} and to the study of conformal field theory more generally - see \cite{DiFrancesco:639405} for a review. 

Most interesting to us is the fact that specifically, conformal primaries are characterised by exponential operators of the form

\begin{equation}
    \mathcal{W}^{}(z,\z,\lambda) = \exp(i\lambda\phi^{}(z,\z)),
\end{equation}
which supply the correct two point function when we evaluate

\begin{equation}
    \lab{\mathcal{W}^{}(z,\z,\lambda)\mathcal{W}^{}(z',\z',\lambda')} = \exp(-\lambda\lambda'\ln|z-z'|^{2}) = \frac{1}{|z-z'|^{2\lambda\lambda'}}.
\end{equation}
Guided by this, we introduce the following definition of dressed asymptotic fields

\begin{equation}
    \widetilde{\mathcal{O}}_{\Lambda_{i}}(u_{i},z_{i},\z_{i}) = \mathcal{W}^{}(z_{i},\z_{i},\eta_{i}e_{i})\mathcal{O}_{\Lambda_{i}}(u_{i},z_{i},\z_{i}).
\end{equation}
With this, the evaluation of the correlator in \mref{eq:2.11} using instead dressed fields gives us

\begin{equation}
\begin{aligned}
    &\lab{\widetilde{\mathcal{O}}_{\Lambda_{1}}(u_{1},z_{1},\z_{1})\dots \widetilde{\mathcal{O}}_{\Lambda_{n}}(u_{n},z_{n},\z_{n})} =\\ &\mathcal{A}^{\rom{QED}}_{n,\rom{soft}}(\lbrace{z_{i},\z_{i}\rbrace})\times\lab{\mathcal{O}_{\Lambda_{1}}(u_{1},z_{1},\z_{1})\dots \mathcal{O}_{\Lambda_{n}}(u_{n},z_{n},\z_{n})}, 
\end{aligned}
\end{equation}
which upon making the identification according to \mref{eq:2.11} yields the expression in \mref{eq:2.7}. 

To briefly comment on what this implies, we see that the leading order factorization of scattering amplitudes in QED implies that the asymptotic theory capturing the scattering amplitudes of QED can be split into two distinct pieces - the putative theory giving rise to the hard part, while the soft $S$-matrix is correctly captured by the simple free boson CFT in two dimensions. Now we turn to the issue of gravity.

We state at the outset that the same idea of dressing by what amounts to a Wilson line works for the gravitational case, with minor changes and a pleasant upshot. Let us first note the fact that the identity \mref{eq:2.12} is generalised to

\begin{equation}
    D_{z}^{2}D_{\z}^{2}\left(|z-z'|^{2}\ln|z-z'|^{2}\right) = \pi\delta(z-z',\z-\z'),
\end{equation}
which tells us of course that the building block of the soft factor in gravity is simply the Green's function of the operator $(D_{z}D_{\z})^{2}$, which is just the biharmonic operator in two dimensions. Now, generalising the QED construction is relatively straightforward, since we observe that the operator product expansion

\begin{equation}
    \ab{\sigma^{}(z)\sigma^{}(z')} = \frac{1}{8\pi^{2}\epsilon}|z-z'|^{2}\ln|z-z'|^{2}
\end{equation}
is recovered from the action

\begin{equation}
    S^{}_{\rom{grav}} = 8\pi\epsilon\int D_{z}^{2}\sigma^{}(z,\z)D_{\z}^{2}\sigma^{}(z,\z)dz\wedge d\z.
\end{equation}
This action was first considered as a dual model for the soft graviton $S$-matrix by the author in \cite{Kalyanapuram:2020epb} (and extended in \cite{Kalyanapuram:2021bvf} to deal with real soft emissions, which we will soon discuss). It was found more recently in \cite{Nguyen:2021qkt} that this action can be directly derived by finding the dynamics of soft modes of the Einstein-Hilbert action on the celestial boundary of Minkowski space. 

We note that this particular form of the action is not manifestly covariant on $\mathbb{CP}^{1}$. However, bringing it into a more covariant form is not especially difficult (or especially illuminating). On the other hand, this form makes a particular property, namely a kind of double copy rather explicit, so we keep it in this form. 

We now consider again, in analogy to the way we proceeded in QED, the quantity

\begin{equation}
    \widetilde{\mathcal{H}}^{\rom{grav}}_{n}(\lbrace{\Lambda_{i},u_{i},z_{i},\z_{i}\rbrace})= \int_{0}^{\infty}\mathcal{H}^{\rom{grav}}_{n}(\lbrace{\omega_{i},z_{i},\z_{i}\rbrace})\prod_{i=1}^{n}\omega_{i}^{\Lambda_{i}}e^{i\eta_{i}u_{i}\omega_{i}}d\omega_{i},
\end{equation}
and assert the existence of asymptotic operators $G_{\Lambda_{i}}(u_{i},z_{i},\z_{i})$ that live on future null infinity as usual and the equality

\begin{equation}\label{eq:2.24}
  \widetilde{\mathcal{H}}^{\rom{grav}}_{n}(\lbrace{\Lambda_{i},u_{i},z_{i},\z_{i}\rbrace}) = \lab{G_{\Lambda_{1}}(u_{1},z_{1},\z_{1})\dots G_{\Lambda_{n}}(u_{n},z_{n},\z_{n})}.
\end{equation}
The dressing in done by defining asymptotic operators as

\begin{equation}
    \widetilde{G}_{\Lambda_{i}}(u_{i},z_{i},\z_{i}) = \mathcal{W}_{\rom{grav}}^{}(z_{i},\z_{i},\kappa\eta_{i}\omega_{i},u_{i})G_{\Lambda_{i}}(u_{i},z_{i},\z_{i}),
\end{equation}
where

\begin{equation}
    \mathcal{W}_{\rom{grav}}^{}(z,\z,\lambda,u) = \exp(\lambda\sigma^{}(z,\z)\partial_{u}).
\end{equation}
It's of course clear from these expressions that the expectation value

\begin{equation}
    \lab{\mathcal{W}^{}_{\rom{grav}}(z,\z,\lambda,u)\mathcal{W}^{}_{\rom{grav}}(z',\z',\lambda',u)} = \exp(\lambda\lambda'|z-z'|^{2}\ln|z-z'|^{2}\partial_{u}\partial_{u'})
\end{equation}
holds. The first observation from this is that the theory that controls soft modes in gravity is manifestly not conformal. This is clearly relevant in determining the validity of the conjecture that gravitational theories are dynamically equivalent to boundary conformal field theories, a point which will receive further consideration in the conclusion.

Making use of the preceding equation, the correlation function of the $\mathcal{W}_{\rom{grav}}(z_{i},\z_{i},\eta_{i}\omega_{i},u_{i})$ is exactly $\mathcal{A}^{\rom{grav}}_{n,\rom{soft}}$; indeed, we have

\begin{equation}
    \begin{aligned}
    &\lab{\widetilde{G}_{\Lambda_{1}}(u_{1},z_{1},\z_{1})\dots \widetilde{G}_{\Lambda_{n}}(u_{n},z_{n},\z_{n})} =\\
    &\mathcal{A}^{\rom{grav}}_{n,\rom{soft}}\left(\lbrace{-i\eta_{i}\partial_{u_{i}},z_{i},\z_{i}\rbrace}\right)\times \lab{G_{\Lambda_{1}}(u_{1},z_{1},\z_{1})\dots G_{\Lambda_{n}}(u_{n},z_{n},\z_{n})},
    \end{aligned}
\end{equation}
correctly reproducing the factorized form of scattering amplitudes in gravity in the modified Mellin basis.

So we have shown that the divergent part of the $S$-matrix in QED and gravity coming from virtual transitions of soft states between external legs is computed equivalently by models which are defined intrinsically on the two-dimensional space $\mathbb{CP}^{1}$. In particular, the theories are essentially free field theories, which are curiously related to one another by the simple replacements

\begin{equation}
    D_{z}\longrightarrow D_{z}^{2},
\end{equation}
and
\begin{equation}
    D_{\z}\longrightarrow D_{\z}^{2}.
\end{equation}
We simply state here for the moment that this is immediately suggestive as being a manifestation of a putative double copy structure between the soft sectors of gauge theory and gravity. We will have more, predominantly speculative, points to discuss regarding this matter in a later section. For now, we move on to studying how the soft theorems due to real soft emissions can be understood in terms of two-dimensional models.

\subsection{Soft Theorems in QED and Gravity}\label{sec:2.3}
In this short section, our main task will be to provide a review of the analytic structure of soft theorems in gauge theory and gravity, once again maintaining an emphasis on QED on the gauge theory side. Specifically, we will recall how the soft theorems are manifested in the celestial basis, namely when the energies are transformed away into the Mellin basis and the momenta are written in terms of coordinates on $\mathbb{CP}^{1}$. In an attempt to be pedagogical, we will first consider the necessary expressions in the momentum basis and then map them into the Mellin basis.

Taking first the case of QED, we start with the $n$ particle scattering amplitude $\mathcal{A}_{n}^{\rom{QED}}$ and ask what happens when the external states (taken to all be electrons or positrons for the sake of simplicity) are allowed to radiate one particle that is of very low energy. Specifically, we require that the radiated energy has energy that is of very low magnitude with respect to the energies of the external states. In such a circumstance, Lorentz invariance and unitarity dictate the form of the new amplitude, which we denote by $\mathcal{A}_{n+\rom{soft}}^{\rom{QED}}$, to be\footnote{We have chosen to instate a factor of $i$ to facilitate comparison with the two-dimensional model that will be shown to reproduce the soft theorem. This choice is relatively unimportant physically (due to cross sections being squared moduli of scattering amplitudes), but will help us stay unconcerned with keeping track of differing factors.}

\begin{equation}
   \mathcal{A}_{n+\rom{soft}}^{\rom{QED}} = \left(\sum_{i=1}^{n}i\eta_{i}e_{i}\frac{p_{i}\cdot \epsilon_{q}}{p_{i}\cdot q}\right)\mathcal{A}_{n}^{\rom{QED}}. 
\end{equation}
Here, $q$ is the momentum of the radiated photon, with the obvious requirement that

\begin{equation}
    q^{0} \ll p^{0}_{i}
\end{equation}
for every external state (labelled by $i$), and $\epsilon_{q}$ is the polarization vector of the radiated photon. Now if we are to move into the celestial basis, the first step is to rewrite the leading soft photon factor, given by 

\begin{equation}
    S_{\rom{QED}}^{(0)} = \sum_{i=1}^{n}i\eta_{i}e_{i}\frac{p_{i}\cdot \epsilon_{q}}{p_{i}\cdot q},
\end{equation}
in terms of coordinates on the celestial sphere. This is a simple exercise; for a positive helicity emission we obtain

\begin{equation}
    S_{\rom{QED}}^{(0),+} = \frac{1}{\omega_{q}}\sum_{i=1}^{n}\frac{i\eta_{i}e_{i}}{z-z_{i}},
\end{equation}
and for a negative helicity state

\begin{equation}
    S_{\rom{QED}}^{(0),-} = \frac{1}{\omega_{q}}\sum_{i=1}^{n}\frac{i\eta_{i}e_{i}}{\z-\z_{i}},
\end{equation}
where $z$ is defined according to the resolution

\begin{equation}
    q = \omega_{q}(1+z\z,z+\z,-i(z-\z),1-z\z).
\end{equation}
Since the soft factor is manifestly independent of the energies of the external states, it simply factors out when the Mellin transform is effected, with one qualification. The soft factor contains a factor of the energy of the radiated photon and is formally divergent. In accordance with this, we define a soft current by introducing

\begin{equation}
    \mathcal{S}^{(0)}_{\rom{QED}} = \lim_{\omega_{q}\rightarrow 0}\omega_{q}S^{(0)}_{\rom{QED}},
\end{equation}
which eliminates the dependence on the energy of the soft state. Our task in the celestial context will then be to determine the soft current in terms of two-dimensional fields. 

This construction goes through for gravity in much the same way. If we take an $n$ particle amplitude $\mathcal{A}_{n}^{\rom{grav}}$ in gravity and ask what happens when the external states are permitted to radiate one soft graviton, the result is fixed entirely analogously by Lorentz invariance and unitarity to give

\begin{equation}
   \mathcal{A}_{n+\rom{soft}}^{\rom{grav}} = \left(\sum_{i=1}^{n}i\kappa\eta_{i}\frac{(p_{i}\cdot \epsilon_{q})^{2}}{p_{i}\cdot q}\right)\mathcal{A}_{n}^{\rom{grav}}. 
\end{equation}
where it is evident that the polarization tensor for an external graviton state is built out of symmetrizing two polarization vectors. The form of the soft factor in the celestial basis is again obtained by direct computation, and we have

\begin{equation}
   \mathcal{S}^{(0),+}_{\rom{grav}} = \sum_{i}^{n}i\kappa\eta_{i}\omega_{i}\frac{\z-\z_{i}}{z-z_{i}},
\end{equation}
for radiation that is of positive helicity and with a negative helicity soft graviton giving a soft factor 

\begin{equation}
   \mathcal{S}^{(0),-}_{\rom{grav}} = \sum_{i}^{n}i\kappa\eta_{i}\omega_{i}\frac{z-z_{i}}{\z-\z_{i}}.
\end{equation}
Notably, in contradistinction to the QED case, the soft current in gravity is manifestly dependent on the energies of the excited states. In the Mellin transformed basis, we naturally obtain the representations

\begin{equation}
   \widetilde{\mathcal{S}}^{(0),+}_{\rom{grav}} = \sum_{i}^{n}\kappa\frac{\z-\z_{i}}{z-z_{i}}\partial_{u_{i}},
\end{equation}
and

\begin{equation}
   \widetilde{\mathcal{S}}^{(0),-}_{\rom{grav}} = \sum_{i}^{n}\kappa\frac{z-z_{i}}{\z-\z_{i}}\partial_{u_{i}}.
\end{equation}

The reader will have noticed the correspondence between results for positive and negative helicity radiation; once a soft theorem is obtained for the radiation of a positive helicity particle, the corresponding negative helicity result is obtained by taking the complex conjugate of all the celestial coordinates. Due to this fact, we will in the following (for the most part) only state results for positive helicity emission, with the corresponding results for negative radiation either implicit or understood to be obtained by complex conjugation. 

Having reviewed the essential facts regarding the leading order soft theorems in QED and gravity, we turn our attention to answering the question we had posed earlier - namely, is it possible to derive the soft theorems as a consequence of symmetries of the two dimension theories we had put forth in the preceding section? As it turns out, we can affirmatively answer this question.

\subsection{Soft Theorems as Ward Identities of Two-Dimensional Models}\label{sec:2.4}
The soft current in the QED is most easily derived by recognizing the symmetries enjoyed by the two-dimensional theory in \mref{eq:2.13}. Specifically, we observe that the action is invariant under the shift symmetry

\begin{equation}
    \phi^{} \longrightarrow \phi^{} + a,
\end{equation}
where $a$ is a constant. In accordance with Noether's theorem, this symmetry leads to two currents and one conserved quantity. Indeed, there are two currents, which are related by complex conjugation (suggesting of course in advance the corresponding relation between positive and negative helicities); we have for the holomorphic current

\begin{equation}
    j_{\rom{QED}}^{}(z) = 8\pi \epsilon D_{z}\phi^{}(z,\z),
\end{equation}
With the current defined in this fashion, the natural next step would be to ask the effect an insertion of one instance of the current would have on the correlators that give rise to scattering amplitudes (working now in the Mellin basis). To answer this, it is best to observe first that due to the fact that we have

\begin{equation}
    \ab{j_{\rom{QED}}^{}(z)\phi^{}(z')} = \frac{8\pi\epsilon}{z-z'},
\end{equation}
we find that the operator product expansion with the Wilson line takes the form

\begin{equation}
    \lab{j_{\rom{QED}}^{}(z)\mathcal{W}^{}_{\rom{QED}}(z',\z',\lambda)} = \frac{i\lambda}{z-z'}\mathcal{W}^{}_{\rom{QED}}(z',\z',\lambda).
\end{equation}
When the current is inserted into a correlation function involving Wilson lines, Wick's theorem can be used repeatedly by making application of the preceding equation to find that we obtain

\begin{equation}
    \begin{aligned}
    &\lab{j_{\rom{QED}}^{}(z)\widetilde{\mathcal{O}}_{\Lambda_{1}}(u_{1},z_{1},\z_{1})\dots \widetilde{\mathcal{O}}_{\Lambda_{n}}(u_{n},z_{n},\z_{n})} =\\
    &\left(i\sum_{i=1}^{n}\frac{\eta_{i}e_{i}}{z-z_{i}}\right)\lab{\widetilde{\mathcal{O}}_{\Lambda_{1}}(u_{1},z_{1},\z_{1})\dots \widetilde{\mathcal{O}}_{\Lambda_{n}}(u_{n},z_{n},\z_{n})}.
    \end{aligned}
\end{equation}
When we have that the correlation function is interpreted as a scattering amplitude for $n$ external particles, what we have is just the celestial presentation of the soft theorem when a single soft photon of positive helicity is radiated. The fact that the photon can be emitted by any of the external states is reflected by the sum in the preceding equation.

In hindsight, the conserved charge can be inferred fairly easily; we only have to note here that both the holomorphic and antiholomorphic currents must obey the global residue theorem. Indeed, this is obviously seen to be the case consistently; the global residue theorem is equivalent to the requirement of conservation of charge. For the sake of being thorough, the conserved charge is simply obtained by the following contour integral around the point at infinity

\begin{equation}
    Q_{\rom{QED}} = \int_{c_{\infty}}D_{z}\phi^{}(z,\z)dz,
\end{equation}
which can be easily checked to simply supply a sum over the quantity $\eta_{i}e_{i}$. The vanishing of the residue at infinity is charge conservation.

When it comes to gravity, we can proceed in two different ways. Either we can follow the route taken in the QED case and find the currents corresponding to these symmetries. A more amusing approach however is to employ the double copy structure alluded to before, wherein we simply square the derivative operators and make a field change $\phi^{}\longrightarrow \sigma^{}$. From this, we have two new currents. The first is

\begin{equation}
    j^{}_{\rom{grav}}(z,\z) = 8\pi\epsilon D^{2}_{z}\sigma^{}(z,\z)
\end{equation}
such that the other current is obtained by making the replacement $D_{z}\rightarrow D_{\z}$. In passing, we note that it is notationally useful to continue to refer to the conjugate current as $\overline{j}^{}_{\rom{grav}}$, although it is really not the complex conjugate of $j^{}_{\rom{grav}}$. Unlike the QED case, the currents in gravity are not holomorphic or antiholomorphic, at least not in a na\"ive sense. This becomes relevant when we will discuss asymptotic symmetries in the next chapter. 

We observe that this time, due to the fact that we have the identity

\begin{equation}
    \ab{j^{}_{\rom{grav}}(z,\z)\sigma^{}(z')} = 8\pi\epsilon\frac{\z-\z'}{z-z'}
\end{equation}
(which really is another way of defining the Green's function of the operator $D_{\z}^{2}$, parenthetically), the Wilson line is acted on by the current according to the operator product expansion

\begin{equation}
    \lab{j_{\rom{grav}}^{}(z,\z)\mathcal{W}^{}_{\rom{grav}}(z',\z',\lambda,u_{i})} = i\lambda\frac{\z-\z'}{z-z'}\partial_{u'}\mathcal{W}^{}_{\rom{grav}}(z',\z',\lambda,u_{i}).
\end{equation}
Once again, this is leveraged in collaboration with a repeated application of Wick's theorem in a correlation function of dressed states to tell us that one insertion of the current into the correlator yields

\begin{equation}
    \begin{aligned}
    &\lab{j_{\rom{grav}}^{}(z,\z)\widetilde{G}_{\Lambda_{1}}(u_{1},z_{1},\z_{1})\dots \widetilde{G}_{\Lambda_{n}}(u_{n},z_{n},\z_{n})} =\\
    &\left(\sum_{i=1}^{n}\frac{\z-\z_{i}}{z-z_{i}}\partial_{u_{i}}\right)\lab{\widetilde{G}_{\Lambda_{1}}(u_{1},z_{1},\z_{1})\dots \widetilde{G}_{\Lambda_{n}}(u_{n},z_{n},\z_{n})},
    \end{aligned}
\end{equation}
which is identified as the soft theorem due to the production of one infrared graviton of positive helicity. Indeed, the negative helicity result is very easily recovered by the simple replacement of the derivative operator with its conjugate. Post facto, we can now see that the currents for soft graviton radiation are the Noether currents due to the invariance of the action controlling the soft modes under the following four parameter deformation of the field $\sigma^{}$.

\begin{equation}
    \sigma^{}(z,\z) \longrightarrow \sigma^{}(z,\z) + a_{0} + a_{1}z + a_{2}\z + a_{3}z\z,
\end{equation}
where the $a_{i}$'s are constants. These symmetries are the two-dimensional representation of the translational part of the Poincar\'e group, and are equivalent to insisting that total four momentum be conserved. 

While this derivation of the leading soft theorem for the emission of a single soft photon or graviton works very well, it fails when we want to derive the soft theorem due to more than one soft emission. We can see this by evaluating the correlator of dressed fields (both in the QED and gravity cases) after the insertion of two, rather than one, soft currents. In the case of QED we have

\begin{equation}
    \begin{aligned}
    &\lab{j_{\rom{QED}}^{}(z)j_{\rom{QED}}^{}(z')\widetilde{\mathcal{O}}_{\Lambda_{1}}(u_{1},z_{1},\z_{1})\dots \widetilde{\mathcal{O}}_{\Lambda_{n}}(u_{n},z_{n},\z_{n})} =\\
    &\mathcal{J}_{\rom{QED}}^{(2)}(z,z')\lab{\widetilde{\mathcal{O}}_{\Lambda_{1}}(u_{1},z_{1},\z_{1})\dots \widetilde{\mathcal{O}}_{\Lambda_{n}}(u_{n},z_{n},\z_{n})}
    \end{aligned},
\end{equation}
where

\begin{equation}
  \mathcal{J}_{\rom{QED}}^{(2)}(z,z') =  \left(i\sum_{i=1}^{n}\frac{\eta_{i}e_{i}}{z-z_{i}}\right)\left(i\sum_{i=1}^{n}\frac{\eta_{i}e_{i}}{z'-z_{i}}\right) + 8\pi\epsilon \frac{1}{(z-z')^{2}},
\end{equation}
with entirely analogous expressions making their appearance when two photons of other helicity configurations are assumed to be radiated\footnote{In the event that the photons are of opposite helicity, note that the term inhomogeneous in $\epsilon$ becomes a delta function.}. We see here that the first term which has no $\epsilon$ dependence is precisely the soft factor that is expected to show up when a conventional calculation is performed. The second term however has $\epsilon$ dependence and is unphysical, in that it contains a spurious double pole. 

We can repeat this calculation for the case of two graviton emissions as well. When we insert two currents for positive graviton radiation we obtain

\begin{equation}
    \begin{aligned}
    &\lab{j_{\rom{grav}}^{}(z,\z)j_{\rom{grav}}^{}(z',\z')\widetilde{G}_{\Lambda_{1}}(u_{1},z_{1},\z_{1})\dots \widetilde{G}_{\Lambda_{n}}(u_{n},z_{n},\z_{n})} =\\
    &\mathcal{J}_{\rom{grav}}^{(2)}(z,z',z',\z')\lab{\widetilde{G}_{\Lambda_{1}}(u_{1},z_{1},\z_{1})\dots \widetilde{G}_{\Lambda_{n}}(u_{n},z_{n},\z_{n})},
    \end{aligned}
\end{equation}
where

\begin{equation}
  \mathcal{J}_{\rom{grav}}^{(2)}(z,z') =  \left(\sum_{i=1}^{n}\frac{\z-\z_{i}}{z-z_{i}}\partial_{u_{i}}\right)\left(\sum_{i=1}^{n}\frac{\z'-\z_{i}}{z'-z_{i}}\partial_{u_{i}}\right) - 16\pi\epsilon \frac{\z-\z'}{(z-z')^{3}},
\end{equation}
where we encounter again a spurious pole, this time of order three. Parenthetically, when a positive and negative helicity current are inserted, the inhomogeneous term will once again be delta function valued.

We first point out that there is nothing especially problematic about these expressions; since we are to understand these correlation functions in the limit of vanishing $\epsilon$, the leading order terms in the corresponding expansions are precisely what we require. However, it is desirable to have expressions for multiple soft theorems which are exact in $\epsilon$ rather than approximate, and it turns out that the best way to achieve this is to consider a slightly modified version of the two-dimensional dual models that we have considered so far. The rest of this section will review this generalization, first considered by the author in \cite{Kalyanapuram:2021bvf}.

The first step to understanding the necessary modification is to understand the undesirable terms in the expressions we already have. This is clear - we need to get rid of the inhomogeneous terms, which themselves come out of contractions of the currents with each other. This is easy to resolve - all we have to do is introduce two, rather than one, fields such that one has a trivial operator product expansion with itself, making it possible to construct currents that do not contract with themselves. Taking the case of QED in particular, we need the following operator product expansions to hold, given two fields $\phi^{}_{1}$ and $\phi^{}_{2}$ -

\begin{equation}
    \ab{\phi^{}_{1}(z,\z),\phi^{}_{1}(z',\z')} = \frac{1}{8\pi^{2}\epsilon}\ln|z-z'|^{2},
\end{equation}

\begin{equation}
    \ab{\phi^{}_{1}(z,\z),\phi^{}_{2}(z',\z')} = \frac{1}{\pi}\ln|z-z'|^{2},
\end{equation}
and

\begin{equation}
    \ab{\phi^{}_{2}(z,\z),\phi^{}_{2}(z',\z')} = 0.
\end{equation}
The action from which such an operator product expansion can be derived is somewhat unusual; we need a theory of two scalar fields, but the conventional one that has a global $O(2)$ symmetry obviously won't work. More concretely, we will use the following model

\begin{equation}\label{eq:2.64}
    \mathcal{I}^{}_{\rom{QED}} = \int g^{ab} D_{z}\phi^{}_{a}(z,\z)D_{\z}\phi^{}_{b}(z,\z) dz \wedge d\z,
\end{equation}
from which we see that

\begin{equation}
    \lab{\phi^{}_{a}(z,\z)\phi^{}_{b}(z',\z')} = (g^{-1})_{ab}\frac{1}{\pi}\ln|z-z'|.
\end{equation}
It's straightforward to read off the elements of the tensor $g^{ab}$ by inspection; we have

\begin{equation}
    g^{11} = 0,
\end{equation}

\begin{equation}
    g^{12} = g^{21} = 1,
\end{equation}
and

\begin{equation}
    g^{22} = -\frac{1}{8\pi\epsilon}.
\end{equation}
Put this way, we now have to define the Wilson line operators and the currents with appropriate modifications. Specifically, we define the following quantities

\begin{equation}\label{eq:2.66}
    \mathcal{W}^{}_{a,\rom{QED}}(z,\z,\lambda) = \exp(i\lambda\phi^{}_{a}(z,\z)),
\end{equation}

\begin{equation}
    j_{a,\rom{QED}}^{}(z) = D_{z}\phi^{}_{a}(z,\z)
\end{equation}
with the conjugate current $\overline{j}_{a,\rom{QED}}^{}$ defined accordingly.

Finally, it only remains to be stated that the dressings of the external fields have to be done using $\mathcal{W}^{(1)}_{1,\rom{QED}}(z_{i},\z_{i},\eta_{i}e_{i})$ and the soft currents to be inserted to produce the soft theorems are the $j_{2,\rom{QED}}^{}$ and $\overline{j}_{2,\rom{QED}}^{}$ for positive and negative helicity radiation respectively. To illustrate this, we observe that for example

\begin{equation}
    \begin{aligned}
    &\lab{j_{2,\rom{QED}}^{}(z)j_{2,\rom{QED}}^{}(z')\widetilde{\mathcal{O}}_{\Lambda_{1}}(u_{1},z_{1},\z_{1})\dots \widetilde{\mathcal{O}}_{\Lambda_{n}}(u_{n},z_{n},\z_{n})} =\\
    &\left(i\sum_{i=1}^{n}\frac{\eta_{i}e_{i}}{z-z_{i}}\right)\left(i\sum_{i=1}^{n}\frac{\eta_{i}e_{i}}{z'-z_{i}}\right)\lab{\widetilde{\mathcal{O}}_{\Lambda_{1}}(u_{1},z_{1},\z_{1})\dots \widetilde{\mathcal{O}}_{\Lambda_{n}}(u_{n},z_{n},\z_{n})},
    \end{aligned}
\end{equation}
where it is to be understood that the operators are dressed using the first component of the Wilson lines defined in \mref{eq:2.66}. It should now be clear that the correct soft theorems are obtained for any number of soft emissions for any helicity. To repeat this for the case of gravity is simply to remark that we now have to use the action given by 

\begin{equation}\label{eq:2.73}
    \mathcal{I}^{}_{\rom{grav}} = \int g^{ab} D^{2}_{z}\sigma^{}_{a}(z,\z)D^{2}_{\z}\sigma^{}_{b}(z,\z) dz \wedge d\z,
\end{equation}
which implies the operator product expansions

\begin{equation}
    \ab{\sigma^{}_{1}(z,\z)\sigma^{}_{1}(z',\z')} = \frac{1}{8\pi^{2}\epsilon}|z-z'|^{2}\ln|z-z'|^{2},
\end{equation}

\begin{equation}
    \ab{\sigma^{}_{1}(z,\z)\sigma^{}_{2}(z',\z')} = \frac{1}{\pi}|z-z'|^{2}\ln|z-z'|^{2},
\end{equation}
and

\begin{equation}
    \ab{\sigma^{}_{2}(z,\z)\sigma^{}_{2}(z',\z')} = 0.
\end{equation}
By generalizing the Wilson lines and currents just as we have done in the QED case, the issue of the inhomogeneous term is again resolved, which is established by observing that

\begin{equation}
    \begin{aligned}
    &\lab{j_{2,\rom{grav}}^{}(z)j_{2,\rom{grav}}^{}(z')\widetilde{G}_{\Lambda_{1}}(u_{1},z_{1},\z_{1})\dots \widetilde{G}_{\Lambda_{n}}(u_{n},z_{n},\z_{n})} =\\
    &\left(\sum_{i=1}^{n}\frac{\z-\z_{i}}{z-z_{i}}\partial_{u_{i}}\right)\left(\sum_{i=1}^{n}\frac{\z'-\z_{i}}{z'-z_{i}}\partial_{u_{i}}\right)\lab{\widetilde{G}_{\Lambda_{1}}(u_{1},z_{1},\z_{1})\dots \widetilde{G}_{\Lambda_{n}}(u_{n},z_{n},\z_{n})},
    \end{aligned}
\end{equation}
where the notation should be clear from context; we have redefined the currents in terms of the new fields in analogy to the modifications for the QED amplitudes. 

\subsection{Application to Infrared Divergences in Yang-Mills Theory}\label{sec:2.5}
So far, we have conspicuously avoided a consideration of the issues of infrared divergences in Yang-Mills theory, having restricted our interest to QED on the gauge theory side. A priori, it may seem like the methods that we have employed are inapplicable to Yang-Mills, due to the complicated nature of its matrix-valued scattering amplitudes. Interestingly, with a little bit of care, it is possible to describe at least part of the soft $S$-matrix in Yang-Mills theory in terms of two-dimensional dual models. The rest of this section is devoted to a discussion of how to go about this - we note in particular that the first part of this entire discussion is essentially contained in \cite{Magnea:2021fvy,Gonzalez:2021dxw}, where the dual model was first proposed.

To construct the dual two-dimensional model for the soft sector of Yang-Mills, the nature of factorization in this theory has to first be described. Given a scattering amplitude involving $n$ gluons in Yang-Mills denoted by $\mathcal{A}^{\rom{YM}}_{n}$, the following factorization is known to occur

\begin{equation}
    \mathcal{A}^{\rom{YM}}_{n} = \mathcal{A}^{\rom{YM}}_{n,\rom{soft}}(\mu)\times\mathcal{A}^{\rom{YM}}_{n,\rom{soft-sing.}}(\mu)\times \mathcal{H}_{n}^{\rom{YM}}.
\end{equation}
Since this is different in appearance somewhat to the factorization structure in QED, each term requires some qualification. First, this factorization is valid in the so-called \emph{dipole approximation}, in which it is assumed that given a representation $R$, the cusp anomalous dimension $\gamma_{R}(\alpha_{YM})$ \cite{Dotsenko:1979wb,Brandt:1981kf,Korchemsky:1985xj,Korchemsky:1985xu,Korchemsky:1987wg} enjoys the splitting

\begin{equation}
    \gamma_{R}(\alpha_{YM}) = C^{(2)}_{R}\gamma(\alpha_{YM}),
\end{equation}
where $C^{(2)}_{R}$ is used to indicate the quadratic Casimir, and $\gamma(\alpha_{YM})$ is representation independent. It is known that this condition fails to hold beyond three loops \cite{Grozin:2017css,Moch:2018wjh,Henn:2019rmi,Henn:2019swt,vonManteuffel:2020vjv}. Under this assumption however, the factorization theorem may be used as-is, which leads us to the following definitions. Starting with the function $\mathcal{A}^{\rom{YM}}_{n,\rom{soft}}$, it is given by the following simple representation

\begin{equation}
    \ln\left(\mathcal{A}^{\rom{YM}}_{n,\rom{soft}}\right) = -\frac{K(\alpha_{YM},\epsilon)}{2}\sum_{i\neq j}T^{a}_{i}\cdot T^{a}_{j}\ln|z_{i}-z_{j}|^{2}, 
\end{equation}
where the $T^{a}_{i}$ are the colour factors of the external states, which we take for simplicity to always belong to the adjoint representation. The object $K(\alpha_{YM},\epsilon)$ is a particular function that depends on the strong coupling constant, which runs with the energy scale. It admits of an expansion in $\epsilon$, and is formally infrared divergent. 

The function $\mathcal{A}^{\rom{YM}}_{n,\rom{soft-sing.}}(\mu)$ comes from the singlet contribution to the soft anomalous dimension, and is independent of the positions $z_{i}$ and $\z_{i}$ on the celestial sphere. However, it does depend on the external energies. In fact, it can be shown that it is simply proportional to powers of the energies of the external states, given roughly by

\begin{equation}
    \mathcal{A}^{\rom{YM}}_{n,\rom{soft-sing.}}(\mu) \sim \prod_{i=1}^{n}\omega_{i}^{-\frac{1}{4}\gamma_{K}(\alpha_{YM})C^{(2)}_{A}}.
\end{equation}
We remark here that from this form of the singlet piece, the only effect that may be anticipated upon motion into the celestial basis is an overall redefinition of the weights $\Lambda_{i}$ that control the Mellin transform. Specfically, we have the following redefinition

\begin{equation}
    \Lambda_{i} \longrightarrow \Lambda_{i} - \frac{1}{4}\gamma_{K}(\alpha_{YM})C^{(2)}_{A} = \Lambda_{i}(\mu),
\end{equation}
which leads to the following amusing renormalization group relation

\begin{equation}
    \frac{d\Lambda_{i}}{d\ln\mu} = -\frac{C^{(2)}_{A}}{4}\gamma'_{K}(\alpha_{YM})\beta_{YM}(\mu).
\end{equation}
It is notable that in clear distinction to the situation in QED and gravity, Yang-Mills celestial amplitudes are manifestly scale-dependent, with the weights undergoing a renormalization group flow controlled precisely by the beta function.

It remains to be said now that there really isn't anything qualitatively new about the last piece, namely $\mathcal{H}_{n}^{\rom{YM}}$; it is just the hard piece of the scattering amplitude in Yang-Mills theory, and is by definition free of any infrared divergences. 

With all this in place, our task is now to construct a two-dimensional model that computes the scattering amplitude given by

\begin{equation}
    \widetilde{\mathcal{A}}^{\rom{YM}}_{n} = \mathcal{A}^{\rom{YM}}_{n,\rom{soft}}\times \widetilde{\mathcal{H}}^{\rom{YM}}_{n}\left(\lbrace{\Lambda_{i}(\mu),z_{i},\z_{i}\rbrace}\right).
\end{equation}
To do this, we once again conjecture a class of asymptotic operators, given this time by the formal symbols $\mathcal{O}_{\Lambda_{i}(\mu)}\left(u_{i},z_{i},\z_{i}\right)$ such that we have the following equivalence

\begin{equation}
    \lab{\mathcal{O}_{\Lambda_{1}(\mu)}\left(u_{1},z_{1},\z_{1}\right)\dots \mathcal{O}_{\Lambda_{n}(\mu)}\left(u_{n},z_{n},\z_{n}\right)} = \widetilde{\mathcal{H}}^{\rom{YM}}_{n}\left(\lbrace{\Lambda_{i}(\mu),z_{i},\z_{i}\rbrace}\right).
\end{equation}
Given this, we need to construct a two-dimensional model of soft gluons that can be used to dress these asymptotic operators and correctly compute the soft part of the $S$-matrix in the dipole approximation as an expectation value. At this point, the relevant model is quite naturally expected to be one of Lie-algebra valued scalars, which is precisely what was suggested in \cite{Magnea:2021fvy,Gonzalez:2021dxw}. However, we point out now, informed by the discussion in the preceding sections, that a model of a single flavour of Lie-algebra valued bosons will once again run into the problem of producing inhomogeneous terms when the double soft theorem has to be found. Accordingly, we consider the following theory

\begin{equation}
    \mathcal{I}^{}_{\rom{YM}} = - \int \gamma^{AB} D_{z}\phi^{a}_{A}(z,\z)D_{\z}\phi^{a}_{B}(z,\z)dz\wedge d\z.  
\end{equation}
In this expression, the index $a$ labels the colour, and is assumed to be traced over. The indices $A$ and $B$ denote the two flavours. Accordingly, the operator product expansion read off from this action is

\begin{equation}
    \ab{\phi^{a}_{A}(z,\z)\phi^{b}_{B}(z',\z')} = (\gamma^{-1})_{AB}\delta^{ab}\ln|z-z'|^{2}.
\end{equation}
Matching now the the requirements

\begin{equation}
    \ab{\phi^{a}_{1}(z,\z)\phi^{b}_{1}(z',\z')} = \frac{K(\alpha_{YM},\epsilon)}{2}\delta^{ab}\ln|z-z'|^{2},
\end{equation}

\begin{equation}
    \ab{\phi^{a}_{1}(z,\z)\phi^{b}_{1}(z',\z')} = \delta^{ab}\ln|z-z'|^{2},
\end{equation}
and
\begin{equation}
    \ab{\phi^{a}_{2}(z,\z)\phi^{b}_{2}(z',\z')} = 0,
\end{equation}
we infer that

\begin{equation}
\gamma^{AB}=\begin{pmatrix}
0 & 1\\ 
1 & -\frac{1}{2}K(\alpha_{YM},\epsilon).
\end{pmatrix}
\end{equation}
We can go ahead and define in analogy to the definitions in the previous sections, Wilson lines and currents, the latter of which are the Noether currents for shift symmetries, this time under $N_{c}$ constants rather than a single one, owing the the colour structure of the action. Indeed, we have

\begin{equation}\label{eq:2.92}
    \mathcal{W}^{}_{A,\rom{YM}}(z,\z) = \exp\left(i T^{a}\phi^{a}_{A}(z,\z)\right),
\end{equation}

\begin{equation}\label{eq:2.93}
    j_{A,\rom{YM}}^{a}(z) = D_{z}\phi^{a}_{A}(z,\z),
\end{equation}
and

\begin{equation}\label{eq:2.94}
    \overline{j}_{A,\rom{YM}}^{a}(\z) = D_{\z}\phi^{a}_{A}(z,\z).
\end{equation}
The dressing is performed as usual by defining

\begin{equation}
    \widetilde{\mathcal{O}}_{\Lambda_{i}(\mu)}\left(u_{i},z_{i},\z_{i}\right) = \exp\left(i T^{a}_{i}\phi^{a}_{1}(z_{i},\z_{i})\right)\mathcal{O}_{\Lambda_{i}(\mu)}\left(u_{i},z_{i},\z_{i}\right).
\end{equation}
Direct evaluation establishes that

\begin{equation}
    \begin{aligned}
    &\lab{\widetilde{\mathcal{O}}_{\Lambda_{1}(\mu)}\left(u_{1},z_{1},\z_{1}\right)\dots \widetilde{\mathcal{O}}_{\Lambda_{n}(\mu)}\left(u_{n},z_{n},\z_{n}\right)} = \\
    &\mathcal{A}^{\rom{YM}}_{n,\rom{soft}}\times \lab{\mathcal{O}_{\Lambda_{1}(\mu)}\left(u_{1},z_{1},\z_{1}\right)\dots \mathcal{O}_{\Lambda_{n}(\mu)}\left(u_{n},z_{n},\z_{n}\right)}.
    \end{aligned}
\end{equation}
The soft theorem due to soft gluon radiation can be obtained by one insertion of the currents $j_{2,\rom{YM}}^{a}(z)$ or $\overline{j}_{2,\rom{YM}}^{a}(\z)$, for positive or negative helicity radiation respectively. This is of course due to the operator product expansions

\begin{equation}
    \lab{j_{2,\rom{YM}}^{a}(z)\mathcal{W}^{}_{1,\rom{YM}}(z_{i},\z_{i})} = \frac{iT^{a}}{z-z_{i}}\mathcal{W}^{}_{1,\rom{YM}}(z_{i},\z_{i}),
\end{equation}
and

\begin{equation}
    \lab{\overline{j}_{2,\rom{YM}}^{a}(\z)\mathcal{W}^{}_{1,\rom{YM}}(z_{i},\z_{i})} = \frac{iT^{a}}{\z-\z_{i}}\mathcal{W}^{}_{1,\rom{YM}}(z_{i},\z_{i}),
\end{equation}
giving the right pole structure for the soft theorem in Yang-Mills theory. When we want to look at multiple soft emissions however, we run into one particular problem, namely the fact that eikonal factorization no longer holds. Note that in the case of say a double soft gluon emission, there are three Feynman diagrams that contribute. Of these, only two can be recovered in terms of a soft theorem in leading order factorization. Specifically, when two soft gluons are radiated (we assume that they are radiated from the same leg in the present discussion for the sake of brevity), the double soft factor becomes

\begin{equation}
    \mathcal{A}^{\rom{YM}}_{n+2\rom{soft}} = \left(\sum_{i}\left(\frac{T^{a}\epsilon_{q}\cdot p_{i}}{q\cdot p_{i}}\frac{T^{b}\epsilon_{q'}\cdot p_{i}}{(q+q')\cdot p_{i}}+\frac{T^{b}\epsilon_{q'}\cdot p_{i}}{q'\cdot p_{i}}\frac{T^{a}\epsilon_{q}\cdot p_{i}}{(q+q')\cdot p_{i}}\right)\right)\mathcal{A}^{\rom{YM}}_{n},
\end{equation}
(in the event that both gluons are emitted by one external leg), when the soft energies of the two gluons are comparable. This limit, known as the simultaneous soft limit, is correctly captured in conventional field theory by the insertion of a Wilson line constructed out of the gauge potential. However, it doesn't seem to be possible to recover this factorization from the two-dimensional model that we have considered in the case that the limit is simultaneous. However, when the limit is not simultaneous, but consecutive, in the sense that the energy of $q'$ is very small compared to that of $q$, the previous expression dramatically simplifies to give us (we take both to be of positive helicity),

\begin{equation}
   \lim_{\omega_{q'}\rightarrow 0}\omega_{q'}\lim_{\omega_{q}\rightarrow 0}\omega_{q} \mathcal{A}^{\rom{YM}}_{n+2\rom{soft}} = \left(\sum_{i,j}\frac{T^{b}}{z'-z_{j}}\frac{T^{a}}{z-z_{i}}\right)\mathcal{A}^{\rom{YM}}_{n},
\end{equation}
to leading order, as we have preserved the most divergent piece. It should be understood that we have suppressed several index variables; the particle indices $i$ and $j$ indicate that the colour matrices act on the amplitude by matrix multiplication along its respective indices. Now this can be extracted from our two-dimensional model by inserting two soft currents, through which the specific matrix structure of the soft theorem is made more explicit. Verifying this means observing first that we have, after the insertion of two soft currents (where we implicitly have the dressing of the external asymptotic operators in using $\mathcal{W}^{}_{1,\rom{YM}}(z_{i},\z_{i})$),

\begin{equation}
    \begin{aligned}
    &\lab{j_{2,\rom{YM}}^{b}(z')j_{2,\rom{YM}}^{a}(z)\widetilde{\mathcal{O}}_{\Lambda_{1}(\mu)}\left(u_{1},z_{1},\z_{1}\right)\dots \widetilde{\mathcal{O}}_{\Lambda_{n}(\mu)}\left(u_{n},z_{n},\z_{n}\right)} = \\
    &\sum_{i=1}^{n}\frac{1}{z-z_{i}}
    \lab{j_{2,\rom{YM}}^{b}(z')\widetilde{\mathcal{O}}_{\Lambda_{1}(\mu)}\left(u_{1},z_{1},\z_{1}\right)\dots T^{a}_{i}\widetilde{\mathcal{O}}_{\Lambda_{i}(\mu)}\left(u_{i},z_{i},\z_{i}\right)\dots \widetilde{\mathcal{O}}_{\Lambda_{n}(\mu)}\left(u_{n},z_{n},\z_{n}\right)},
    \end{aligned}
\end{equation}
which can be further simplified to yield a sum of two terms - 

\begin{equation}
    \begin{aligned}
   &\sum_{i=1}^{n}\frac{1}{z-z_{i}}
    \lab{j_{2,\rom{YM}}^{b}(z')\widetilde{\mathcal{O}}_{\Lambda_{1}(\mu)}\left(u_{1},z_{1},\z_{1}\right)\dots T^{a}_{i}\widetilde{\mathcal{O}}_{\Lambda_{i}(\mu)}\left(u_{i},z_{i},\z_{i}\right)\dots \widetilde{\mathcal{O}}_{\Lambda_{n}(\mu)}\left(u_{n},z_{n},\z_{n}\right)}\\
    &= \mathcal{C}_{\rom{double}} + \mathcal{C}_{\rom{contact}},
    \end{aligned}
\end{equation}
where
\begin{equation}
\begin{aligned}
    \mathcal{C}_{\rom{double}} = &\sum_{i\neq j}^{n}\bigg(\frac{1}{z'-z_{j}}\frac{1}{z-z_{i}}
    \bigg \langle \widetilde{\mathcal{O}}_{\Lambda_{1}(\mu)}\left(u_{1},z_{1},\z_{1}\right)\dots T^{a}_{i}\widetilde{\mathcal{O}}_{\Lambda_{i}(\mu)}\left(u_{i},z_{i},\z_{i}\right)\dots T^{b}_{j}\widetilde{\mathcal{O}}_{\Lambda_{j}(\mu)}\left(u_{j},z_{j},\z_{j}\right)\\
    &\dots\widetilde{\mathcal{O}}_{\Lambda_{n}(\mu)}\left(u_{n},z_{n},\z_{n}\right)\bigg\rangle\bigg),
\end{aligned}
\end{equation}
and
\begin{equation}
\begin{aligned}
    \mathcal{C}_{\rom{double}} =& \sum_{i=1}^{n}\bigg(\frac{1}{z'-z_{i}}\frac{1}{z-z_{i}}
    \bigg \langle\widetilde{\mathcal{O}}_{\Lambda_{1}(\mu)}\left(u_{1},z_{1},\z_{1}\right)\dots T^{b}_{i}T^{a}_{i}\widetilde{\mathcal{O}}_{\Lambda_{i}(\mu)}\left(u_{i},z_{i},\z_{i}\right)\\
    &\dots \widetilde{\mathcal{O}}_{\Lambda_{n}(\mu)}\left(u_{n},z_{n},\z_{n}\right)\bigg \rangle\bigg).
\end{aligned}
\end{equation}
We see that the correct form of the consecutive double soft theorem is reproduced by the introduction of two soft currents. The reader will observe that the divergent term due to the splitting of one soft gluon is not captured in this manner. In omitting this contribution, we have followed the convention laid out in \cite{Feige:2013zla,Feige:2014wja} at leading order factorization. The splitting of a soft gluon is dynamically controlled by the full theory, and is absorbed into the hard part of the $S$-matrix. The part of the $S$-matrix captured by soft factorization is then restricted to the kinds of terms we have presented here.

%% file: sec3asymptotic.tex
\section{Dual Models and Asymptotic Symmetries}\label{sec:3}
One insight obtained from the modern study of soft theorems in gauge theory and gravity has been to realize that the pole structure due to soft radiation is actually equivalent to the statement that scattering amplitudes in these theories are invariant under a class of symmetries that act asymptotically, on null infinity. Suggestively, such symmetries are referred to in the literature as \emph{asymptotic symmetries}. 

For QED, the relevant symmetries are alternatively known as large gauge transformations. Recall that in the case of QED, gauge transformations are represented by multiplying asymptotic operators (such as creation and annihilation operators) by an overall phase, in the case that the gauge transformations are global. Large gauge transformations differ from this in that they are local on null infinity, and depend on coordinates on the celestial sphere. Put concretely, given an asymptotic operator $\mathcal{O}$ in QED, a large gauge transformation has the effect

\begin{equation}
    \delta_{\varepsilon} \mathcal{O}(z,\z) = ie \varepsilon(z,\z)\mathcal{O}(z,\z),
\end{equation}
where $e$ is the charge of the external state and the function $\varepsilon(z,\z)$ controls the nature of the large gauge transformation. There is considerable freedom in choosing it however; generically it is any analytic function of $z$ and $\z$. This means in particular that it can be decomposed into a holomorphic and antiholomorphic piece, which will be of use when we derive the asymptotic symmetry from our two-dimensional model. 

It's clear how this generalizes to the case of Yang-Mills (now that we have reviewed how the Yang-Mills case is handled in general, with can treat it alongside the QED case henceforth); an asymptotic operator on the null boundary in Yang-Mills transforms under large gauge transformations according to

\begin{equation}\label{eq:3.2}
    \delta_{\varepsilon}\mathcal{O}(z,\z) = iT^{a}\varepsilon(z,\z)\mathcal{O}(z,\z),
\end{equation}
where it is to be understood that $T^{a}$ is the colour charge of the asymptotic operator. The large gauge transformations are once again labelled by a holomorphic and antiholomorphic function that arise out of the soft theorems.

The case of gravity is somewhat different; indeed, it was the understanding that in the case of gravity soft theorems lead to the corresponding asymptotic symmetries that really set off the modern interest in soft theorems. Specifically, the asymptotic realization of soft theorems is the transformation of asymptotic operators in terms of what are known as \emph{supertranslations}. We'll get to a concrete definition of the group of supertranslations shortly, but here we state that they were already known as the group of diffeomorphisms of general relativity for asymptotically flat spacetimes (see \cite{Bondi:1962px,PhysRev.128.2851,Penrose:1965am,Newman:1968uj}).

Supertranslations, as the name suggests, are translations on the celestial sphere. Unlike ordinary translations however supertranslations are local in the sense that they can be expanded in terms of coordinates on the celestial sphere. Specifically, under the action of a supertranslation, an external state $G(z,\z)$ in a gravitational theory transforms according to

\begin{equation}
    \delta_{P}G(z,\z,u) = P(z,\z)\partial_{u}G(z,\z,u). 
\end{equation}
In other words, a supertranslation has the effect of `translating' the retarded time parameter $u$ as

\begin{equation}
    u \longrightarrow u + P(z,\z).
\end{equation}
Generically, $P$ can be expanded in powers of $z$ and $\z$, but a specific class of such modes are captured by the soft theorems, a set of supertranslations that happens to be closed under the action of the Lorentz group. It will be our task to extract the operators controlling these modes in terms of our two-dimensional models. 

\subsection{Soft Current Modes in Gauge Theory}\label{sec:3.1}
While the main focus of this part of our work will be in tying together the two-dimensional representations of the soft modes in gravity to what we already known about asymptotic symmetries, it is instructive to carry out that analysis for gauge theory as well, although it might seem somewhat redundant. For the sake of completeness, we do this for QED and Yang-Mills independently.

We recall first that the single soft emission theorem in QED was due to the insertion of one soft current. In a celestial basis, it is known that the leading holomorphic soft current is defined by the limit

\begin{equation}
    j^{}_{2,\rom{QED}}(z) = \lim_{\Lambda\rightarrow 1}(\Lambda-1)\mathcal{O}_{\Lambda,+}(z),
\end{equation}
where it is to be understood that the asymptotic operator $\mathcal{O}_{\Lambda,+}(z)$ prepares one photon on the celestial sphere of positive helicity. In the limit that the weight goes to one, the Mellin transform is dominated by modes which are of low momentum. This actually happens to be true for a tower of integers, and leads to a tower of soft operators - a fact which we will have occasion to review in the next section. 

Analogously, the antiholomorphic soft current can be related to external states according to

\begin{equation}
    \overline{j}^{}_{2,\rom{QED}}(\z) = \lim_{\Lambda\rightarrow 1}(\Lambda-1)\mathcal{O}_{\Lambda,-}(\z),
\end{equation}
where this time the asymptotic operator creates one photon that is of negative helicity. At this time, we observe that what we have is actually a direct equivalence, which holds between a specific limit of an asymptotic operator, which is merely conjectured and an operator in a truly two-dimensional theory that may be explicitly constructed. It remains to be seen how useful this particular observation will be in providing a better description of asymptotic operators in terms of a holographic theory - future work will hopefully reveal this.

With these formal definitions in place, we can observe that the following operator product expansions hold, between dressed asymptotic states and the soft currents, namely

\begin{equation}
    j^{}_{2,\rom{QED}}(z) \widetilde{\mathcal{O}}_{\Lambda_{i}}(z_{i},\z_{i}) \sim  \frac{i\eta_{i}e_{i}}{z-z_{i}}\widetilde{\mathcal{O}}_{\Lambda_{i}}(z_{i},\z_{i}), 
\end{equation}
and its negative helicity counterpart obtained by complex conjugation of the celestial variables. Obviously $e_{i}$ vanishes if the asymptotic operator creates a photon. Making contact with the relation to large gauge transformations is obtained by pointing out that the field equation governing the two-dimensional soft mode is

\begin{equation}
    D_{z}D_{\z}\phi^{}_{1}(z,\z) = 0,
\end{equation}
which just means that $\phi^{}_{1}$ is a sum of a holomorphic and an antiholomorphic piece, a resolution which is obtained directly from the definitions of the currents. Post facto, we have accurately referred to the currents as holomorphic and antiholomorphic. Keeping this in mind, the holomorphic current can be resolved into a collection of modes, denoted by $j^{}_{\rom{QED},p}$, and defined according to the contour prescription

\begin{equation}\label{eq:3.10}
    j^{}_{\rom{QED},p} = \int_{c_{z_{0}}}z^{p}j^{}_{2,\rom{QED}}(z)dz,
\end{equation}
where $c_{z_{0}}$ indicates a small circle around $z_{0}$ clockwise, some arbitrary point in the complex plane. The point $z_{0}$ can be kept quite general - it is fixed when we consider the commutator of the mode with some other operator in the theory. Indeed, noting that the commutator between two holomorphic operators can be defined in two dimensions by

\begin{equation}\label{eq:3.12}
 [A,B](w) =  \int_{c_{w_{0}}}A(z)B(w)dz, 
\end{equation}
with an analogous definition for antiholomorphic operators, we obtain the relation

\begin{equation}
    [j^{}_{\rom{QED},p},\widetilde{\mathcal{O}}_{\Lambda_{i}}(z_{i},\z_{i},u_{i})] = 2\pi\eta_{i}e_{i}z^{p}\widetilde{\mathcal{O}}_{\Lambda_{i}}(z_{i},\z_{i},u_{i})
\end{equation}
for the holomorphic modes; the antiholomorphic modes' commutation relations with the asymptotic operators are found by replacing $z\rightarrow \z$ in the preceding equation.

These are precisely the commutation relations required between asymptotic states and the generators of large gauge transformations. As we made sure to point out previously, what we have just done is not too different from what is already known. However, it is qualitatively different in the sense that we have determined the generators in terms of two-dimensional fields directly. Furthermore, asymptotic operators need to be dressed, once again using two-dimensional fields. Large gauge transformations are then in a sense defined in a manner that is independent of the bulk fields themselves. 

Doing this with Yang-Mills follows the same logic; Yang-Mills modes defined in analogy to \mref{eq:3.10} as $j^{}_{\rom{YM},p}$ lead to the commutation relation

\begin{equation}
    [j^{a}_{\rom{YM},p},\widetilde{\mathcal{O}}_{\Lambda_{i}(\mu)}(z_{i},\z_{i},u_{i})] = 2\pi\eta_{i}z^{p}T^{a}_{i}\widetilde{\mathcal{O}}_{\Lambda_{i}(\mu)}(z_{i},\z_{i},u_{i}),
\end{equation}
which when combined with their antiholomorphic cousins and compared to the expansion of the arbitrary function in \mref{eq:3.2} are the generators of large gauge transformations in Yang-Mills theory. To complete the comparison with the existing literature on these modes, we can verify that they are indeed the correct generators of large gauge transformation by computing the commutator of two modes, namely

\begin{equation}
    [j^{a}_{\rom{YM},p},j^{b}_{\rom{YM},q}] = \pm2\pi i f^{abc} j^{c}_{\rom{YM},p+q}, 
\end{equation}
where the $\pm$ has been indicated as in our convention, the modes when acting on an asymptotic state record the signature variables $\eta_{i}$. That being said, these commutation relations are in agreement with the ones in \cite{Banerjee:2020vnt}. 

While the case of gauge theory is simple due to the fact that the splitting into holomorphic and antiholomorphic soft currents is automatic, this resolution is a little trickier in gravity. We'll see now how the supertranslation modes are obtained from the two-dimensional models of soft gravitons, but it is good to first have a basic review of the presentation of supertranslations (and another class of transformations known as superrotations) on the celestial sphere.

\subsection{Supertranslations and Superrotations on Null Infinity}\label{sec:3.2}
Suppose we have a class of operators $G_{\Lambda_{i}}(z_{i},\z_{i},u_{i})$ on future null infinity, which itself is parametrized by three parameters, namely a null coordinate $u$, which runs over the real axis, and two complex coordinates $z$ and $\z$ which are valued on the celestial sphere. In accordance with the notation we have already employed and for the sake of simplicity, it suffices to assume that these asymptotic operators create gravitational states, although this restriction is not especially important.

Now if we want to `translate' such operators on the celestial sphere in a manner that is local, there are three natural ways to do this. The simplest of course is to translate the $u$ coordinate, weighted by the location of the operator on the celestial sphere. The other two ways involve translating along the celestial coordinates themselves. Concretely, we mean the following three transformations

\begin{equation}
    \delta G(z_{i},\z_{i},u_{i}) = z^{p+1}_{i}\z^{q+1}_{i}\partial_{u_{i}}G(z_{i},\z_{i},u_{i}),
\end{equation}

\begin{equation}
    \delta G(z_{i},\z_{i},u_{i}) = z^{p}_{i}\partial_{z_{i}}G(z_{i},\z_{i},u_{i})+\frac{1}{2}z^{p}_{i}\partial_{u_{i}}G(z_{i},\z_{i},u_{i}),
\end{equation}
and

\begin{equation}
    \delta G(z_{i},\z_{i},u_{i}) = \z^{p}_{i}\partial_{\z_{i}}G(z_{i},\z_{i},u_{i})+\frac{1}{2}\z^{p}_{i}\partial_{u_{i}}G(z_{i},\z_{i},u_{i}).
\end{equation}
The first of these transformations, labelled by two integers $p$ and $q$ are known as supertranslations, a reflection of the fact that they literally translate the operator along the retarded coordinate. However, they depend on the local coordinates of the operator on the celestial sphere, generalizing traditional translations which are global and independent of the $z_{i}$ ad $\z_{i}$. The operators effecting these transformations are denoted by $P_{p,q}$

The next two transformations are called superrotations, and are a generalization of Lorentz rotations to local rotations. Holomorphic and antiholomorphic superrotations are denoted by $L_{p}$ and $\overline{L}_{p}$. In particular, (connected) Lorentz transformations are generated by the operators $\lbrace{L_{-1},L_{0},L_{1},\overline{L}_{-1},\overline{L}_{0},\overline{L}_{1}\rbrace}$. 

Superrotations and supertranslations interact with each other to form a closed algebra; two supertranslations commute - 

\begin{equation}
    [P_{p,q},P_{p',q'}] = 0,
\end{equation}
while a supertranslation and a superrotation close into a supertranslation as evidenced by

\begin{equation}
    [L_n, P_{p,q}] = \left(\frac{n+1}{2} - p\right) P_{p+n,q}, 
\end{equation}
and

\begin{equation}
    [\overline{L}_n, P_{p,q}] = \left(\frac{n+1}{2} - q\right) P_{p,q+n}. 
\end{equation}
This combined algebra of superrotations and supertranslations is known in the literature as the extended BMS algebra \cite{Barnich:2010ojg}, and is know understood to be the diffeomorphism group of spacetimes which are asymptotically flat. In particular, the invariance of the gravitational $S$-matrix under the extended BMS group is has been shown to be equivalent to the first two soft theorems in gravity, namely the leading order and subleading order soft theorems of gravitational emission. In the context of this relation, there is a special subgroup of the extended BMS group, which will be our main concern, as the generators of this group can be shown to directly arise out of the modes of soft currents in gravity. We will briefly review this subgroup and then proceed to actually constructing it out of our two-dimensional model.

The subgroup in question consists of the generators $\lbrace{P_{-1,q},P_{0,q},P_{p,0},P_{p,-1}\rbrace}$,where $p,q\in\mathbb{Z}$, along with the six generators of the Lorentz group that we have already mentioned. Checking that this algebra is closed is a matter of using the commutators in the last three equations to make sure; indeed, we have

\begin{equation}
    \begin{aligned}
     &[L_{-1}, P_{0,p}] = 0 \;\;\;\; &[L_0, P_{0,p}] = \frac{1}{2} P_{0,p}\;\;\;\;&[L_1, P_{0,p}] = P_{1,p}\\ 
     &[L_{-1}, P_{1,p}] = -P_{0,p}\;\;\;\; &[L_0, P_{1,p}] = -\frac{1}{2} P_{1,p}\;\;\;\;&[L_1, P_{1,p}] = 0,
\end{aligned}
\end{equation}
with the antiholomorphic analogues holding as well. Clearly, the algebra closes, and is a proper subalgebra of the full extended BMS group. We will now see how these generators - the ones arising out of the supertranslation part of the BMS group - can be realized as modes of soft currents defined in our two-dimensional models.

\subsection{Two-Dimensional Soft Currents and Supertranslations}\label{sec:3.3}
The first hint that the analytic structure has to be handled carefully in the case of soft theorems in gravity comes from the fact that given the action for the soft modes in two dimensions, the field equations obeyed are as follows

\begin{equation}
    D_{z}^{2}D_{\z}^{2}\sigma^{}_{a}(z,\z) = 0,
\end{equation}
which tells us that the functions $D_{z}D_{\z}\sigma^{}_{a}$ are harmonic, and can be expanded as a sums of holomorphic and antiholomorphic functions. Thus we have

\begin{equation}
    D_{z}D_{\z}\sigma^{}_{a}(z,\z) = F'_{a}(z) + G'_{a}(\z),
\end{equation}
which has the general solution

\begin{equation}
    \sigma^{}_{a}(z,\z) = \z F_{a}(z) + zG_{a}(\z) + f_{a}(z) + g_{a}(\z)
\end{equation}
for generic analytic functions $F$, $G$, $f$ and $g$. Recall that the soft currents, which like in the gauge theory case we now define in terms of asymptotic operators that create gravitons of positive helicity

\begin{equation}
    j^{}_{2,\rom{grav}}(z,\z) = \lim_{\Lambda\rightarrow 1}(\Lambda-1) \widetilde{G}_{\Lambda,+}(z,\z)
\end{equation}
are equivalently defined in terms of the second $D_{z}$ derivative of the auxiliary field $\sigma^{}_{2}$, which obeys a biharmonic equation as its Euler-Lagrange equation. From the expansion of the solution of the biharmonic equation that we have just obtained, the second $z$ derivative of the field can be expressed in terms of two linearly independent currents in the basis $\lbrace{1,\z\rbrace}$ according to the resolution

\begin{equation}
   j^{}_{a,\rom{grav}}(z,\z) = j^{0}_{a,\rom{grav}}(z)+\z j^{1}_{a,\rom{grav}}(z). 
\end{equation}
Comparing with the form of the soft graviton theorems for positive and negative helicity radiation, we can infer that the following operator product expansion must hold

\begin{equation}\label{eq:3.31}
    j^{1}_{2,\rom{grav}}(z)\widetilde{G}_{\Lambda_{i}}(z_{i},\z_{i},u_{i}) \sim \frac{\kappa}{z-z_{i}}\partial_{u_{i}}\widetilde{G}_{\Lambda_{i}}(z_{i},\z_{i},u_{i}).
\end{equation}
We now want to express this part of the soft graviton current directly in terms of the auxiliary field $\sigma^{}_{2}(z,\z)$ rather than as a term implicit in the current's expansion. This is done by first noting that we have the following identity,

\begin{equation}\label{eq:3.33}
    \lab{D_{\z}D_{z}^{2}\sigma^{}_{2}(z,\z)\sigma^{}_{1}(z',\z')} = \frac{1}{z-z'}.
\end{equation}
Since this seen to be the contribution of that part of the positive helicity soft graviton current proportional to $\z$, we obtain the identification

\begin{equation}\label{eq:3.35}
    j^{1}_{2,\rom{grav}}(z) = D_{\z}D_{z}^{2}\sigma^{}_{2}(z,\z).
\end{equation}
From this, we can resolve the purely holomorphic contribution of the positive helicity soft graviton current by performing a subtraction according to

\begin{equation}\label{eq:3.37}
    j^{0}_{2,\rom{grav}}(z) = D_{z}^{2}\sigma^{}_{2}(z,\z) - \z D_{\z}D_{z}^{2}\sigma^{}_{2}(z,\z).
\end{equation}
For the sake of completeness, we note what happens when the holomorphic soft graviton current acts on a dressed asymptotic state,

\begin{equation}\label{eq:3.39}
    j^{0}_{2,\rom{grav}}(z)\widetilde{G}_{\Lambda_{i}}(z_{i},\z_{i},u_{i}) \sim -\kappa\frac{\z_{i}}{z-z_{i}}\partial_{u_{i}}\widetilde{G}_{\Lambda_{i}}(z_{i},\z_{i},u_{i}).
\end{equation}
Combined with the definitions from \mref{eq:3.35} to \mref{eq:3.37}, the operator product expansions between the currents and the asymptotic states have been defined intrinsically in a two-dimensional fashion. The reader will observe that in combination with the operation in \mref{eq:3.31}, these two currents actually generate a class of supertranslations when the denominators are expanded around the points $z_{i}$ (or $\z_{i}$ for the negative helicity soft theorem). We now show how this comes about concretely.

We claim that given the holomorphic currents, the modes satisfy 

\begin{equation}
    P_{p,0} = -\frac{1}{2\pi i\kappa}\oint_{c_{w}} \left(D_{z}^{2}\sigma^{}_{2}(z,\z) - \z D_{\z}D_{z}^{2}\sigma^{}_{2}(z,\z)\right) dz,
\end{equation}
and

\begin{equation}
    P_{p,-1} = \frac{1}{2\pi i\kappa}\oint_{c_{w}} \left(D_{\z}D_{z}^{2}\sigma^{}_{2}(z,\z)\right) dz,
\end{equation}
where the point $w$ is encircled by a circle in the anticlockwise direction. $w$ is picked in terms of the operator with which the commutation relations of the modes are computed. Indeed, we verify this once again by using \mref{eq:3.12} to compute the commutators

\begin{equation}
    [P_{p,0},\widetilde{G}_{\Lambda_{i}}(z_{i},\z_{i},u_{i})] = \z_{i}z^{p}_{i}\partial_{u_{i}}\widetilde{G}_{\Lambda_{i}}(z_{i},\z_{i},u_{i}),
\end{equation}
and

\begin{equation}
    [P_{p,-1},\widetilde{G}_{\Lambda_{i}}(z_{i},\z_{i},u_{i})] = z^{p}_{i}\partial_{u_{i}}\widetilde{G}_{\Lambda_{i}}(z_{i},\z_{i},u_{i}),
\end{equation}
verifying the claim. We have similarly the antiholomorphic counterparts $P_{0,p}$ and $P_{-1,p}$ in terms of the two-dimensional field as follows

\begin{equation}
    P_{0,p} = -\frac{1}{2\pi i\kappa}\oint_{c_{w}} \left(D_{\z}^{2}\sigma^{}_{2}(z,\z) - z D_{z}D_{\z}^{2}\sigma^{}_{2}(z,\z)\right) d\z,
\end{equation}
and

\begin{equation}
    P_{-1,p} = \frac{1}{2\pi i\kappa}\oint_{c_{w}} \left(D_{z}D_{\z}^{2}\sigma^{}_{2}(z,\z)\right) d\z,
\end{equation}
and the attendant commutation relations

\begin{equation}
    [P_{0,p},\widetilde{G}_{\Lambda_{i}}(z_{i},\z_{i},u_{i})] = z_{i}\z^{p}_{i}\partial_{u_{i}}\widetilde{G}_{\Lambda_{i}}(z_{i},\z_{i},u_{i}),
\end{equation}
and

\begin{equation}
    [P_{-1,p},\widetilde{G}_{\Lambda_{i}}(z_{i},\z_{i},u_{i})] = \z^{p}_{i}\partial_{u_{i}}\widetilde{G}_{\Lambda_{i}}(z_{i},\z_{i},u_{i}).
\end{equation}

We close this section by making some comments on the so-called leading order\footnote{Memory effects beyond leading order have also been studied; references include \cite{Pasterski:2015tva,Compere:2019odm,Himwich:2019qmj,Mao:2020vgh}} memory effect, and how it arises in the two-dimensional context.  Most famously stated in the context of gravity \cite{Zeldovich:1974gvh,Braginsky:1985vlg,PhysRevD.44.R2945,PhysRevD.45.520,Blanchet:1992br,Favata:2010zu,Tolish:2014bka,Tolish:2014oda,Zhang:2017rno,Zhang:2017jma,Zhang:2018srn}, it states that distances measured undergo variation as one moves along null infinity. Specifically, the difference is nontrivial when evaluated across the whole of null infinity. This difference, not observed in ordinary Minkowski space, arises due to the fact that massless particles scattering in the bulk deposit gravitational radiation on null infinity, which is collected in the form of `gravitational memory'\footnote{For discussions on experimental signatures of gravitational memory, see \cite{Favata:2008yd,vanHaasteren:2009fy,Seto:2009nv,Wang:2014zls,NANOGrav:2015xuc,Lasky:2016knh,Nichols:2017rqr,Yang:2018ceq,Talbot:2018sgr,Johnson:2018xly,Hubner:2019sly,Boersma:2020gxx,Hubner:2021amk,Divakarla:2021xrd}.}. 

At particularly elegant form of the gravitational memory is obtained by the so-called Braginsky-Thorne formula \cite{Braginsky}, which states that we have for the change in the metric (working in momentum space) along null infinity

\begin{equation}
    \ab{\Delta h^{\mu\nu}(q)} = \left(\sum_{i=1}^{n}\frac{p^{\mu}_{i}p^{\nu}_{i}}{p_{i}\cdot \hat{q}}\right)^{TT},
\end{equation}
where the $TT$ indicates that the transverse and traceless piece has to be extracted. We have placed the metric in angle brackets to indicate the fact that it is a classical object. The hat indicates that the soft frequency has been stripped away.

Obviously, this is simply the leading order soft factor with the helicity polarizations stripped away. Indeed, what this tells us is that when the field $h^{\mu\nu}$ is regarded as an operator in the standard field theoretic sense, one has the relation 

\begin{equation}
    \frac{\epsilon^{+}_{\mu}\epsilon^{+}_{\nu}\bra{\rom{out}}{\Delta h^{\mu\nu}(q)}\mathcal{S}\ket{\rom{in}}}{\bra{\rom{out}}\mathcal{S}\ket{\rom{in}}} = \frac{\lab{j_{\rom{grav}}^{}(z,\z)\widetilde{G}_{\Lambda_{1}}(u_{1},z_{1},\z_{1})\dots \widetilde{G}_{\Lambda_{n}}(u_{n},z_{n},\z_{n})}}{\lab{\widetilde{G}_{\Lambda_{1}}(u_{1},z_{1},\z_{1})\dots \widetilde{G}_{\Lambda_{n}}(u_{n},z_{n},\z_{n})}},
\end{equation}
and an analogue for negative radiation. Indeed, this correctly places the memory effect as an infrared-safe quantity that is computed equivalently by our two-dimensional model. We won't go through it here to avoid repetitiveness, but precisely the same fact holds for QED as well \cite{Bieri:2013hqa}. The gauge field undergoes a shift along null infinity, which is computed by the soft factor for QED \cite{Pasterski:2015zua}. It can thus be recast as an infrared safe quantity by generalizing the previous formula to the two-dimensional model for QED. 

%% file: sec4beyond.tex
\section{Dual Models for Soft Theorems Beyond Leading Order}\label{sec:4}
So far, we have considered the so-called leading order soft theorems, which are characterized by the soft factors being formally divergent. While not being of great phenomenological importance, there are soft theorems of higher order\footnote{See the \cite{Low:1954kd,GellMann:1954kc,Low:1958sn,Burnett:1967km} for derivations of the soft theorem in QED at subleading order. For the subleading soft theorem in gravity, see \cite{Cachazo:2014fwa}.}, which among other things are fixed by gauge invariance. In this section, we will consider the presentation of these higher order soft theorems both in gauge theory and gravity from the point of view of celestial amplitudes. In particular, our focus will be to conjecture a class of two-dimensional theories that we hope will serve to provide the first step towards a proper holographic representation of the soft $S$-matrix beyond leading order.

In section \ref{sec:4.1}, we start with a description of the soft theorems beyond leading order and recall how they are defined in the celestial context. Two-dimensional models are then used to derive, in turn, the subleading soft theorem in QED and Yang-Mills, the subleading soft graviton theorem and the subsubleading soft graviton theorem in sections \ref{sec:4.2}, \ref{sec:4.3} and \ref{sec:4.4} respectively. We then conclude this part of our analysis by putting these results together and commenting on a class of double copy operations that relate the two-dimensional models and currents that supply the higher order soft theorems in section \ref{sec:4.5}.

\subsection{The Celestial Hierarchy of Soft Theorems}\label{sec:4.1}
There are several ways of deriving soft theorems, at leading as well as beyond leading order in constrained theories such as gauge theory and gravity. In particular, invariance under gauge transformations and diffeomorphisms are known to fix the form of the soft theorems essentially uniquely. While this derivation is simple and conceptually pleasing, what we are interested in the present work is how these theorems manifest themselves and can be isolated in the context of celestial amplitudes. To see that, we have to go back to the definition of celestial amplitude, namely the Mellin transform.

Suppose you have a scattering amplitude (gauge theory or gravity) $\mathcal{A}_{n}(\lbrace{\omega_{i},z_{i},\z_{i}\rbrace})$ and its Mellin transform $\widetilde{\mathcal{A}}_{n}(\lbrace{\Lambda_{i},u_{i},z_{i},\z_{i}\rbrace})$, we have the relation 

\begin{equation}
    \widetilde{\mathcal{A}}_{n}(\lbrace{\Lambda_{i},u_{i},z_{i},\z_{i}\rbrace}) = \int \mathcal{A}_{n}(\lbrace{\omega_{i},z_{i},\z_{i}\rbrace})\prod_{i=1}^{n}\omega^{\Lambda_{i}}_{i}e^{i\eta_{i}\omega_{i}u_{i}}d\omega_{i}.
\end{equation}
The emission of a single soft particle, be it a photon, gluon or graviton, is accompanied by the introduction of a factor of the energy of the soft factor. Specifically, we expect (in momentum space), that the soft expansion follows, which as it turns out that it does, the hierarchy

\begin{equation}
    \mathcal{A}_{n+1}(\omega,z,\z\lbrace{\omega_{i},z_{i},\z_{i}\rbrace}) = \lim_{\omega\rightarrow 0}\left(\sum_{i=0}^{n}\omega^{i-1}S^{(i)}(z,\z,\lbrace{z_{i},\z_{i},\omega_{i}\rbrace})\right)\mathcal{A}_{n}(\lbrace{\omega_{i},z_{i},\z_{i}\rbrace}),
\end{equation}
where $\omega$ is the energy of the soft particle and the $z$ and $\z$ are its coordinates on the celestial sphere. The soft factors at the $i$th order, denoted by $S^{(i)}$, are independent of the energy of the soft particle. 

From the form of the soft expansion, we see that at each order, the soft theorem contributes a factor of the form

\begin{equation}
    \widetilde{S}^{(i)}(\Lambda,u=0,\lbrace{\Lambda_{i},z_{i},\z_{i},u_{i}\rbrace}) \sim \frac{1}{\Lambda+i} S^{(i)}(\lbrace{\Lambda_{i},z_{i},\z_{i},-i\eta_{i}\partial_{u_{i}}\rbrace}).
\end{equation}
From this, the soft factor in the celestial basis is really just the coefficient of the amplitude after taking the residue of $\Lambda$ at $-i$. 

The fact that the soft theorems can be found this way was shown to all orders in gauge theory and gravity in \cite{Guevara:2019ypd} by Guevara. Since we will only have occasion to use the theorems, rather than having to derive them from scratch, we choose to simply record the results from \cite{Guevara:2019ypd}. Starting with gauge theory, we have the following form of the leading soft theorem.

\begin{equation}
    \rom{Res}_{\Lambda = 0}\widetilde{A}^{\rom{QED}}_{n+\rom{soft}}(\Lambda,u=0,\lbrace{\Lambda_{i},u_{i},z_{i},\z_{i}\rbrace}) = \sum_{i=1}^{n}\frac{\eta_{i}e_{i}}{z-z_{i}}\widetilde{A}^{\rom{QED}}_{n}(\lbrace{\Lambda_{i},u_{i},z_{i},\z_{i}\rbrace}),
\end{equation}
for a positive helicity soft particle with the obvious analogue holding for a negative helicity particle. The subleading soft theorem is obtained by taking the residue at $\Lambda = -1$ to give

\begin{equation}
\begin{aligned}
     &\rom{Res}_{\Lambda = -1}\widetilde{A}^{\rom{QED}}_{n+\rom{soft}}(\Lambda,u=0,\lbrace{\Lambda_{i},u_{i},z_{i},\z_{i}\rbrace}) =\\ &\sum_{i=1}^{n}\frac{\eta_{i}}{z-z_{i}}\left(2h_{i}-1-(\z-\z_{i})D_{\z_{i}}\right)e_{i}\mathcal{P}_{-1}\widetilde{A}^{\rom{QED}}_{n}(\lbrace{\Lambda_{i},u_{i},z_{i},\z_{i}\rbrace}),
\end{aligned}
\end{equation}
when a positive helicity gluon is being radiated and

\begin{equation}
    \begin{aligned}
     &\rom{Res}_{\Lambda = -1}\widetilde{A}^{\rom{QED}}_{n+\rom{soft}}(\Lambda,u=0,\lbrace{\Lambda_{i},u_{i},z_{i},\z_{i}\rbrace}) =\\ &\sum_{i=1}^{n}\frac{\eta_{i}}{\z-\z_{i}}\left(2\overline{h}_{i}-1-(z-z_{i})D_{z_{i}}\right)e_{i}\mathcal{P}^{-1}_{i}\widetilde{A}^{\rom{QED}}_{n}(\lbrace{\Lambda_{i},u_{i},z_{i},\z_{i}\rbrace}),
    \end{aligned}
\end{equation}
when a negative helicity graviton is radiated. The operator $\mathcal{P}^{-1}_{k}$ has the effect of shifting $\Lambda_{i}$ to $\Lambda_{i}-1$. The variables $h_{i}$ and $\overline{h}_{i}$ are defined as

\begin{equation}
    h_{i} = \frac{\Lambda_{i}-s_{i}}{2},
\end{equation}
and
\begin{equation}
    \overline{h}_{i} = \frac{\Lambda_{i}+s_{i}}{2},
\end{equation}
where $s_{i}$ is the helicity of the $i$th particle. 

When we move beyond the subleading order in gauge theory, the soft theorems can no longer be readily broken down into holomorphic and antiholomorphic pieces. This is really due to the fact that gauge invariance no longer fixes the soft theorems beyond subleading order, and the soft momentum mixes with the external momenta in a way that doesn't make any kind of analyticity obvious. Accordingly, at the level of gauge theory our attention will be restricted to the subleading theorem. 

Moving now to the case of gravity, the interesting soft theorems beyond leading order are the subleading and subsubleading order theorems, which are $\mathcal{O}(\omega^{0})$ and $\mathcal{O}(\omega^{1})$ respectively. This means that the relevant residues are at $\Lambda = -1$ and $\Lambda = -2$ respectively. Of course, we have at leading order the fact that

\begin{equation}
    \rom{Res}_{\Lambda = 0}\widetilde{A}^{\rom{grav}}_{n+\rom{soft}}(\Lambda,u=0,\lbrace{\Lambda_{i},u_{i},z_{i},\z_{i}\rbrace}) = \sum_{i=1}^{n}\kappa\frac{\z-\z_{i}}{z-z_{i}}\partial_{u_{i}}\widetilde{A}^{\rom{grav}}_{n}(\lbrace{\Lambda_{i},u_{i},z_{i},\z_{i}\rbrace}),
\end{equation}
and the antiholomorphic counterpart. For the residue at $\Lambda=-1$ we have for a positive helicity graviton

\begin{equation}
\begin{aligned}
     &\rom{Res}_{\Lambda = -1}\widetilde{A}^{\rom{grav}}_{n+\rom{soft}}(\Lambda,u=0,\lbrace{\Lambda_{i},u_{i},z_{i},\z_{i}\rbrace}) =\\ &\sum_{i=1}^{n}\kappa\frac{1}{z-z_{i}}\left(2h_{i}(\z-\z_{i})-(\z-\z_{i})^{2}D_{\z_{i}}\right)\widetilde{A}^{\rom{grav}}_{n}(\lbrace{\Lambda_{i},u_{i},z_{i},\z_{i}\rbrace}),
\end{aligned}
\end{equation}
and

\begin{equation}
\begin{aligned}
     &\rom{Res}_{\Lambda = -1}\widetilde{A}^{\rom{grav}}_{n+\rom{soft}}(\Lambda,u=0,\lbrace{\Lambda_{i},u_{i},z_{i},\z_{i}\rbrace}) =\\ &\sum_{i=1}^{n}\kappa\frac{1}{\z-\z_{i}}\left(2\overline{h}_{i}(z-z_{i})-(z-z_{i})^{2}D_{z_{i}}\right)\widetilde{A}^{\rom{grav}}_{n}(\lbrace{\Lambda_{i},u_{i},z_{i},\z_{i}\rbrace}),
\end{aligned}
\end{equation}
for a negative helicity graviton emission. The soft theorem at subsubleading order is obtained by evaluating the residue at $\Lambda = -2$. Again, the positive and negative helicity cases are given by the expressions

\begin{equation}
\begin{aligned}
     &\rom{Res}_{\Lambda = -2}\widetilde{A}^{\rom{grav}}_{n+\rom{soft}}(\Lambda,u=0,\lbrace{\Lambda_{i},u_{i},z_{i},\z_{i}\rbrace}) =\\ &\left(\sum_{i=1}^{n}\left(\kappa\frac{1}{z-z_{i}}\left(-2h_{i}(2h_{i}-1)(\z-\z_{i})+4(\z-\z_{i})^{2}h_{i}D_{\z_{i}} -(\z-\z_{i})^{3}D^{2}_{\z_{i}}\right)\mathcal{P}^{-1}_{i}\right)\right)\\
     &\times\widetilde{A}^{\rom{grav}}_{n}(\lbrace{\Lambda_{i},u_{i},z_{i},\z_{i}\rbrace}),
\end{aligned}
\end{equation}
and

\begin{equation}
\begin{aligned}
     &\rom{Res}_{\Lambda = -2}\widetilde{A}^{\rom{grav}}_{n+\rom{soft}}(\Lambda,u=0,\lbrace{\Lambda_{i},u_{i},z_{i},\z_{i}\rbrace}) =\\ &\left(\sum_{i=1}^{n}\left(\kappa\frac{1}{\z-\z_{i}}\left(-2\overline{h}_{i}(2\overline{h}_{i}-1)(z-z_{i})+4(z-z_{i})^{2}\overline{h}_{i}D_{z_{i}} -(z-z_{i})^{3}D^{2}_{z_{i}}\right)\mathcal{P}^{-1}_{i}\right)\right)\\
     &\times\widetilde{A}^{\rom{grav}}_{n}(\lbrace{\Lambda_{i},u_{i},z_{i},\z_{i}\rbrace}),
\end{aligned}
\end{equation}
respectively. 

The forms of the soft theorems in gauge theory and gravity up to this order suggest the possibility of a kind of double copy. This is especially clear if we move back to the momentum representation, in which the subleading soft factor for gauge theory and gravity become

\begin{equation}
    S^{(1),a}_{\rom{QED}} = \sum_{i=1}^{n}\frac{\epsilon_{\mu}q_{\nu}e_{i}J^{\mu\nu}_{i}}{p_{i}\cdot q},
\end{equation}
and
\begin{equation}
    S^{(1)}_{\rom{grav}} = \sum_{i=1}^{n}\frac{\epsilon\cdot p_{i}\epsilon_{\mu}q_{\nu}J^{\mu\nu}_{i}}{p_{i}\cdot q},
\end{equation}
while the subsubleading soft theorem for gravity takes the form

\begin{equation}
    S^{(2)}_{\rom{grav}} = \sum_{i=1}^{n}\frac{\epsilon_{\rho}\epsilon_{\mu}q_{\nu}q_{\lambda}J^{\mu\nu}_{i}J^{\rho\lambda}_{i}}{p_{i}\cdot q},
\end{equation}
where $J_{i}$ is the angular momentum operator defined by

\begin{equation}
    J^{\mu\nu}_{i} = p^{[\mu}\frac{\partial}{\partial p_{\nu]}}.
\end{equation}
Inspecting the numerators, it is tempting to suggest that at the level of soft theorems, the following schematic double copy relations should manifest in some form

\begin{equation}\label{eq:4.18}
    (\rom{leading})_{\rom{QED}}\otimes (\rom{leading})_{\rom{QED}} = (\rom{leading})_{\rom{grav}},
\end{equation}

\begin{equation}\label{eq:4.19}
    (\rom{subleading})_{\rom{QED}}\otimes (\rom{leading})_{\rom{QED}} = (\rom{subleading})_{\rom{grav}},
\end{equation}
and

\begin{equation}\label{eq:4.20}
    (\rom{subleading})_{\rom{QED}}\otimes (\rom{subleading})_{\rom{QED}} = (\rom{subsubleading})_{\rom{grav}},
\end{equation}
As it turns out, these relations are realized in the two-dimensional setting at the level of the soft currents. The specific manner in which this takes place will be discussed in section \ref{sec:4.5}. For now, we will study a class of two-dimensional theories which can correctly capture the higher order soft theorems we have just stated.

\subsection{The Subleading Soft Photon Theorem}\label{sec:4.2}
In this section, for the sake of simplicity we work with QED, and state the generalization to Yang-Mills, which really only entails a shift in notation at the end. This is just to avoid an unreasonable proliferation of indices.

The first soft theorem beyond leading order that we will study in more detail is the subleading soft theorem in gauge theory. We first note that we encounter two kinds of functions of the external punctures $z_{i}$ and $\z_{i}$ in the case of the subleading soft gluon theorem. The first is a singularity as the soft particle approaches one of the external states. The second is the same kind of function that shows up in the case of the leading order soft graviton theorem, in which the collinear singularity is subdued. Informed by this correspondence, and due to the fact that this time, the positive and negative helicity sectors necessarily need to be handled independently, we consider an action of the form

\begin{equation}
    \mathcal{I}^{(1)}_{\rom{QED}} = -\int\left(D^{2}_{z}V^{z}_{1}(z,\z)D^{2}_{\z}V^{z}_{2}(z,\z) + D^{2}_{z}V^{\z}_{1}(z,\z)D^{2}_{\z}V^{\z}_{2}(z,\z)\right)dz\wedge d\z.
\end{equation}
This time, we have four fields - $V^{z}_{1}$, $V^{z}_{2}$, $V^{\z}_{1}$ and $V^{\z}_{1}$, where the superscripts are just meant to suggest the helicity sectors for which they will be employed. Before listing the operator product expansions from this action, we first point out the global symmetry group. Diagonalizing each term quadratic in the fields will give four terms, two of which have the opposite helicities, but none of which mix different fields. Put differently, the global symmetry of each pair of coupled fields is the group that preserves the matrix

\begin{equation}
\eta^{AB}=\begin{pmatrix}
0 & 1\\ 
1 & 0
\end{pmatrix}.
\end{equation}
The global symmetry group is thus $O(1,1)\times O(1,1)$, which remains unbroken. If required, this symmetry can be broken to produce nontrivial operator product expansions for the $V^{A}_{1}$ or $V^{A}_{2}$ fields with themselves, but we won't consider such generalizations here. Indeed, the only nontrivial operator product expansions are

\begin{equation}
    \ab{V^{z}_{1}(z,\z)V^{z}_{2}(z',\z')} = \frac{1}{\pi}|z-z'|^{2}\ln|z-z'|^{2},
\end{equation}
and

\begin{equation}
    \ab{V^{\z}_{1}(z,\z)V^{\z}_{2}(z',\z')} = \frac{1}{\pi}|z-z'|^{2}\ln|z-z'|^{2}.
\end{equation}
The relevant operator product expansions we require to construct the dressing operators and currents for the subleading soft theorems in QED are

\begin{equation}
    \ab{D_{z}^{2}V^{z}_{1}(z,\z)V^{z}_{2}(z',\z')} = \frac{1}{\pi}\frac{\z-\z'}{z-z'},
\end{equation}

\begin{equation}
    \ab{D_{z}^{2}V^{z}_{1}(z,\z)D_{\z'}V^{z}_{2}(z',\z')} = -\frac{1}{\pi}\frac{1}{z-z'},
\end{equation}

\begin{equation}
    \ab{D_{\z}^{2}V^{\z}_{1}(z,\z)V^{\z}_{2}(z',\z')} = \frac{1}{\pi}\frac{z-z'}{\z-\z'},
\end{equation}
and

\begin{equation}
    \ab{D_{\z}^{2}V^{\z}_{1}(z,\z)D_{z'}V^{\z}_{2}(z',\z')} = -\frac{1}{\pi}\frac{1}{\z-\z'}.
\end{equation}
We now have to find the right linear combination of fields and their derivatives that can act as operators which provide the right dressings for the asymptotic states. Of course, this will be the generalization of the dressing that we used in the leading case, but the form of the subleading soft theorem naturally suggests a slightly more complicated form of the dressing operator. The operators that we have in mind, let's denote them by $Q^{V}_{+}$ and $Q^{V}_{-}$ will be expected to satisfy the following operator product expansions

\begin{equation}
    D_{z}^{2}V^{z}_{1}(z,\z)Q^{V}_{+}(h_{i},z_{i},\z_{i}) = \frac{1}{\pi}\frac{1}{z-z_{i}}\left(2h_{i}-1 - (\z-\z_{i})D_{\z_{i}}\right),
\end{equation}
and

\begin{equation}
    D_{\z}^{2}V^{\z}_{1}(z,\z)Q^{V}_{-}(\overline{h}_{i},z_{i},\z_{i}) = \frac{1}{\pi}\frac{1}{\z-\z_{i}}\left(2\overline{h}_{i}-1 - (z-z_{i})D_{z_{i}}\right).
\end{equation}
The form of these expansions fixes the operators $Q^{V}_{\pm}$; we have

\begin{equation}
    Q^{V}_{+}(h_{i},z_{i},\z_{i}) = -\left((2h_{i}-1)D_{\z_{i}}V^{z}_{2}(z_{i},\z_{i}) + V^{z}_{2}(z_{i},\z_{i})D_{z_{i}}\right),
\end{equation}
and

\begin{equation}
    Q^{V}_{-}(\overline{h}_{i},z_{i},\z_{i}) = -\left((2\overline{h}_{i}-1)D_{z_{i}}V^{\z}_{2}(z_{i},\z_{i}) + V^{\z}_{2}(z_{i},\z_{i})D_{\z_{i}}\right).
\end{equation}
Dressing the asymptotic operators that create external states is then done by constructing suitably defined dressing operators. This is achieved by introducing the subleading Wilson line

\begin{equation}
    \mathcal{W}^{V}_{\rom{QED}}(e_{i},h_{i},\overline{h}_{i},z_{i},\z_{i}) = \exp\left(ie_{i}\eta_{i}(Q^{V}_{+}(h_{i},z_{i},\z_{i})+Q^{V}_{-}(h_{i},z_{i},\z_{i}))\mathcal{P}^{-1}_{i}\right),
\end{equation}
and then introducing dressed states denoted as

\begin{equation}
    \widetilde{\mathcal{O}}^{V}_{\Lambda_{i}}(z_{i},\z_{i},u_{i}) =  \mathcal{W}^{V}_{\rom{QED}}(e_{i},h_{i},\overline{h}_{i},z_{i},\z_{i})\widetilde{\mathcal{O}}^{V}_{\Lambda_{i}}(z_{i},\z_{i},u_{i}).
\end{equation}
The specific form of the action we have chosen ensures that the correlation function of the Wilson line operators themselves vanishes. This can be modified as required if one chooses; we have chosen not to do so as we are not interested in the soft $S$-matrix itself beyond leading order, with our focus being placed on the soft theorems. Indeed, it is easy to see now that an insertion of one instance of the soft current

\begin{equation}
    j^{V}_{+}(z,\z) = D_{z}^{2}V^{z}_{1}(z,\z),
\end{equation}
or
\begin{equation}
    j^{V}_{-}(z,\z) = D_{\z}^{2}V^{\z}_{1}(z,\z),
\end{equation}
into the correlation function $\lab{\widetilde{\mathcal{O}}^{V}_{\Lambda_{1}}(z_{1},\z_{1},u_{1})\dots \widetilde{\mathcal{O}}^{V}_{\Lambda_{n}}(z_{n},\z_{n},u_{n})}$ is equivalent to the subleading soft theorem for a positive and negative helicity photon respectively.

In the context of the subleading soft theorem in gauge theory, the authors in \cite{Banerjee:2020vnt} analyzed the decomposition of the soft current generating the subleading soft theorem into mode functions. Due to the analytic structure of the subleading soft theorem, we can import the methods we had employed earlier in the study of the leading order soft theorem in gravity. Specifically, we introduce the following currents

\begin{equation}
    K^{V}_{+}(z,\z) = D_{z}^{2}D_{\z}V^{z}_{1}(z,\z),
\end{equation}

\begin{equation}
    J^{V}_{+}(z,\z) = D_{z}^{2}V^{z}_{1}(z,\z) - \z D_{\z}D_{z}^{2}V^{z}_{1}(z,\z),
\end{equation}
with corresponding expressions for $K^{V}_{-}$ and $J^{V}_{-}$ in terms of the conjugate fields and derivatives. In contrast to the gravitational case, these expressions lead to contact terms when the operators product expansions with the dressed operators are evaluated. Concretely, we have

\begin{equation}
\begin{aligned}
     & K^{V}_{+}(z,\z) \widetilde{\mathcal{O}}^{V}_{\Lambda_{i}}(z_{i},\z_{i},u_{i}) \sim\\ 
     &\frac{ie_{i}}{\pi}\left(\pi\delta^{2}(z-z_{i},\z-\z_{i})(2h_{i}-1)-\frac{1}{z-z_{i}}D_{\z_{i}}\right)\mathcal{P}^{-1}_{i}\widetilde{\mathcal{O}}^{V}_{\Lambda_{i}}(z_{i},\z_{i},u_{i}),
\end{aligned}
\end{equation}
and
\begin{equation}
\begin{aligned}
     & J^{V}_{+}(z,\z) \widetilde{\mathcal{O}}^{V}_{\Lambda_{i}}(z_{i},\z_{i},u_{i}) \sim\\  
     &\frac{ie_{i}}{\pi}\left(-\z\pi\delta^{2}(z-z_{i},\z-\z_{i})(2h_{i}-1)+\frac{(2h_{i}-1+\z_{i}D_{\z_{i}})}{z-z_{i}}\right)\mathcal{P}^{-1}_{i}\widetilde{\mathcal{O}}^{V}_{\Lambda_{i}}(z_{i},\z_{i},u_{i}).
\end{aligned}
\end{equation}
Fortunately, the contact terms really aren't a problem - when we compute the residues of these expressions about $z_{i}$, the contact terms identically vanish. Accordingly, it is straightforward to reproduce the results of \cite{Banerjee:2020vnt}, which we now state. Defining mode operators the usual way, we have the following commutation relations

\begin{equation}
    [K^{V}_{+,p},\widetilde{\mathcal{O}}^{V}_{\Lambda_{i}}(z_{i},\z_{i},u_{i})] = \frac{ie_{i}}{\pi}z^{p}_{i}D_{\z_{i}}\mathcal{P}^{-1}_{i}\widetilde{\mathcal{O}}^{V}_{\Lambda_{i}}(z_{i},\z_{i},u_{i}),
\end{equation}

\begin{equation}
    [K^{V}_{-,p},\widetilde{\mathcal{O}}^{V}_{\Lambda_{i}}(z_{i},\z_{i},u_{i})] = \frac{ie_{i}}{\pi}\z^{p}_{i}D_{z_{i}}\mathcal{P}^{-1}_{i}\widetilde{\mathcal{O}}^{V}_{\Lambda_{i}}(z_{i},\z_{i},u_{i}),
\end{equation}

\begin{equation}
    [J^{V}_{+,p},\widetilde{\mathcal{O}}^{V}_{\Lambda_{i}}(z_{i},\z_{i},u_{i})] = \frac{ie_{i}}{\pi}z^{p}_{i}(2h_{i}-1+\z_{i}D_{\z_{i}})\mathcal{P}^{-1}_{i}\widetilde{\mathcal{O}}^{V}_{\Lambda_{i}}(z_{i},\z_{i},u_{i}),
\end{equation}

\begin{equation}
    [J^{V}_{-,p},\widetilde{\mathcal{O}}^{V}_{\Lambda_{i}}(z_{i},\z_{i},u_{i})] = \frac{ie_{i}}{\pi}\z^{p}_{i}(2\overline{h}_{i}-1+z_{i}D_{z_{i}})\mathcal{P}^{-1}_{i}\widetilde{\mathcal{O}}^{V}_{\Lambda_{i}}(z_{i},\z_{i},u_{i}).
\end{equation}
The collection of operators $\lbrace{K^{V}_{+,p},J^{V}_{+,p},K^{V}_{-,p},J^{V}_{-,p}\rbrace}$ turns out to \emph{not} form a closed algebra. This is due to the fact that once one of these operators is acted on an external state, the application of another one results in a total reduction of the weight of the asymptotic operator by 2, rather than one. This of course continues to be true for nested commutators of any rank. Closing the algebra consequently requires the inclusion of all possible commutators involving these operators.

In closing this section, we just state what is required to generalize all of this to the Yang-Mills setting. Specifically, we have to promote the vector fields we have used to Lie-algebra valued vector fields, and replace the charges with the generators corresponding to the colour charges of the external states. The rest of the calculation is entirely analogous, only that we have to keep in mind that the soft theorem will now act as matrix multiplication on each asymptotic state individually, rather than multiplying the amplitude as a whole.

\subsection{The Subleading Soft Graviton Theorem}\label{sec:4.3}
Let us now return to the case of gravity, and in particular the case of the soft graviton theorem at subleading order. In gauge theory, the subleading soft theorem is known to be a consequence of the fact that electric dipole charge must be conserved across the celestial sphere; in gravity, the conservation of an analogous class of charges, namely superrotation charges, lead to the subleading soft theorem.

One similarity between the subleading theorems in gauge theory and gravity is the fact that in both cases, the asymptotic charges corresponding to the relevant conservation laws are parametrized by vector fields on the celestial sphere. A comparison of the forms of the two soft theorems when celestial coordinates are employed however suggests that the two-dimensional models controlling them cannot quite be identical, although it is reasonable to expect that a simple generalization should suffice. In the following, we will see that that is indeed the case.

Let us take an action that assumes the following form

\begin{equation}
    \mathcal{I}^{(1)}_{\rom{grav}} = \int \left(D^{3}_{z}Y^{z}_{1}(z,\z)D^{3}_{\z}Y^{z}_{2}(z,\z)+ D^{3}_{z}Y^{\z}_{1}(z,\z)D^{3}_{\z}Y^{\z}_{2}(z,\z)\right)dz\wedge d\z.
\end{equation}
Once again, the global symmetry of this action is $O(1,1)\times O(1,1)$, with the two helicity sectors decoupled by construction. This time, the operator product expansions are determined by the fact that the Green's function of the \emph{triharmonic} operator $D_{z}^{3}D_{\z}^{3}$ is found by the computation

\begin{equation}
    D_{z}^{3}D_{\z}^{3}\left(|z|^{4}\ln|z|^{2}\right) = 4\pi\delta^{2}(z,\z),
\end{equation}
which tells us that the two nontrivial operator product expansions among the various fields involved in the preceding action are

\begin{equation}
    \ab{Y^{z}_{1}(z,\z)Y^{z}_{2}(z',\z')} = \frac{1}{4\pi}|z-z'|^{4}\ln|z-z'|^{2},
\end{equation}
and

\begin{equation}
    \ab{Y^{\z}_{1}(z,\z)Y^{\z}_{2}(z',\z')} = \frac{1}{4\pi}|z-z'|^{4}\ln|z-z'|^{2},
\end{equation}
with all the others vanishing identically. Taking a hint from the calculations we have already done, and inspecting the form of the subleading soft graviton, we see that the natural choices for the soft current as triple holomorphic or antiholomorphic derivatives of the first vector field. This can be verified by taking note of the following two operator product expansions

\begin{equation}
    \ab{D_{z}^{3}Y^{z}_{1}(z,\z)Y^{z}_{2}(z',\z')} = \frac{1}{2\pi}\frac{(\z-\z')^{2}}{z-z'},
\end{equation}
and

\begin{equation}
    \ab{D_{z}^{3}Y^{z}_{1}(z,\z)D_{\z}Y^{z}_{2}(z',\z')} = -\frac{1}{\pi}\frac{(\z-\z')}{z-z'},
\end{equation}
and their conjugate counterparts. Indeed, we see that these are precisely the analytic objects we require to construct charges that lead ultimately to the subleading soft graviton theorem. 

The dressing of asymptotic states is done by preparing two operators, which this time we denote by the symbols $Q^{Y}$ and $\Q^{Y}$, corresponding to positive and negative helicity gravitons respectively. The required operator product expansions are (we have modified the normalization somewhat)

\begin{equation}\label{eq:4.51}
j^{Y}(z,\z)Q^{Y}(h_{i},z_{i},\z_{i}) \sim     \frac{1}{2\pi}\frac{1}{z-z_{i}}\left(2h_{i}(\z-\z_{i})-(\z-\z_{i})^{2}D_{\z_{i}}\right),
\end{equation}

\begin{equation}
\jj^{Y}(z,\z)\Q^{Y}(h_{i},z_{i},\z_{i}) \sim     \frac{1}{2\pi}\frac{1}{\z-\z_{i}}\left(2\h_{i}(z-z_{i})-(z-z_{i})^{2}D_{z_{i}}\right),
\end{equation}
where

\begin{equation}
    j^{Y}(z,\z) = D^{3}_{z}Y^{z}_{1}(z,\z),
\end{equation}
and

\begin{equation}
    \jj^{Y}(z,\z) = D^{3}_{\z}Y^{\z}_{1}(z,\z).
\end{equation}
The charges in this case are directly determined by the conditions to be exactly analogous to the charges in the gauge theory case. However, this time we have to keep in mind that the higher order derivatives introduced coefficients which were not present before; we have accordingly

\begin{equation}
    Q^{Y}(h_{i},z_{i},\z_{i}) = -\left(h_{i}D_{\z_{i}}Y^{z}_{2}(z_{i},\z_{i}) + Y^{z}_{2}(z_{i},\z_{i})D_{z_{i}}\right),
\end{equation}
and

\begin{equation}
    \Q^{Y}(\h_{i},z_{i},\z_{i}) = -\left(\h_{i}D_{z_{i}}Y^{\z}_{2}(z_{i},\z_{i}) + Y^{\z}_{2}(z_{i},\z_{i})D_{\z_{i}}\right).
\end{equation}
One difference between the subleading soft theorems in gauge theory and gravity is the fact that the soft factor does not effect the transformations $\Lambda_{i}\rightarrow \Lambda_{i}-1$ in the case of gravity. Consequently, the Wilson line in the case of gravity is slightly simpler; we find that

\begin{equation}
    \mathcal{W}^{Y}_{\rom{grav}}(\kappa,h_{i},\h_{i},z_{i},\z_{i}) = \exp\left(i\kappa\left(Q^{Y}(h_{i},z_{i},\z_{i})+\Q^{V}(\h_{i},z_{i},\z_{i})\right)\right),
\end{equation}
is the appropriate operator that should be used to dress the asymptotic states $G_{\Lambda_{i}}(z_{i},\z_{i},u_{i})$. Indeed, inserting one soft current into a correlation function built out of

\begin{equation}
    \widetilde{G}_{\Lambda_{i}}(z_{i},\z_{i},u_{i}) = \mathcal{W}^{Y}_{\rom{grav}}(\kappa,h_{i},\h_{i},z_{i},\z_{i}) G_{\Lambda_{i}}(z_{i},\z_{i},u_{i}),
\end{equation}
is equivalent to the subleading soft graviton theorem. With this construction, it is now our task to ask if we can reproduce the results of \cite{Banerjee:2020kaa}, namely the extraction of the generators of the $\overline{SL(2,\mathbb{C})}\times SL(2,\mathbb{C})$ algebra from the generator of the subleading soft theorem. To do this, we have to repeat the analysis carried out in the previous section, just with a little more care. We observe that given a solution $\phi(z,\z)$ of the triharmonic equation, it can be expanded as

\begin{equation}
    \phi(z,\z) = \sum_{i=0}^{2}\left(\z^{i}F_{i}(z)+z^{i}G_{i}(z)\right),
\end{equation}
where the $F_{i}$ and $G_{i}$ are analytic functions of $z$ and $\z$ respectively. Consequently, if we are to expand the soft operators in powers of the variable $z$, we will need to recursively repeat the calculation we have basically already done. For the sake of completeness, let us go through the process step by step.

Consider first the operator

\begin{equation}
    K^{Y}(z) = \frac{1}{2}D^{2}_{\z}D^{3}_{z}Y^{z}_{1}(z,\z). 
\end{equation}
The operator product expansion of this when acting on an asymptotic state can be recovered by differentiating \mref{eq:4.51} to give

\begin{equation}
     K^{Y}(z)\widetilde{G}_{\Lambda_{i}}(z_{i},\z_{i},u_{i}) \sim -\frac{1}{2\pi}\frac{1}{z-z_{i}}D_{\z_{i}}\widetilde{G}_{\Lambda_{i}}(z_{i},\z_{i},u_{i}) + \rom{contact},
\end{equation}
where we have indicated the operator product expansion up to contact terms (we will continue to do so in the rest of the calculation and state the contact terms separately at the end of the section for completeness.)

The linear term is recovered by removing the quadratic piece and differentiating. We define

\begin{equation}
    J^{Y}(z) = \frac{1}{2}D_{\z}\left(D_{z}^{3}Y^{z}_{1}(z,\z)-\z^{2}K^{Y}(z)\right.
\end{equation}
This leads to the following operator product expansion

\begin{equation}
\begin{aligned}
     J^{Y}(z)\widetilde{G}_{\Lambda_{i}}(z_{i},\z_{i},u_{i}) =  \frac{1}{2\pi}\left(\frac{h_{i}+\z_{i}D_{\z_{i}}}{z-z_{i}}\right)\widetilde{G}_{\Lambda_{i}}(z_{i},\z_{i},u_{i}) + \rom{contact}.
\end{aligned}
\end{equation}
Finally, the piece independent of $\z$ is obtained by the construction

\begin{equation}
    L^{Y}(z) = D^{3}_{z}Y^{z}_{1}(z,\z) - z^{2}K^{Y}(z) - 2zJ^{Y}(z),
\end{equation}
which results, finally, in the following operator product expansion when acting on a dressed asymptotic state

\begin{equation}
\begin{aligned}
     L^{Y}(z)\widetilde{G}_{\Lambda_{i}}(z_{i},\z_{i},u_{i}) =  \frac{1}{2\pi}\left(\frac{-2h_{i}\z_{i}-\z^{2}_{i}D_{\z_{i}}}{z-z_{i}}\right)\widetilde{G}_{\Lambda_{i}}(z_{i},\z_{i},u_{i}) + \rom{contact}.
\end{aligned}
\end{equation}
Naturally, the antiholomorphic conjugate generators $\K^{V}$, $\J^Y$ and $\el^{Y}$ are defined in the same way with all the helicities and position variables conjugated. The modes, defined the usual way, are inferred then to obey the following set of commutation relations with the external dressed states (we list all of them in what follows) - 

\begin{equation}
    [K^{Y}_{p},\widetilde{G}_{\Lambda_{i}}(z_{i},\z_{i},u_{i})] = -\frac{1}{2\pi}z_{i}^{p}D_{\z_{i}}\widetilde{G}_{\Lambda_{i}}(z_{i},\z_{i},u_{i}),
\end{equation}

\begin{equation}
    [J^{Y}_{p},\widetilde{G}_{\Lambda_{i}}(z_{i},\z_{i},u_{i})] = \frac{1}{2\pi}z_{i}^{p}(h_{i}+\z_{i}D_{\z_{i}})\widetilde{G}_{\Lambda_{i}}(z_{i},\z_{i},u_{i}),
\end{equation}

\begin{equation}
    [J^{Y}_{p},\widetilde{G}_{\Lambda_{i}}(z_{i},\z_{i},u_{i})] = \frac{1}{2\pi}z_{i}^{p}(-2h_{i}\z_{i}-\z^{2}_{i}D_{\z_{i}})\widetilde{G}_{\Lambda_{i}}(z_{i},\z_{i},u_{i}),
\end{equation}
with conjugate commutation relations holding between the modes for the antiholomorphic generators. Upon inspection, the reader will observe that the zero mode generators, namely $K^{Y}_{0}$, $J^{Y}_{0}$, $L^{Y}_{0}$ and their antiholomorphic counterparts actually generate a particular class of diffeomorphisms. Specifically, the holomorphic operators make the transformation

\begin{equation}
    \z \longrightarrow \z + \sum_{i=0}^{2}\z^{i}F_{i}(z),
\end{equation}
and the conjugate operators effect

\begin{equation}
    z \longrightarrow z + \sum_{i=0}^{2}z^{i}F_{i}(\z).
\end{equation}
It is known that these actually form a basis for the Lorentz generators in $4$ dimensions - which makes sense; the subleading soft graviton theorem is known to be a realization of the Ward identity corresponding to superrotations, of which Lorentz transformations form the smallest closed subgroup. 

Finally, let us list the contact terms that we had alluded to earlier. We have for the three generators the following

\begin{equation}
    K^{Y}(z)\widetilde{G}_{\Lambda_{i}}(z_{i},\z_{i},u_{i})|_{\rom{contact}} \sim \frac{h_{i}}{2}\delta^{2}(z-z_{i},\z-\z_{i})\widetilde{G}_{\Lambda_{i}}(z_{i},\z_{i},u_{i}),
\end{equation}
 
 \begin{equation}
 \begin{aligned}
      &J^{Y}(z)\widetilde{G}_{\Lambda_{i}}(z_{i},\z_{i},u_{i})|_{\rom{contact}} \sim\\ 
      &-\frac{1}{2}\left(h_{i}\z_{i}\dfunction + h_{i}\z\dfunction\right)\widetilde{G}_{\Lambda_{i}}(z_{i},\z_{i},u_{i})\\
       &-\frac{1}{2}\left(\z_{i}^{2}\dfunction D_{\z_{i}}+h_{i}\z^{2}D_{\z}\dfunction\right)\widetilde{G}_{\Lambda_{i}}(z_{i},\z_{i},u_{i}),
 \end{aligned}
 \end{equation}
 and
 
  \begin{equation}
 \begin{aligned}
      &L^{Y}(z)\widetilde{G}_{\Lambda_{i}}(z_{i},\z_{i},u_{i})|_{\rom{contact}} \sim\\ 
      &-\frac{1}{2}\left(h_{i}\z_{i}\dfunction + 2h_{i}\z\dfunction\right)\widetilde{G}_{\Lambda_{i}}(z_{i},\z_{i},u_{i})\\
       &-\frac{1}{2}\left(\z_{i}^{2}\dfunction D_{\z_{i}}+3h_{i}\z^{2}D_{\z}\dfunction\right)\widetilde{G}_{\Lambda_{i}}(z_{i},\z_{i},u_{i})\\
       &-\frac{1}{2}\left(h_{i}z\z_{i}D_{\z}\dfunction + h_{i}\z^{2}D_{\z}\dfunction\right)\widetilde{G}_{\Lambda_{i}}(z_{i},\z_{i},u_{i})\\
       &-\frac{1}{2}\left(\z_{i}^{2}\z D_{\z}\dfunction D_{\z_{i}}+h_{i}\z^{3}D^{2}_{\z}\dfunction\right)\widetilde{G}_{\Lambda_{i}}(z_{i},\z_{i},u_{i}).
 \end{aligned}
 \end{equation}
Fortunately for us, despite the complexity of these terms, they vanish generically; the soft particle generally is not assumed to be collinear, in which case all the terms are zero. Otherwise, they always vanish when modes are computed by performing contour integrals. Either way, they don't seem to be of much physical significance, although it remains to be seen whether or not this is really true. Now, we will move on to consider the most complicated of the soft theorems in this work, namely the soft graviton theorem at subsubleading order.
 
\subsection{The Subsubleading Soft Graviton Theorem}\label{sec:4.4}
We move on now to the final case of the soft graviton theorem. By now, we can anticipate relatively easily what the form of the action will have to be. However, it is worth pointing out a subtle fact, which won't be especially important for our purposes, but still worth knowing. The subsubleading soft graviton theorem is known to arise from a Ward identity which is labelled by a tensor valued object on the celestial sphere. Indeed, this is the simplest reflection of the fact that this theorem arises really out of two copies of the subleading soft gluon theorem. 

In this vein, we consider two pairs of fields, $X^{zz}_{i}$ and $X^{\z\z}_{i}$, which form the diagonal elements of a second rank tensor on the celestial sphere and propose the following higher derivative action coupling them together

\begin{equation}
    \mathcal{I}^{(2)}_{\rom{grav}} = \int \left(D^{4}_{z}X^{zz}_{1}(z,\z)D^{4}_{\z}X^{zz}_{2}(z,\z) + D^{4}_{z}X^{\z\z}_{1}(z,\z)D^{4}_{\z}X^{\z\z}_{2}(z,\z)\right)dz\wedge d\z.
\end{equation}
This time, the kinetic term is dominated by a \emph{tetraharmonic} operator, or the fourth power of the Laplacian on the celestial sphere. The Green's function for this object is probably easy to guess at this point; we have the identity

\begin{equation}
    D^{4}_{z}D^{4}_{\z}\left(|z-z'|^{6}\ln|z-z'|^{2}\right) = 36\pi\delta^{2}(z-z',\z-\z').
\end{equation}
Taking note of the structure of the subsubleading soft factor in gravity, we observe that we need to obtain three types of terms, namely ones which are cubic, quadratic and linear in the separation between the soft particle position and the external puncture. Constructing the right charge that will produce these terms is achieved by taking note of the fact that we have the following identities

\begin{equation}
    D^{4}_{z}\left(|z-z_{i}|^{6}\ln|z-z_{i}|^{2}\right) = 6\frac{(\z-\z_{i})^{3}}{z-z_{i}}, 
\end{equation}

\begin{equation}
    D_{\z_{i}}D^{4}_{z}\left(\frac{(\z-\z_{i})^{3}}{z-z_{i}}\right) = -18\frac{(\z-\z_{i})^{2}}{z-z_{i}},
\end{equation}
and

\begin{equation}
    D^{2}_{\z_{i}}D^{4}_{z}\left(\frac{(\z-\z_{i})^{3}}{z-z_{i}}\right) = 36\frac{\z-\z_{i}}{z-z_{i}}.
\end{equation}
These expressions correctly reproduce all the moving parts of the positive helicity subsubleading soft theorem (with the negative helicity counterparts obtained readily by complex conjugation). To build the charges, in addition to the preceding three identities, the nontrivial operator product expansions

\begin{equation}
    \ab{X^{zz}_{1}(z,\z)X^{zz}_{2}(z',\z')} = \frac{1}{36\pi}|z-z'|^{4}\ln|z-z'|^{2},
\end{equation}
and

\begin{equation}
    \ab{X^{\z\z}_{1}(z,\z)X^{\z\z}_{2}(z',\z')} = \frac{1}{36\pi}|z-z'|^{4}\ln|z-z'|^{2},
\end{equation}
lead us to suggest the following

\begin{equation}
    Q^{X}(h_{i},z_{i},\z_{i}) = -\left(\frac{2}{6}h_{i}(2h_{i}-1)D^{2}_{\z_{i}}X^{zz}(z_{i},\z_{i})+\frac{4}{3}h_{i}D_{\z_{i}}X^{zz}_{1}(z_{i},\z_{i})D_{\z_{i}} +X^{zz}_{2}(z_{i},\z_{i})D_{\z_{i}}^{2}\right),
\end{equation}
and

\begin{equation}
    \Q^{X}(\h_{i},z_{i},\z_{i}) = -\left(\frac{2}{6}\h_{i}(2\h_{i}-1)D^{2}_{z_{i}}X^{\z\z}(z_{i},\z_{i})+\frac{4}{3}\h_{i}D_{z_{i}}X^{\z\z}_{1}(z_{i},\z_{i})D_{z_{i}} +X^{\z\z}_{2}(z_{i},\z_{i})D_{z_{i}}^{2}\right).
\end{equation}
We dress asymptotic operators using a generalized form of the Wilson line in terms of these charges. This time, just like the case of the subleading soft theorem in gauge theory, the soft factor depresses the weights by $1$ when it acts on an amplitude. As a result of this, we need to define the Wilson line as

\begin{equation}
    \mathcal{W}^{X}_{\rom{grav}}(h_{i},\h_{i},z_{i},\z_{i}) = \exp\left(i\kappa\left(Q^{X}(h_{i},z_{i},\z_{i})+\Q^{X}(\h_{i},z_{i},\z_{i})\right)\mathcal{P}^{-1}_{i}\right),
\end{equation}
and dressed operators by setting

\begin{equation}
    \widetilde{G}^{X}_{\Lambda_{i}}(z_{i},\z_{i},u_{i}) = \mathcal{W}^{X}_{\rom{grav}}(h_{i},\h_{i},z_{i},\z_{i})G_{\Lambda_{i}}(z_{i},\z_{i},u_{i}).
\end{equation}
It can now be verified by direct computation that we have

\begin{equation}
    \begin{aligned}
    &\lab{D^{4}_{z}X^{zz}_{1}(z,\z)\widetilde{G}^{X}_{\Lambda_{1}}(z_{1},\z_{1},u_{1})\dots \widetilde{G}^{X}_{\Lambda_{n}}(z_{n},\z_{i},u_{n})} = \\
    &\left(\sum_{i=1}^{n}\left(\frac{\kappa}{6\pi}\frac{1}{z-z_{i}}\left(-2h_{i}(2h_{i}-1)(\z-\z_{i})+4(\z-\z_{i})^{2}h_{i}D_{\z_{i}} -(\z-\z_{i})^{3}D^{2}_{\z_{i}}\right)\mathcal{P}^{-1}_{i}\right)\right)\\
    &\times \lab{G_{\Lambda_{1}}(z_{1},\z_{1},u_{1})\dots G_{\Lambda_{n}}(z_{n},\z_{i},u_{n})},
    \end{aligned}
\end{equation}
which is the positive subsubleading soft theorem for gravitons. The negative helicity result is recovered by making the replacement $D^{4}_{z}X_{1}^{zz}(z,\z)\rightarrow D^{4}_{\z}X^{\z\z}_{1}(z,\z)$.

The asymptotic symmetry algebra generated by the subsubleading soft theorem is obtained by isolating the currents that are proportional to $\z^{i}$, where $i$ runs from zero to $3$. This is of course due to the nature of the solutions of the tetraharmonic equation. We start with the term that is proportional to $z^{3}$; note that by defining

\begin{equation}
    J^{X}(z) = \frac{1}{6}D_{\z}^{3}D_{z}^{4}X^{zz}_{1}(z,\z),
\end{equation}
we have the following expansion in terms of the dressed asymptotic states

\begin{equation}
    J^{X}(z)\widetilde{G}^{X}_{\Lambda_{i}}(z_{i},\z_{i},u_{i}) \sim -\frac{\kappa}{6\pi}\frac{1}{z-z_{i}}D_{\z_{i}}^{2}\mathcal{P}^{-1}_{i}\widetilde{G}^{X}_{\Lambda_{i}}(z_{i},\z_{i},u_{i}) + \rom{contact}.
\end{equation}
This is identified with the operator $S^{3}$ in the notation of \cite{Banerjee:2021cly}. To obtain the $S^{2}$, we define

\begin{equation}
    K^{X}(z) = \frac{1}{2}D_{\z}^{2}\left(D^{4}_{z}X_{1}^{zz}(z,\z) - \frac{z^{3}}{6}D_{\z}^{3}D_{z}^{4}X^{zz}_{1}(z,\z)\right).
\end{equation}
This isolates the part of the operator product expansion in the subsubleading term that is proportional to $\z^{2}$; we have

\begin{equation}
    K^{X}(z)\widetilde{G}^{X}_{\Lambda_{i}}(z_{i},\z_{i},u_{i}) \sim \frac{\kappa}{6\pi}\frac{4h_{i}D_{\z_{i}}+3\z_{i}D_{\z_{i}}^{2}}{z-z_{i}}\mathcal{P}^{-1}_{i}\widetilde{G}^{X}_{\Lambda_{i}}(z_{i},\z_{i},u_{i}) + \rom{contact}.
\end{equation}
Extracting the linear term is then done by making another subtraction - 

\begin{equation}
    L^{X}(z) = D_{\z}\left(D^{4}_{z}X_{1}^{zz}(z,\z) - \z^{2}K^{X}(z)\right),
\end{equation}
from which we find that

\begin{equation}
    L^{X}(z)\widetilde{G}^{X}_{\Lambda_{i}}(z_{i},\z_{i},u_{i}) \sim -\frac{\kappa}{6\pi}\frac{2h_{i}(2h_{i}-1)+8h_{i}\z_{i}D_{\z_{i}}+3\z^{2}_{i}D_{\z_{i}}^{2}}{z-z_{i}}\mathcal{P}^{-1}_{i}\widetilde{G}^{X}_{\Lambda_{i}}(z_{i},\z_{i},u_{i}) + \rom{contact}.
\end{equation}
The homogeneous term is then found by taking what is left,

\begin{equation}
    M^{X}(z) = D^{4}_{z}X_{1}^{zz}(z,\z) - \z L^{X}(z),
\end{equation}
which gives us,

\begin{equation}
    M^{X}(z)\widetilde{G}^{X}_{\Lambda_{i}}(z_{i},\z_{i},u_{i}) \sim \frac{\kappa}{6\pi}\frac{2h_{i}(2h_{i}-1)\z_{i}+4h_{i}\z_{i}^{2}D_{\z_{i}}+\z^{3}_{i}D_{\z_{i}}^{2}}{z-z_{i}}\mathcal{P}^{-1}_{i}\widetilde{G}^{X}_{\Lambda_{i}}(z_{i},\z_{i},u_{i}) + \rom{contact}.
\end{equation}
The modes of these currents, which are denoted the usual way supply the following commutators when evaluated with asymptotic fields

\begin{equation}
    [J^{X}_{p},\widetilde{G}^{X}_{\Lambda_{i}}(z_{i},\z_{i},u_{i})] = -\frac{\kappa}{6\pi}z_{i}^{p}D_{\z_{i}}^{2}\mathcal{P}^{-1}_{i}\widetilde{G}^{X}_{\Lambda_{i}}(z_{i},\z_{i},u_{i}),
\end{equation}

\begin{equation}
    [K^{X}_{p},\widetilde{G}^{X}_{\Lambda_{i}}(z_{i},\z_{i},u_{i})] = \frac{\kappa}{6\pi}z_{i}^{p}(4h_{i}D_{\z_{i}}+3\z_{i}D_{\z_{i}}^{2})\mathcal{P}^{-1}_{i}\widetilde{G}^{X}_{\Lambda_{i}}(z_{i},\z_{i},u_{i}),
\end{equation}

\begin{equation}
    [L^{X}_{p},\widetilde{G}^{X}_{\Lambda_{i}}(z_{i},\z_{i},u_{i})] = -\frac{\kappa}{6\pi}z_{i}^{p}(2h_{i}(2h_{i}-1)+8h_{i}\z_{i}D_{\z_{i}}+3\z^{2}_{i}D_{\z_{i}}^{2})\mathcal{P}^{-1}_{i}\widetilde{G}^{X}_{\Lambda_{i}}(z_{i},\z_{i},u_{i}),
\end{equation}
and

\begin{equation}
    [M^{X}_{p},\widetilde{G}^{X}_{\Lambda_{i}}(z_{i},\z_{i},u_{i})] = \frac{\kappa}{6\pi}z_{i}^{p}(2h_{i}(2h_{i}-1)\z_{i}+4h_{i}\z_{i}^{2}D_{\z_{i}}+\z^{3}_{i}D_{\z_{i}}^{2})\mathcal{P}^{-1}_{i}\widetilde{G}^{X}_{\Lambda_{i}}(z_{i},\z_{i},u_{i}).
\end{equation}
The algebra formed by these generators was studied in \cite{Banerjee:2021cly}, by computing not only the commutation relations amongst themselves, but with the Lorentz and subleading soft algebra as well. As it turns out, the algebra is not closed and all commutators must be included in defining the algebra. The algebra is considerably complicated with the inclusion of the antiholomorphic modes, which mix with the holomorphic ones when a single commutation relation is evaluated. 

Lastly, we will just make a small comment on the nature of the contact terms mentioned earlier in the context of the operator product expansions of the currents with the fields. Like in the case of the subleading theorem, the contact terms are unimportant insofar as the modes are concerned, as they drop out of the contour integrals used to evaluate them. The contact terms in subsubleading case are unnecessarily complicated (the most singular terms in which involve fourth order derivatives of the delta function.), so we won't repeat their expressions here. 

\subsection{Comments on Double Copy Aspects}\label{sec:4.5}
In this last section, we would like to say a few words about a class of double copy structures that have shown up in our study of two-dimensional models of soft theorems. First, let us briefly recall what the double copy is, and why we think it is rather natural that this structure show up in the context of our two-dimensional models.

The double copy is a conjecture that in the broadest sense suggests that analytical objects in a class of gravitational theories can be obtained from similar objects in gauge theory by some kind of squaring operation. This is a fairly generic statement, since the object that is being squared can really be different given the theory. Accordingly, it is often necessary to specify the kind of double copy being considered, since the nature of the various manifestations thereof can be rather different.

Take the simple case of tree level amplitudes in gauge theory and gravity. Here, the double copy manifests in a particularly striking way, by means of the relation due to Kawai, Lewellen and Tye, now known as the KLT relation \cite{Kawai:1985xq}. The KLT relation tells us that given a two bases of colour ordered amplitudes in the gauge theory, spanned by $(n-3)!\times (n-3)!$ amplitudes for $n$ particle scattering, the amplitude for $n$ graviton scattering is obtained by fusing the two vectors of amplitudes using a kinematic object known as the KLT kernel\footnote{The field theory KLT relation descends from its string theory counterpart linking open and closed string amplitudes. There, the kernel is known to be computed by a generalization of the biadjoint scalar theory \cite{Mizera:2016jhj,Mizera:2017cqs,Mizera:2019gea}}. The KLT relation has since been generalized to theories with spin \cite{Johansson:2015oia,delaCruz:2016wbr,Brown:2018wss,Johansson:2019dnu,Bautista:2019evw}, soft theorems \cite{Aoude:2019xuz}, scalar theory with polynomial interactions \cite{Kalyanapuram:2020axt}, celestial amplitudes \cite{Kalyanapuram:2020aya} and winding string amplitudes \cite{Gomis:2021hxa}.

Another, more commonly discussed, variety of the double copy relation is the one originally due to Bern, Carassco and Johansson\footnote{The literature here is vast; see \cite{Borsten:2015pla,Bern:2019prr,Borsten:2020bgv} for reviews and \cite{Bern:2012cd,Bern:2017ucb,Bern:2019prr} and references therein (and thereof) for state of the art applications.} \cite{Bern:2005hs,Bern:2010ue,Bern:2010yg}, known eponymously as the BCJ relation. Their claim rests on the following putative expansion of the \emph{integrand} of a gauge theory amplitude for $n$ particle scattering at $L$ loops - 

\begin{equation}
    \mathcal{I}_{n,L} = \sum_{\Gamma}\frac{n_{\Gamma}c_{\Gamma}}{D_{\Gamma}},
\end{equation}
where for a given trivalent graph $\Gamma$ the $c_{L}$ and $n_{L}$ are colour and kinematical factors. The BCJ double copy now asserts that in the event such that for every triple $(\Gamma_{s},\Gamma_{t},\Gamma_{u})$ the relation

\begin{equation}
    c_{\Gamma_{s}}+c_{\Gamma_{t}}+c_{\Gamma_{u}} = 0,
\end{equation}
holds the relation

\begin{equation}
    n_{\Gamma_{s}}+n_{\Gamma_{t}}+n_{\Gamma_{u}} = 0,
\end{equation}
is obeyed as well, the replacement $c_{\Gamma}\longrightarrow n_{\Gamma}$ for every $\Gamma$ yields the $n$ particle integrand for graviton scattering at $L$ loops. 

One common feature shared by both of the preceding double copy constructions is that they are manifestly kinematic in nature. They involve relating a colour structure on the side of the gauge theory to a kinematical structure on the gravity side, such that a kind of `squaring' operation provides a map from one to the other. Informed by this, we now ask if there is some analogue of the double copy that appears in the two-dimensional models we have studied in this work. We note first the double copy prescription already pointed out in the leading soft theorems. Specifically, the following na\"ive replacements

\begin{equation}
    (D_{z},D_{\z},e_{i})\longrightarrow(D_{z}^{2},D_{\z}^{2},\kappa\omega_{i}),
\end{equation}
and

\begin{equation}
    \phi^{}_{a}(z,\z)\longrightarrow \sigma^{}_{a}(z,\z),
\end{equation}
map the soft $S$-matrix in QED to the soft $S$-matrix in gravity. The squaring operation here is applied to the derivatives on the celestial sphere rather than to specific colour factors, while the charges are mapped to the external energies of the particles. 

The double copy structure beyond leading order are a little less natural as far as appearance is concerned. Take the cases of the subleading soft theorems in QED and gravity. In QED, the two-dimensional model is built out of a pair of vector fields on the sphere that obey the biharmonic equation, while it is a pair of vectors obeying a triharmonic equation that defines the subleading soft theorem in gravity. Obviously, there isn't any obvious double copy that relates the two. More importantly, there is no obvious `replacement' rule that helps us move from a QED description to the soft theorem in gravity at subleading order. The same appears to be the case for the subsubleading soft theorem in gravity; no clean replacement rule yields the dressing operator in the subsubleading case from dressing operators in QED.

An inexact double copy structure between the \emph{soft currents} however is easier to infer. Note that the subleading soft charges in QED and gravity are related by the replacements

\begin{equation}
    (D_{z}^{2},D_{\z}^{2},V_{1}^{z},V_{1}^{\z})\longrightarrow(D_{z}^{3},D_{\z}^{3},Y_{1}^{z},Y_{1}^{\z}),
\end{equation}
with the following analogous map from the subleading QED soft current and the subsubleading soft current in gravity

\begin{equation}
    (D_{z}^{2},D_{\z}^{2},V_{1}^{z},V_{1}^{\z})\longrightarrow(D_{z}^{4},D_{\z}^{4},X_{1}^{zz},X_{1}^{\z\z}).
\end{equation}
These realize at the level of the soft currents the double copy relations mentioned in \mref{eq:4.18} to \mref{eq:4.20}. While this is at most a schematic connection, this is perhaps the case due to the fact that there are multipole constructions that encode the double copy structure enjoyed by soft theorems (as discussed in \cite{Bautista:2019tdr}). The precise connection between the schematic double copy we have pointed out in the brief discussion above and \cite{Bautista:2019tdr} is likely worth further study.

More interesting however is asking whether or not this possible presentation of a double copy (which is rigorous at least at leading order) is related to the previously discussed double copies due to KLT or BCJ. Indeed, at first glance it appears that this is not the case; the two-dimensional models are related by squaring operations on the derivatives, which the KLT or BCJ relations map colour and kinematic factors in some well defined way. 

To make some concrete statements on a potential relation between these ideas, it is worth recalling some recent progress on the technical interpretation of the double copies we have discussed. Let us start with the case of the KLT relations. The KLT relations are known to arise rather directly from string theory - they are actually realized at the level of string amplitudes at genus zero, where they relate open string amplitudes to closed string amplitudes. The field theory version of the KLT relations are obtained by taking the limit $\alpha'\rightarrow 0$. 

While this serves as an indication that the double copy at tree level is fundamentally a string theoretic phenomenon, it leaves open the question of what happens at higher loops, about which it is the BCJ relation\footnote{The BCJ and KLT double copies are actually known to be \emph{equivalent} at tree level, which was proven making use of the formalism due to Cachazo, He and Yuan \cite{Cachazo:2013gna,Cachazo:2013hca,Cachazo:2013iaa,Cachazo:2013iea,Cachazo:2014nsa,Cachazo:2014xea,Baadsgaard:2015voa,Baadsgaard:2015ifa,Baadsgaard:2015hia,Baadsgaard:2016fel} (CHY) in \cite{Bjerrum-Bohr:2016axv}.} that makes a statement. Progress along these lines was made in \cite{Mizera:2019gea}, where it was shown using the CHY formalism that when scattering amplitudes are written in a form that manifests the KLT relation, the kinematic Jacobi identity follows as a direct consequence of a global residue theorem. Since the global residue theorem can be applied to Riemann surfaces of any genus, this manner of deriving the BCJ relations removed the restriction to tree level, lending further credence to the possibility that the double copy is really of string theoretic origin\footnote{The duality between the global residue theorem and BCJ numerators was derived by using moduli space half integrands derived from perturbative superstring theory at tree level. All-loop generalizations of these expressions were found by the present author in \cite{Kalyanapuram:2021xow,Kalyanapuram:2021vjt}.}. In this light, it may seem at first glance that the celestial double copy structures in this work have no relation to the more conventional double copies already understood in the literature.

However, this conclusion might be too quick, for the following reasons. In \cite{Casali:2020uvr}, the authors observed that tree level scattering amplitudes in the celestial basis could be written in a form that manifested the double copy, much like the CHY formula manifests the double copy in the momentum basis. Furthermore, in \cite{Nguyen:2021qkt}, the authors showed that the two-dimensional model for soft gravitons can actually be derived directly from the action of four dimensional gravity by carefully evaluating it on the boundary. Showing that this is simply a presentation of the more standard double copy would entail repeating this analysis for QED as well and deriving the two-dimensional conformal field theory of free bosons directly from four dimensional QED. This would establish the celestial double copy at leading order as an infrared version of the standard KLT relation. Beyond this, considering the fact that in \cite{Adamo:2019ipt} the soft currents beyond leading order were directly extracted from CHY vertex operators and keeping in mind the structure found in \cite{Casali:2020uvr}, it would be interesting to see how this plays out at higher orders.

%% file: sec5conclusion.tex
\section{Discussion and Further Issues}\label{sec:5}
In this article, we have systematically constructed models on $\mathbb{CP}^{1}$ - which parametrizes each point along null infinity - to characterize soft theorems in QED and gravity, considerably extending work on this problem which applied exclusively to the leading order soft theorems in these theories. Specifically, we saw that the asymptotic charges - the Ward identities of which lead to these theories - can be dualized in terms of fields which live naturally on the celestial sphere. This duality was expressed by mapping the soft charges to Noether currents of the two-dimensional models we constructed and the hard charges to exponential operators that serve as dressings of the external hard states. In doing so, we have interpreted soft theorems as fundamentally two-dimensional phenomena, deriving them in what is essentially a holographic framework. 

The natural step beyond what we have analyzed in this work is to extend the two-dimensional picture of gauge theory and gravity beyond the soft sector to include (in the case of gauge) theory the collinear sector and perhaps the hard interactions as well. 

Additionally, the models studied in this work leave open a number of rather interesting questions that might serve as valuable problems in their own right. While a detailed study these are beyond the scope of this particular paper, we present a survey of these ideas in the following extended discussion.

\subsection*{Fermionic Soft Theorems}
Throughout this paper, we have focused our attention on soft theorems that arise due to the emission of bosons, specifically those of spin-$1$ and spin-$2$. However, it is known that there are a class of soft theorems due to the emission of infrared fermions in theories with supersymmetry which bear several qualitative similarities with the soft photon and graviton theorems. Consequently, it might be possible to reconstruct these results as a consequence of the dynamics of two-dimensional fields. Let us briefly review first how these soft theorems are presented conventionally. We begin with the soft fermion theorem due to spin-$\frac{1}{2}$ emission.

Denote a generic scattering amplitude in momentum space by $\mathcal{M}_{n}$. Let us assume furthermore that the theory has a global $\mathcal{N}=1$ supersymmetry. Following the radiation of one soft photino of helicity $\frac{1}{2}$, the new amplitude, denoted by $\mathcal{M}_{n+1}$, obeys the following relation

\begin{equation}
    \lim_{\omega\rightarrow 0}\omega\mathcal{M}_{n+1} = \sum_{i=1}^{n}\frac{e_{i}\eta_{i}}{z-z_{i}}\mathcal{F}_{i}\mathcal{M}_{n}.
\end{equation}
Notably, unlike in the case of the soft photon theorem, this theorem differs due to the presence of the operators $\mathcal{F}_{i}$. For an external state labelled by $i$, the operator $\mathcal{F}_{i}$ is a supersymmetry operator. Generically, it transforms the state $i$ from a boson to a fermion or vice versa, in line with the fact that a fermion cannot interact with two fermions or two bosons. The external state accoringly must undergo a reversal of statistics.  

Now the important point to note is that although the soft factor for the soft photino theorem is \emph{qualitatively} rather different than the soft photon factor (it is operator valued for one thing), quantitatively it is essentially identical. Indeed, if we make the simple replacement

\begin{equation}
    e_{i} \longrightarrow e_{i}\mathcal{F}_{i},
\end{equation}
to the dressing operators of the dual Coulomb gas model controlling the soft photon at leading order, the Ward identities of the theory yield precisely the soft photino theorem. This however is an incomplete description due to the fact that unlike in the case of the soft photon theorem, the soft photino theorem for negative helicity radiation will not be obtained by making the replacement $D_{z}\longrightarrow D_{\z}$. This is due to the fact that when a negative helicity soft photino is radiated, one must also make the replacement $\mathcal{F}_{i}\longrightarrow \mathcal{F}^{\dagger}_{i}$. 

We run into a similar problem when we want to find two-dimensional duals for the soft theorem due to radiation of low energy gravitinos - the soft photino theorem naturally generalizes to supply a soft gravitino theorem of the form

\begin{equation}
    \lim_{\omega\rightarrow 0}\omega\mathcal{M}_{n+1} = \sum_{i=1}^{n}\kappa\eta_{i}\omega_{i}\frac{\z-\z_{i}}{z-z_{i}}\mathcal{F}_{i}\mathcal{M}_{n}.
\end{equation}
We see here that an entirely analogous problem is encountered. Indeed, it would seem most na\"ively that the generalization from the soft graviton to the soft gravitino theorem proceeds by performing the replacement of

\begin{equation}
    \omega_{i} \longrightarrow \omega_{i}\mathcal{F}_{i},
\end{equation}
which is not sufficient to describe the full soft gravitino theorem due to its missing the contribution due to the radiation of a negative helicity gravitino, which contains the operators $\mathcal{F}_{i}^{\dagger}$.

Solving this question perhaps rests on viewing the soft theorems for fermions in a manner similar to how we view the higher order soft theorems in gauge theory and gravity. There, the positive and negative helcity sectors were handled independently, with the dressing operator carrying both contributions decoupled from one another. The soft currents, then defined in terms of separate fields, could be used to extract the relevant soft theorems. One way to import this approach would be to have recourse to the chiral splitting of the two-dimensional fields $\phi$ and $\sigma$. Lets look at how to do this in turn.

In the case of the soft photon theorem (we work for the time being using a single component field $\phi$ to save ourselves of the profusion of indices), recall the operator product expansion

\begin{equation}
    \langle{\phi(z,\z)\phi(z',\z')\rangle} = \frac{1}{\pi}\ln|z-z'|^{2}.
\end{equation}
Chiral splitting proceeds by noting that the logarithm maps addition on to multiplication due to the trivial identity

\begin{equation}
    \ln|z-z'|^{2} = \ln(z-z') + \ln(\z-\z').
\end{equation}
This motivates a specific decomposition of the field into holomorphic and holomorphic parts (guided alternatively by the fact that it is harmonic on $\mathbb{CP}^{1}$) as follows

\begin{equation}
    \phi(z,\z) = \varphi(z) + \widetilde{\varphi}(\z).
\end{equation}
These fields, decoupled from each other, give rise to operator product expansions that go as $\sim\ln(z-z')$ and $\sim\ln(\z-\z')$ respectively. By decoupling the soft sectors in this fashion, the Wilson line can be refined by writing

\begin{equation}
    \mathcal{W}_{\rom{SQED}}(z_{i},\eta_{i}e_{i},\mathcal{F}_{i}) = \exp\left(ie_{i}\eta_{i}(\varphi(z)\mathcal{F}_{i} + \widetilde{\varphi}(\z)\mathcal{F}_{i}^{\dagger})\right).
\end{equation}
Using this Wilson line, no change needs to be made to the soft currents; when the soft currents $D_{z}\phi(z,\z)$ and $D_{\z}\phi(z,\z)$ are inserted into a correlation function of Wilson lines of the preceding form, the soft photino theorems are obtained as the Ward identities, as desired.

Splitting the soft field for gravity can also be done in a chiral fashion, although it's a little more involved due to the trickier analytic structure of the Greens function. A good way to proceed is to frame the question in a way that makes the goals more evident. Specifically, since we want a chiral splitting that isolates the positive and negative helicity pieces in a decoupled fashion, what we are really looking for is a splitting of the form

\begin{equation}
    \sigma(z,\z) = \theta(z,\z) + \widetilde{\theta}(z,\z),
\end{equation}
where

\begin{equation}
    D_{\z}^{2}\theta(z,\z) = 0,
\end{equation}
and

\begin{equation}
    D_{z}^{2}\widetilde{\theta}(z,\z) = 0.
\end{equation}
This is conveyed by breaking up the Greens function making use of the identity that states 

\begin{equation}
    |z-z'|^{2}\ln|z-z'|^{2} = |z-z'|^{2}\ln(z-z') + |z-z'|^{2}\ln(\z-\z').
\end{equation}
Indeed, the first term will be computed by the OPE due to $\theta$ and the latter term will arise due to the $\widetilde{\theta}$ OPE. Each of these fields will then be decoupled from one another due to the combined constraints on second order derivatives. As a side note, this splitting is natural from the point of view of solutions of the biharmonic equation, making reference to the expansion in 

This splitting enables a refinement of the Wilson line to encode the correct chiral form of the soft expansion in the radiation of a soft gravitino. By introducing the Wilson line

\begin{equation}
    \mathcal{W}_{\rom{sgrav}}(u_{i},z_{i},\z_{i},\mathcal{F}_{i}) = \exp\left(i\kappa\eta_{i}(\theta(z_{i},\z_{i})\mathcal{F}_{i} + \widetilde{\theta}(z_{i},\z_{i})\mathcal{F}_{i}^{\dagger})\partial_{u_{i}}\right),
\end{equation}
the Ward identities due to the currents $D_{z}^{2}\sigma(z,\z)$ and $D_{\z}^{2}\sigma(z,\z)$ automatically yield the soft theorems due to positive and negative helicity soft gravitino radiation without any further modifications.

The procedure described here to prescribe two-dimensional models for gravity is attractive, but it isn't clear if this is the entire story. Specifically, it is worth further study to check whether or not the multiple soft gravitini emission theorems are reproduced by making the appropriate modifications according to section \ref{sec:2.4}. It will also perhaps prove valuable to search for soft theorems due to soft fermionic radiation beyond leading order and ask whether or not the methods of the present work will be helpful is describing such theorems.

\subsection*{Global Symmetries and Conserved Quantities}
One theme of the present work, which among other things served to make clear the relation of the two-dimensional models presented here to the conventional realization of soft theorems as Ward identities, was the appearance of global symmetries associated to certain shifts of the two-dimensional fields which kept the corresponding actions invariant. In particular, these global symmetries made the conservation of certain observables a consequence, rather than an assumption.

In QED, the leading order action was invariant under the shift of the asymptotic field $\phi$ by a constant, while the relevant shift was along a four parameter group in the basis $\lbrace{1,z,\z,z\z\rbrace}$ in the case of gravity. In the language of conserved currents, these are shifts obeyed

\begin{equation}
    D_{z}f(z,\z) = D_{\z}f(z,\z) = 0,
\end{equation}
in QED and

\begin{equation}
    D_{z}^{2}f(z,\z) = D^{2}_{\z}f(z,\z) = 0,
\end{equation}
in gravity, at leading order in the soft expansion. These turned out to the precisely the shifts which preserved the structure of the so-called soft charges at the level of the Ward identities, which had the impact of ensuring that global charge conservation and momentum conservation held identically (see Appendix \ref{app} for an extensive discussion of this point).

These natural implications of the global symmetries at the level of the leading order soft theorems don't serve us especially well when we move beyond leading order. Indeed, due to the higher derivative nature of the corresponding two-dimensional models, the global symmetries are enhanced, leading to invariance under functions that satisfy the equations (using the variable $N$ an in the Introduction)

\begin{equation}
    D^{N}_{z}f(z,\z) = D^{N}_{\z}f(z,\z) = 0,
\end{equation}
which has the general solution

\begin{equation}
    f(z,\z) = \sum_{i,j=0}^{N-1}c_{ij}z^{i}\z^{j}.
\end{equation}
where the $c_{ij}\in\mathbb{C}$. For example, in the case of the subleading soft graviton theorem, where $N=3$, this corresponds to the basis $\mathcal{B}_{3} = \lbrace{1,z,\z,z^{2},\z^{2},z\z,(z\z)^{2}\rbrace}$. Along the lines of the arguments used to insist on charge conservation in the Coulomb gas model, such shifts have the effect of implying conservation laws on the scattering amplitude when applied to the exponential operators that dress the external states. In the Ward identity language, an example of such a conservation law in the case of the subleading soft graviton theorem would be to state that the operator

\begin{equation}
    \mathcal{D} = \sum_{i=1}^{n}\mathcal{J}^{(1)}_{f}(z_{i},\z_{i}),
\end{equation}
for any $f$ expanded in the basis $\mathcal{B}_{3}$ annihilates the amplitude. It seems to be the case that such conservation laws are implied not only in the language of two-dimensional models, but are intrinsic to the Ward identity formalism as well. At the same time, it appears that they have so far received little, if any, attention. Understanding the nature of such conservation laws is likely important at least insofar as constructing consistent amplitudes are concerned. More speculatively, they might serve as bootstrap constraints to kinematically construct scattering amplitudes as solutions of the conservation laws.

On a related note, it might really be the case that the dual models for higher order soft theorems can be derived directly by performing expansions of the bulk fields on the null boundary, appropriating and extended the calculations performed in \cite{Nguyen:2021qkt}. In addition to providing direct proof that the actions in this work are indeed accurate, the calculations performed may shed light on the asymptotic symmetry interpretations of the invariance under the global shifts we have discussed here.

\subsection*{Dual Models for Scalars}
In all our discussions so far, we have ultimately relied on the simplifications afforded by gauge invariance and diffeomorphism invariance to guide our understanding of soft theorems and the attendant dual models. In the absence of these essentially constraining symmetries, soft theorems end up being considerably more complicated. Worse yet, their forms indicate that a proper symmetry interpretation might not be available.

This question was studied in relative detail by Campiglia, Coito and Mizera in \cite{Campiglia:2017dpg}, where the soft theorems due to scalar emission leading to a soft factor given by

\begin{equation}
    \mathcal{S}_{\rom{scalar}} = \sum_{i=1}^{n}\frac{g}{p_{i}\cdot q},
\end{equation}
was recast as a Ward identity on the celestial sphere. Here, $g$ is the coupling constant the external field (say, a fermion) has with the scalar. The problem here is that there is no sense of holomorphicity or antiholomorphicity that one can exploit, in stark contrast with the soft theorems in gauge theory and gravity. Indeed, it seems to be the case that even when expressed as a Ward identity, the soft charge does not seem to encode any kind of asymptotic transformation. This is in contrast to the case of QED and gravity, where the smeared soft charges could readily be identified as representing large gauge transformations and supertranslations respectively.

As far as dual models are concerned, it doesn't seem at the outset that there is any guiding principle which will enable the construction of two-dimensional models two recover the soft $S$-matrix for scalars, which can be shown to take the general form

\begin{equation}
    \ln\left(\mathcal{A}^{\rom{scalar}}_{n,\rom{soft}}\right) = \sum_{i<j}\frac{g^{2}}{|z_{i}-z_{}|^{2}}\ln|z_{i}-z_{j}|^{2},
\end{equation}
where it is assumed in accordance with our general theme in this work that the external states are massless. The reader will note now that the transform given by $\sim\frac{1}{|z|^{2}}\ln|z|^{2}$ is not evidently the Greens function of any operator. Curiously, a brute force computation of the inverse of this transform can be found by applying a Fourier transform. Once this is done (we don't repeat the result, sparing the reader the form of the expression), one obtains a complicated combination of special functions. Fundamentally, it doesn't seem to be the case that there is a local quantum field theory on the two-dimensional celestial sphere that can be used to derive the soft $S$-matrix (and the soft theorems) due to scalar radiation in a holographic fashion.

It may seem that the situation with soft scalar theorems is hopeless, but it may be the case that along the lines of \cite{Campiglia:2017dpg}, there may be a possibility of deriving dual descriptions of such theorems by taking heed of the fact that one class of soft scalar theorems are due to dilatonic fields. More concretely, perhaps there is insight to be had from infrared dynamics in string models, where the graviton belongs to an $\mathcal{N}=0$ multiplet alongside the $B$ field and dilaton. Guided by the asymptotic symmetry already known in the case of gravity, there are perhaps interesting phenomena that show up in soft scalar theorems in string models \cite{Hata:1992it,DiVecchia:2015oba,DiVecchia:2015jaq,DiVecchia:2017gfi}.

\subsection*{Massive External States and Loop Corrected Soft Theorems}
Finally, we elaborate on a class of soft theorems which must be considered in realistic theories, namely those that arise when the external states are massive, rather than massless as we have assumed in the present work. To appreciate this problem, we have to go back to the momentum space representation of the soft theorems, observing that for QED, the soft factor takes the form

\begin{equation}
    \mathcal{S}_{\rom{QED}} = \sum_{i=1}^{n}\frac{p_{i}\cdot \epsilon}{p_{i}\cdot q},
\end{equation}
while for gravity one has

\begin{equation}
\mathcal{S}_{\rom{grav}} = \sum_{i=1}^{n}\frac{(p_{i}\cdot \epsilon)^{2}}{p_{i}\cdot q}.
\end{equation}
Notably, when the momenta are expanded on timelike infinity, the expressions are considerably more unilluminating. We won't really perform the expansion here (the reader can have a look at \cite{Strominger:2017zoo}); we simply note that to describe the soft $S$-matrix and soft theorems, one has to define the following smearing functions (we assume that the radiated particle is of positive helicity)

\begin{equation}
    G^{(1)}(\rho_{i},z_{i};z,\z) = D_{\z}\left(\frac{p_{i}\cdot \epsilon}{p_{i}\cdot q}\right),
\end{equation}
and

\begin{equation}
    G^{(2)}(\rho_{i},z_{i};z,\z) = D_{\z}^{2}\left(\frac{p_{i}\cdot \epsilon}{p_{i}\cdot q}\right).
\end{equation}
In doing so, we have made use of the fact that at timelike infinity, the momentum of an external state is labelled by three variables in addition to energy - a radial variable $\rho$ and the sphere variables $(z,\z)$. Indeed, the Wilson lines must now be refined by rewriting

\begin{equation}
    \phi(z_{i},\z_{i}) \longrightarrow \int dz\wedge d\z G^{(1)}(\rho_{i},z_{i};z,\z)\phi(z,\z),
\end{equation}
for QED and 

\begin{equation}
    \sigma(z_{i},\z_{i}) \longrightarrow \int dz\wedge d\z G^{(2)}(\rho_{i},z_{i};z,\z)\sigma(z,\z),
\end{equation}
for gravity. It can now be checked (it's a simple exercise) that the insertion of a soft current corresponding to positive helicity radiation (identically) supplies the soft theorem when the external states are massive. 

This trick, which really relies on the nature of the Greens functions of the dual models characterizing the leading order theorems, might require modification if we want to generalize it to the higher order soft theorems. To the author's knowledge, this has not yet been attempted, and would certainly be rather interesting to analyze. 

The study of soft theorems in the case of massive external states would be of particular relevance if we want to describe the nature of the dual models that explain loop corrections that are known to arise at logarithmic order in the soft expansion. Specifically, it was observed in \cite{Sahoo:2018lxl} that for QED, the soft theorem was modified at higher loop order, and resulted in logarithmic corrections to the subleading soft theorem to give

\begin{equation}
    \mathcal{M}_{n+1} = \ln\omega \mathcal{S}^{\rom{QED}}_{\rom{ln}}\mathcal{M}_{n}+\dots,
\end{equation}
where the dots signify other terms in the soft expansion. The factor $\mathcal{S}^{\rom{QED}}_{\rom{ln}}$ takes the schematic form

\begin{equation}
    \mathcal{S}^{\rom{QED}}_{\rom{ln}} = \sum_{i,j}f_{ij},
\end{equation}
where the sum now runs over pairs of external states. We won't repeat the functions that make up this expansion, the reader is referred to \cite{Sahoo:2018lxl}, but they are most interesting when the external states are massive. It was suggested in \cite{Campiglia:2019wxe} that the Ward identity that gives rise to this theorem ultimately involved correcting the hard piece of the charge due to the subleading soft theorem. Synthesizing these theorems with the framework in the present study will likely involved a more detailed discussion of both massive states and the dressing operators used to study the subleading soft photon theorem.

%% file: sec6appendix.tex
\section{A Review of Soft Theorems as Ward Identities}\label{app}
In this appendix, we provide a comprehensive review of the realization of soft theorems as Ward identities satisfied by the $S$-matrix in gauge theory and gravity. In particular, we will point out how these Ward identities in a broad sense anticipate the holographic models that we have presented here. 

Since in our study of soft theorems we have chosen to focus on the scattering of massless particles, much of the analysis is carried out at null infinity. As a result, before spelling out in detail the structure of the Ward identities now known to be equivalent to the soft theorems, we briefly recall how quantization is carried out on null infinity.

\subsection{Ward Identities and Field Expansions at Null Infinity}\label{app:a1}
Let's set up the basic framework in which the duality between Ward identities and soft theorems has been understood to operate. We have to go back to the generic definition of the $S$-matrix, as an operator $\mathcal{S}$ that is sandwiched between in and out states, denoted by $\ket{\rom{in}}$ and $\bra{\rom{out}}$ respectively. The main idea now is that there exists a class of charges, denoted generically by $Q$, which commute with the $S$-matrix - 

\begin{equation}
   \bra{\rom{out}} [Q,\mathcal{S}]\ket{\rom{in}}=0.
\end{equation}
As it turns out, charges can be defined in a manner that makes the foregoing equation equivalent to a soft theorem in gauge theory or gravity. Operationally, this is done by preparing an operator that enjoys the decomposition

\begin{equation}
    Q = Q^{\rom{hard}} + Q^{\rom{soft}}.
\end{equation}
The commutator with the hard operator, when properly defined, supplies the soft theorem and the soft operator creates the soft particle in the asymptotic state. As a result, the identity

\begin{equation}\label{eq:a.3}
    \bra{\rom{out}} [Q^{\rom{hard}},\mathcal{S}]\ket{\rom{in}}= -\bra{\rom{out}} [Q^{\rom{soft}},\mathcal{S}]\ket{\rom{in}},
\end{equation}
is identified with a soft theorem. 

Making this process concrete generally proceeds in two steps. Working with massless particles throughout, the creation and annihilation operators corresponding to various interacting particles are first defined in terms of the bulk fields. Since this process, known as asymptotic quantization \cite{Ashtekar:1979xeo,Ashtekar:1981bq,Ashtekar:1981hw,Ashtekar:1981sf,Ashtekar:2018lor}, has to be carried out on null infinity parametrized by $(u,z,\z)$, the fields themselves are expanded at large-$r$ according to

\begin{equation}\label{eq:a.4}
    O(r,u,z,\z) = \frac{O(u,z,\z)}{r} + \mathcal{O}\left(\frac{1}{r^{2}}\right).
\end{equation}
The leading contribution $O(u,z,\z)$, known as radiative data is used to define the creation and annihilation operators post quantization. Then the soft theorems written in celestial coordinates are mapped to corresponding charges, which themselves are defined using creation and annihilation operators, which is then easily expressed in terms of the radiative data. 

While this is a very effective \emph{ab initio} method of relating soft theorems to Ward identities, such dualities were not historically determined in this fashion. Indeed, the realization of soft theorems as Ward identities was a consequence of studying a class of charges known to be conserved in gauge theories and gravity, the Ward identities of which happened to be precisely certain soft theorems which had been discovered independently.

Understanding how the radiative data are expressed after quantization is instructive to seeing more clearly how certain conserved charges actually encode soft modes in field theory. Consider first the gauge theory case of quantum electrodynamics. Here, we define the radiative data of the electromagnetic potential in terms of a large $r$ expansion according to the prescription

\begin{equation}
    A_{\mu}(u,z,\z) = \lim_{r\rightarrow \infty}rA_{\mu}(r,u,z,\z).
\end{equation}
This just means that when the electromagnetic potential obeys a fall-off condition that is consistent with \mref{eq:a.4}, the leading term in the expansion provides the radiative data that must be quantized. Quantizing radiative data is relatively straightforward (see for example exercise 4 in \cite{Strominger:2017zoo}), and proceeds by simply performing a saddle point expansion of the usual quantized electromagnetic field, and picking out the leading term. Indeed, the result is familiar; we have the following expansion for $A_{z}$

\begin{equation}
    A_{z}(u,z,\z) = -\frac{i\sqrt{2}e}{8\pi^{2}}\int_{0}^{\infty}d\omega\left(a^{\rom{out}}_{+}e^{-i\omega u}-a^{\rom{out},\dagger}_{-}e^{i\omega u}\right),
\end{equation}
where the operator $a^{\rom{out}}_{\pm}$ destroys a photon of positive or negative helicity in the out state. In particular, integrating over the null coordinate $u$ sets $\omega$ to vanish, which is the statement that the photon created is a soft particle. Indeed, it is in this form that the electromagnetic radiative data shows up in the definition of the soft charge(s). Needless to say, an entirely analogous expansion for $A_{\z}$ exchanging the positive and negative helicity sectors holds as well. Parenthetically, the exchange of in and out states is achieved by quantizing on past null infinity, labelled by the advanced coordinate $v$ instead.

The gravitational radiative data is determined in terms of a quantity known as the shear tensor. The shear tensor is characterized by two components, denoted  by $C_{zz}$ and $C_{\z\z}$ is defined in terms of the metric perturbation $h_{\mu\nu}$ expanded on null infinity

\begin{equation}
    C_{zz} = \kappa\lim_{r\rightarrow \infty}\frac{1}{r}h_{zz}(u,z,\z),
\end{equation}
with a corresponding definition for $C_{\z\z}$, since the expansion of the metric perturbation around Minkowski begins at $r$. In terms of creation and annihilation operators, the field $C_{zz}$ once quantized reads as

\begin{equation}
    C_{zz} = \frac{-i}{4\pi^{2}}\int_{0}^{\infty}d\omega\left(b^{\rom{out}}_{+}e^{-i\omega u}-b^{\rom{out},\dagger}_{-}e^{i\omega u}\right),
\end{equation}
where the operator $b^{\rom{out}}_{\pm}$ prepares graviton states of the indicated helicity in the out sector of a given scattering process. It goes without saying that the soft modes are once again extracted by effecting the integral over the retarded time $u$, anticipating the form of the soft charges when the soft theorems are derived as Ward identities. 

\subsection{Charge Conservation and the Leading Soft Theorem in QED}\label{app:a2}
The duality between Ward identities and soft theorems in quantum field theory is most easily illustrated by studying the simplest case of the leading order soft theorem in gauge theory. Recapitulating, we know that when a single photon of positive helicity is radiated in a scattering process, the scattering amplitude is multiplied by an overall factor that results in the equality

\begin{equation}
    \lim_{\omega\rightarrow 0} \omega \bra{\rom{out}}a^{\rom{out}}_{\omega,z}\mathcal{S}\ket{\rom{in}} = \left(\sum_{i=1}^{n}\frac{\eta_{i}e_{i}}{z-z_{i}}\right)\bra{\rom{out}}\mathcal{S}\ket{\rom{in}},
\end{equation}
where $z$ designates the direction of the soft photon on the celestial sphere and the variables $\eta_{i}$ take values $\pm 1$ for incoming and outgoing particles respectively. Consequently, our task is now to relate the left side of this equality to the commutator of a soft charge, which creates the soft photon, with the $S$-matrix, and the right side with the so-called hard charge, which when acting on external states yields their respective charges.  

Bringing the leading soft theorem in QED into the form of a Ward identity proceeds by noting that there is a conserved quantity on null infinity that is decomposed in the following fashion

\begin{equation}
    Q_{\varepsilon} = Q^{\rom{soft}}_{\varepsilon} + Q^{\rom{hard}}_{\varepsilon}, 
\end{equation}
where in terms of the asymptotic fields we have the expressions

\begin{equation}
    Q^{\rom{soft}}_{\varepsilon} = 2\int d^{2}z N(z,\z)D_{z}D_{\z}\varepsilon(z,\z),
\end{equation}
and

\begin{equation}
    Q^{\rom{hard}}_{\varepsilon} = \int d^{2}z \varepsilon(z,\z)j_{u}(z,\z).
\end{equation}
A couple of qualifications need to be made regarding these definitions. First, we observe that the charges themselves are parametrized by a function $\varepsilon$ defined on the celestial sphere. Accordingly, if this charge is to commute with the $S$-matrix for generic $\varepsilon$, the upshot is the existence of an infinite number of conserved quantities corresponding to each choice of the latter function. Indeed, it is a specific choice of this function that turns out to be equivalent to the soft theorem. 

The current $j_{u}(z,\z)$ is the current of charged matter in the theory. It may be quite generic, and depends on the microscopic structure of the underlying charged fields. Most relevant is the action of the current on external states; when it operates on a state $\ket{z'}$ with direction $(z',\z')$ on the celestial sphere we have

\begin{equation}
    j_{u}(z,\z)\ket{z'} = Q\dfunction\ket{z'},
\end{equation}
where $Q$ is the charge of the state. For more details on how the current is constructed, the reader is referred to \cite{Strominger:2017zoo}. Finally, the function $N$ is defined implicity in terms of the field strength. Specifically, we have 

\begin{equation}
    e^{2}D_{z}N = \int_{-\infty}^{\infty}du F_{uz}(u,z,\z).
\end{equation}
Expressed in terms of creation and destruction operators, it becomes easy to see that this is actually the desired soft operator that creates low energy photons - 

\begin{equation}
    e^{2}D_{z}N = -\frac{\sqrt{2}e}{8\pi^{2}}\lim_{\omega\rightarrow 0^{+}}\omega a^{\rom{out}}_{+}.
\end{equation}
It turns out that the conservation of the charge $Q_{\varepsilon}$ really comes directly as a consequence of the classical motions of equation, and due immediately to the fact that field strengths match at antipodes in spacetime. The upshot however is that the charge must commute with the $S$-matrix, which means that for any $\varepsilon$ we have

\begin{equation}
    \bra{\rom{out}} [Q_{\varepsilon},\mathcal{S}]\ket{\rom{in}} = 0.
\end{equation}
The actions of the hard and soft contributions can now be evaluated independently according to the definitions we have just provided. The hard piece yields the simple result

\begin{equation}
    \bra{\rom{out}} [Q^{\rom{hard}}_{\varepsilon},\mathcal{S}]\ket{\rom{in}} = \left(\eta_{i}e_{i}\varepsilon(z_{i},\z_{i})\right)\bra{\rom{out}} \mathcal{S}\ket{\rom{in}},
\end{equation}
while the soft contribution provides exactly the $S$-matrix element with one soft emission smeared with the function $\varepsilon(z,\z)$ - 

\begin{equation}
    -\bra{\rom{out}} [Q^{\rom{soft}}_{\varepsilon},\mathcal{S}]\ket{\rom{in}} = \frac{2\sqrt{2}}{8\pi^{2}e}\lim_{\omega\rightarrow 0^{+}}\omega\int d^{2}z'\bra{\rom{out}} a^{\rom{out}}_{+}(z',\z')\mathcal{S} \ket{\rom{in}}D_{z}D_{\z}\varepsilon(z',\z').
\end{equation}
Making the choice 

\begin{equation}
    \varepsilon(z',\z') = \frac{1}{z-z'},
\end{equation}
it is easily verified that the hard contribution is the soft factor multiplying the $S$-matrix, while the soft contribution is simply the $S$-matrix element with one soft emission in the case of a positive helicity photon. As a result, the Ward identity corresponding to the commutation of $Q_{\varepsilon}$ with the $S$-matrix turns out to be identical to the leading order soft photon theorem, which as we have stated earlier, is really determined entirely as a result of unitarity and Lorentz invariance. 

The theme of such charges being labelled by an infinity of smooth functions on the sphere appears to be recurring, and as such has inspired the approach we have taken in this work (particularly in the context of having to construct two-dimensional models for soft theorems beyond leading order). The connection to the dual models is actually expressed rather naturally if one has recourse to the global symmetries enjoyed by the soft charges described in this section and the dual model proposed for the corresponding soft theorem. Note that if we look at the Coulomb gas model for the leading soft photon theorem, the action is invariant under the global shift 

\begin{equation}
    \varphi_{a}(z,\z) \longrightarrow \varphi_{a}(z,\z) + c_{a},
\end{equation}
where the $c_{a}$ are just $c$-numbers. This is the well known shift symmetry of the free two-dimensional boson, and is known to imply as a consequence the global conservation of charge. This symmetry of the dual model is clearly a d=symmetry of the soft part $Q^{\rom{soft}}_{\varepsilon}$ of the charge as well, if we map it to the shift

\begin{equation}
    \varepsilon(z,\z) \longrightarrow \varepsilon(z,\z) + c, 
\end{equation}
where $c$ is a constant. Since the Ward identity itself must hold for any choice of the smearing function, the effect the above transformation has on the Ward identity is obtain by taking the variation of \mref{eq:a.3} to give

\begin{equation}
    \bra{\rom{out}} [Q^{\rom{hard}}_{c},\mathcal{S}]\ket{\rom{in}}= -\bra{\rom{out}} [Q^{\rom{soft}}_{c},\mathcal{S}]\ket{\rom{in}},
\end{equation}
where $Q^{\rom{soft}}_{c} = 0$ and the action of the hard charge $Q^{\rom{hard}}_{c}$ once evaluated gives

\begin{equation}
    \left(\sum_{i=1}^{n}\eta_{i}e_{i}\right)\bra{\rom{out}} \mathcal{S}\ket{\rom{in}} = 0,
\end{equation}
which for generic matrix elements implies the global conservation of charge. We note accordingly that the identification of the two-dimensional dual model as the correct description of the soft theorem is supported by the sharing of global symmetries. We will repeat this analysis for the leading soft theorem in gravity and make a few comments at the level of the higher order soft theorems in the following sections.

As we will see soon, the subleading soft theorems in gauge theory as well as gravity are labelled by vector valued functions on the sphere, in accordance with which we defined our two-dimensional actions for these theorems. For now, let us first move on to understanding how the leading soft theorem in gravity is a result of the conservation of supertranslation charges.

\subsection{Supertranslations and the Leading Soft Graviton Theorem}\label{app:a3}
To the author's knowledge, the conservation of charges at null infinity in massless theories like QED and gravity were first studied by Newman and Penrose in \cite{Newman:1968uj}, where a tower of conserved quantities on null infinity were obtained directly from the field equations. It turns out that a specific class of conserved quantities correspond to the invariance of the $S$-matrix under transformations known as \emph{supertranslations}, which are local shifts of the retarded time parameter $u$. The supertranslation charge in an asymptotically flat spacetime is measured by the limiting value of a quantity known as the Bondi mass aspect, which is formally obtained by the limit

\begin{equation}
    m_{B}(u,z,\z) = \frac{1}{2}\lim_{r\rightarrow\infty}r h_{uu}(u,z,\z),
\end{equation}
where of course $h_{uu}$ is a perturbation around the Minkowski solution. Now it turns out that the Bondi mass aspect is not an independent object in its own right, and can be determined in terms of the radiative data of spacetime (namely the shear tensor and matter radiative data) as a result of the Einstein field equations. Indeed, this constraint is captured by the equation of motion governing the time evolution of $m_{B}$ on null infinity, which reads 

\begin{equation}
    \partial_{u}m_{B} = \frac{1}{4}\left(D_{z}^{2}N_{\z\z}+ D_{\z}^{2}N_{zz}\right) - T_{uu},
\end{equation}
where $T_{uu}$ is the $uu$ component of the stress-energy tensor, measuring the energies of the asymptotic states, and the tensors $N_{zz}$ and $N_{\z\z}$, called the Bondi news, are defined as time derivatives of the shear, whence we have

\begin{equation}
    N_{zz} = \partial_{u}C_{zz},
\end{equation}
and a corresponding definition for $N_{\z\z}$. 

The Bondi mass aspect is determined now by direct integration of the equation of motion. The supertranslation charge is defined by setting the limits of integration to $\pm\infty$ and smearing with an arbitrary function $f(z,\z)$ on the celestial sphere. This procedure resolves the supertranslation charge into two distinct parts as a consequence of the form of the equation of motion to provide

\begin{equation}
    Q_{f} = Q^{\rom{soft}}_{f} + Q^{\rom{hard}}_{f},
\end{equation}
where the soft term, in anticipation of the form that the soft theorem has to take, is defined as that part of the charge that is linear in the Bondi news tensor as follows

\begin{equation}
    Q^{\rom{soft}}_{f} = \int_{-\infty}^{\infty}du\int d^{2}z f(z,\z) \frac{1}{4}\left(D_{z}^{2}N_{\z\z}+ D_{\z}^{2}N_{zz}\right),
\end{equation}
and the hard part is defined by integrating over the stress-energy tensor component over $u$ and smearing with $f$.

Relating this to the soft theorem is achieved by computing how the soft and hard charges act on external states, in exact analogy to what was done before in the case of QED. The action of the hard charge is again most easily recovered - one simply notes that the $uu$ component is the energy density; a short calculation (see for example \cite{H:2018ktv}) shows that we have

\begin{equation}
     -\bra{\rom{out}} [Q^{\rom{hard}}_{f},\mathcal{S}]\ket{\rom{in}} = \left(\sum_{i=1}^{n}\eta_{i}f(z_{i},\z_{i})\omega_{i}\right)\bra{\rom{out}} \mathcal{S}\ket{\rom{in}}.
\end{equation}
It is clear that the soft theorem is obtained if we make the following identification for the function $f$

\begin{equation}
    f(z',\z') = \frac{\z-\z'}{z-z'},
\end{equation}
if we want to describe the case of positive helicity gravitons. Naturally, the negative helicity case is obtained by complex conjugating this $f$. 

Indeed, it becomes clear after a short computation that this is the correct choice for the function $f$ as far as the soft part is considered as well. Specifically, employing the definition of the Bondi news tensor in terms of the shear, the soft charge once expanded in terms of creation and destruction operators contains only soft graviton modes. Indeed, it can be shown that due to the identity

\begin{equation}
    D_{\z}^{2}\left(\frac{\z-\z'}{z-z'}\right) = \pi\dfunction,
\end{equation}
the contribution of the soft charge reduces to 

\begin{equation}
    \bra{\rom{out}} [Q^{\rom{soft}}_{f},\mathcal{S}]\ket{\rom{in}} = \frac{1}{16\pi^{2}}\lim_{\omega\rightarrow 0}\omega\bra{\rom{out}}b^{\rom{out}}_{+}(z,\z)\mathcal{S}\ket{\rom{in}},
\end{equation}
which is the statement that a single soft graviton of negative helicity is added to the out state. The Ward identity is now easily recognized as the soft theorem, in accordance with what we had claimed. 

The relation that this presentation of supertranslations has to the two-dimensional representation that we discussed in section \ref{sec:3} is the following. It can be shown that the Bondi mass (specifically, the time derivative thereof) generates translations on null infinity along the retarded time parameter. Along with the fact that the smearing function $\epsilon$ can be expanded into arbitrary holomorphic and anitholomorphic pieces, such translations can be made manifestly local on the celestial spheres along null infinity - transformations which are identified precisely with the supertranslations we have already discussed.

We point out at this stage that we already have an indication as to the kind of action that soft modes would be described by dynamically on the celestial sphere. The local contribution of the soft mode in the QED case was obtained by the first derivative of the smearing function, while the corresponding local expression in the case of gravity was recovered by a second order derivative. In fact, if we regard the smearing functions as fields in their own right, the respective presentation of these objects in the soft charges are actually equivalent to the soft currents that we had found in our two-dimensional models.

This apparent duality is further strengthened by pointing out that in total analogy to the previously considered case of the leading soft photon theorem, the soft charge $Q^{\rom{soft}}_{f}$ shares the global symmetries obeyed by the dual model for the corresponding soft theorems. Recall that the dual model for the leading order soft graviton theorem was given by a pair of scalars $\sigma_{a}$ that obeyed the biharmonic equation. These symmetries were simply shifts of the fields according to the prescription

\begin{equation}
    \sigma_{a}(z,\z) \longrightarrow \sigma_{a}(z,\z) + c_{a,1}+c_{a,2}z + c_{a,3}\z + c_{a,4}z\z,
\end{equation}
which is equivalent to saying that the four momentum, written out in a basis labelled by $\lbrace{1,z,\z,z,\z\rbrace}$ is always conserved in a given scattering process. The same result is inferred by noting that it is precisely the foregoing class of shift symmetries which are obeyed by the soft charge $Q^{\rom{soft}}_{f}$. Indeed, this is the case when the following transformation

\begin{equation}
    f(z,\z)\longrightarrow f(z,\z) + c_{1}+c_{2}z + c_{3}\z + c_{4}z\z = f(z,\z) + \delta f,
\end{equation}
is made given some arbitrary function $f(z,\z)$ on the celestial sphere. That the soft charge for a given $f$ is preserved by this transformation is checked by simply inspection. Together with the fact that the Ward identity due to $Q_{f}$ must hold irrespective of the choice of the smearing function, one obtains that

\begin{equation}
     \bra{\rom{out}} [Q^{\rom{hard}}_{\delta f},\mathcal{S}]\ket{\rom{in}} = \left(\sum_{a=1}^{4}\sum_{i=1}^{n}\eta_{i}c_{a}p^{a}_{i}\right)\bra{\rom{out}} \mathcal{S}\ket{\rom{in}},
\end{equation}
where $p^{a}_{i} = \omega_{i}(1,z_{i},\z_{i},z_{i}\z_{i})$ must vanish. The mutual independence of the constants $c_{a}$ then implies four momentum conservation.

\subsection{Local Dipoles and the Subleading Soft Theorem in QED}\label{app:a4}
We have seen that the soft photon theorem at leading order in QED is a direct consequence of the conservation of a specific class of charges on null infinity. In particular, these charges are labelled by an analytic function on the celestial sphere, with a specific choice of the function supplying the equivalence of the corresponding Ward identity to the soft theorem. Asking how this generalizes to the subleading soft photon theorem turns out to be an interesting exercise, which we will now consider.

First, we need to observe that the power counting of the soft theorems in energy is relevant to isolating out the subleading contribution. More concretely, not that the leading soft theorem is of order $\mathcal{O}\left(\frac{1}{\omega}\right)$, in accordance with which the projection operator we used was

\begin{equation}
    \mathcal{P}_{-1} = \lim_{\omega\rightarrow 0}\omega,
\end{equation}
which when acted upon the amplitude easily isolated the leasing soft contribution, as the others vanished in the limit $\omega\rightarrow 0$. To extract the subleading part, we observe that the leading contribution needs to be projected \emph{out}. This is done by noting that the operation $\omega\partial_{\omega}$ simply inverts the sign on the leading contribution, and has the effect of annihilating the subleading term, which is independent of $\omega$. Accordingly, we infer that the projection operator that extracts the subleading contribution is simply

\begin{equation}
    \mathcal{P}_{0} = 1 + \omega\partial_{\omega}.
\end{equation}
Now since we are working on null infinity, we need to ask (and answer) how this is expressed in the $(u,z,\z)$ basis. It is easily seen that the operator $u\partial_{u}$ does the job, as a simple integration by parts reveals. This suggests the form of the soft charge, in which we now expect the electromagnetic potential to appear with the operator $u\partial_{u}$ acting on it.

To figure out the complete form of the charge, we observe first that the subleading soft theorem is nonlocal (as is any soft theorem) in the variables $z$ and $\z$, which parametrize the direction of the soft photon. Making it local requires taking two derivatives using $D_{z}$ (or $D_{\z}$ for a negative helicity emission), which turns each term in the soft theorem into a delta function or a derivative thereof. Consequently, in analogy to the leading order case, we expect that the soft charge is accompanied this time by two derivatives along $D_{z}$. Lorentz invariance dictates that such a term must be multiplied by a vector function to remain invariant. It turns out that this is indeed the case

With these general comments in place, let us simply state the form of the hard charge, which can be found by just demanding that it supply the soft current (it can also be found by a rather tedious expansion along null infinity; see \cite{Campiglia:2016hvg} for the calculation). It takes the form

\begin{equation}
    Q^{\rom{hard}}_{V} = \int du d^{2}z\left(uD_{\z}V_{z}j_{u} + V_{z}j_{\z}\right) + \rom{c.c}. 
\end{equation}
A short calculation using the expansion of the current in terms of creation and annihilation operators then shows that when acting on asymptotic states, the hard charge supplies the following result

\begin{equation}
    -\bra{\rom{out}} [Q^{\rom{hard}}_{V},\mathcal{S}]\ket{\rom{in}} = \left(\sum_{i=1}^{n}K_{V}(z_{i},\z_{i})\right)\bra{\rom{out}} \mathcal{S}\ket{\rom{in}},
\end{equation}
where we have in terms of the vector $V$ the following expansion for the function $K_{V}$ as follows (where we have assumed that the external states are scalars; the generalization to spinning states is made by introducing terms proportional to $\frac{s_{i}}{\omega_{i}}$ for states of sin $s_{i}$),

\begin{equation}
    K_{V}(z_{i},\z_{i}) = \eta_{i}e_{i}\left(D_{\z}V_{z}(z_{i},\z_{i})\partial_{\omega_{i}}+V_{z}(z_{i},\z_{i})\partial_{\z_{i}}\right) + \rom{c.c},
\end{equation}
which tells us that once we make the following special choice

\begin{equation}
    V_{z}(z',\z') = \frac{\z-\z'}{z-z'},
\end{equation}
and

\begin{equation}
    V_{\z}(z',\z') = 0,
\end{equation}
what we obtain from the preceding commutator is precisely the soft factor for the subleading soft photon theorem. In accordance with this, it should come as no surprise that the soft charge takes the simple form

\begin{equation}
    Q^{\rom{soft}}_{V} = -2\int du d^{2}z\left(uD_{\z}V_{z}\partial_{u}D_{\z}A_{z} + \rom{c.c}\right).
\end{equation}
This is precisely the form that we had anticipated earlier; the operator $u\partial_{u}$ yields the desired projection operator and in combination with the identity

\begin{equation}
    D_{\z}^{2}\left(\frac{\z-\z'}{z-z'}\right) = \pi\dfunction,
\end{equation}
for the choice of the vector field that we have made herein, the commutator of the $S$-matrix with the soft charge gives exactly the matrix element for the scattering amplitude with one soft emission and the leading term projected out. The Ward identity corresponding to $Q_{V}$ is then exactly the subleading soft photon theorem.

The physical interpretation of the Ward identity generalizes that of the Ward identity for the leading soft theorem. In this case, the charge $Q_{V}$ parametrized by the vector field $V$ turns out to measure the local analogue of the dipole charge density on null infinity. Specifically, the arbitrariness of the field $V$ means that the Ward identity is equivalent to demanding that dipole electric charge is conserved not only globally, but at every angle on the celestial sphere. 

\subsection{Superrotations and the Subleading Soft Graviton Theorem}\label{app:a5}
The conservation of dipole charge at every angle finds its analogue in the context of gravity when we look at the Ward identity corresponding to the subleading soft graviton theorem. Here, the relevant Ward identity indicates the conservation of the measured superrotation charge at every angle on the celestial sphere. Superrotations, as we have mentioned earlier, translate locally not only along $u$, but effect shifts along the celestial sphere as well. Unlike supertranslations however, superrotation currents do not mix holomorphic and antiholomorphic transformations. 

Superrotations, like supertranslations, are derived by studying the expansion of the metric at null infinity and isolating out those terms responsible for measuring the amount of angular momentum passing though a given region (just as how the Bondi mass measures the outgoing energy). The correct object to study turns out to be what is known as the Bondi angular momentum aspect, denoted by $N_{z}$ (or $N_{\z}$), and shows up as a subleading contribution to the $uz$ (or $u\z$) component of the metric perturbation around the Minkowski solution. They are determined by the Einstein field equations \cite{Strominger:2017zoo} to satisfy the equation

\begin{equation}
    \partial_{u}N_{z} = \frac{1}{4}\left(D_{z}^{3}C_{\z\z} - D_{z}D_{\z}^{2}C_{zz}\right) -u\partial_{u}D_{z}m_{B} - T_{uz},
\end{equation}
where $T_{uz}$ is the subleading ($\mathcal{O}\left(\frac{1}{r^{2}}\right)$) contribution from the $uz$ component of the stress-energy tensor, measuring the flux of angular momentum in the direction labelled by $uz$. It is the conservation of the Bondi angular momentum aspect across every angle on the celestial sphere that represents the Ward identities of superrotations, which generalize ordinary rotations to arbitrary local rotations on null infinity. Indeed, we have the following definitions of the soft and hard charges, this time labelled by a vector field $Y$ - 

\begin{equation}
    Q^{\rom{soft}}_{Y} = -\frac{1}{2}\int du d^{2}z uD_{\z}^{3}Y_{\z}\partial_{u}C_{zz} + \rom{c.c},
\end{equation}
and

\begin{equation}
    Q^{\rom{hard}}_{Y} = \int dud^{2}z \left(Y_{z}T_{u\z}+ uD_{\z}Y_{z}T_{uu}\right) + \rom{c.c}.
\end{equation}
We remark that the appearance of derivative at third order in the soft charge echo the kind of action that will be needed to describe the dynamics of subleading soft modes in a two-dimensional model. Indeed, we saw that it was a sixth order kinetic action, namely one dominated by a triharmonic operator that determined the two-dimensional model. 

Evaluating the effect of the hard charge on asynmptotic states is really a matter of rewriting the stress energy tensor components in terms of boundary values of fields; the result is in terms of the vector $Y$, which turns out to be

\begin{equation}
    -\bra{\rom{out}} [Q^{\rom{hard}}_{Y},\mathcal{S}]\ket{\rom{in}} = \left(\sum_{i=1}^{n}\mathcal{J}^{(1)}_{Y}(z_{i},\z_{i})\right)\bra{\rom{out}} \mathcal{S}\ket{\rom{in}},
\end{equation}
where the operator $\mathcal{J}^{(1)}_{Y}$ is given by the following expansion (the reader can find a full derivation of this in the paper \cite{Campiglia:2014yka} due to Laddha and Campiglia) - 

\begin{equation}
    \mathcal{J}^{(1)}_{Y}(z_{i},z_{i}) = \eta_{i}\left(Y_{z}(z_{i},\z_{i})\partial_{\z_{i}}+\frac{1}{2}D_{\z}Y_{z}(z_{i},\z_{i})\omega_{i}\partial_{\omega_{i}}\right) + \rom{c.c}.
\end{equation}
Again, we have made the simplifying assumption that the external states are scalars, dropping the term proportional to the helicities of the interacting particles. The reader will observe that this form of the operator $J_{V}$ is precisely what we need in order to recover the subleading soft theorem in gravity. Once again, the positive and negative helicity sectors appear simultaneously; a special choice of the vector $Y$ is required to supply a specific soft theorem. Indeed, the positive helicity theorem is obtained by making the simple choice

\begin{equation}
    Y_{z}(z',\z') = \frac{1}{2}\frac{(\z-\z')^{2}}{z-z'},
\end{equation}
and

\begin{equation}
    Y_{\z}(z,'z) = 0.
\end{equation}
Just as in the case of QED, the commutator of the soft part of the charge with the $S$-matrix is easily reduced to the matrix element for the scattering process in the presence of one subleading soft graviton, keeping in mind that we have the identity

\begin{equation}
    D_{\z}^{3}\left(\frac{1}{2}\frac{(\z-\z')^{2}}{z-z'}\right) = \pi\dfunction,
\end{equation}
due to which the soft charge simply contains soft modes projected into the subleading sector using $\mathcal{P}_{0}$ due to the operator $u\partial_{u}$. The result of course is the establishment of the equality of the Ward identity due to $Q_{Y}$ and the subleading soft graviton theorem.

It's worth pointing out that the soft charge in every case we have discussed so far (and in the case of the subsubleading soft graviton theorem) are precisely the soft currents that we derived in the two-dimensional models presented in this work. In fact, the dual models are precisly those two-dimensional theories that remain invariant under the global shift transformations of the smearing functions that preserve the form of the soft charges in each case. We saw how the duality played out explicitly in the leading order theorems, and they may be summarized in the subleading theorems by the qualified statements

\begin{equation}
    (D_{z}^{2}V^{z}_{1},D_{\z}^{2}V^{\z}_{1}) \longleftrightarrow Q_{V},
\end{equation}
and

\begin{equation}
    (D_{z}^{3}Y^{z}_{1},D_{\z}^{3}Y^{\z}_{1}) \longleftrightarrow Q_{Y},
\end{equation}
where we have indicated that both helicity sectors are contained in the corresponding soft charges. Similarly, a qualified duality between the hard charges and the dressing operators is obtained by observing that they are related essentially by the Mellin transform - 

\begin{equation}
K_{V}(z_{i},\z_{i}) \overset{\rom{Mellin}}{\longrightarrow} ie_{i}\eta_{i}(Q^{V}_{+}(h_{i},z_{i},\z_{i})+Q^{V}_{-}(h_{i},z_{i},\z_{i}))\mathcal{P}^{-1}_{i},
\end{equation}

\begin{equation}
\mathcal{J}^{(1)}_{V}(z_{i},\z_{i}) \overset{\rom{Mellin}}{\longrightarrow} i\kappa\eta_{i}(Q^{Y}_{+}(h_{i},z_{i},\z_{i})+Q^{Y}_{-}(h_{i},z_{i},\z_{i})).
\end{equation}
The duality is qualified in the sense that we have not indicated any precise relation between the fields that build the theory, simply that there is an analytic correspondence between the structure of the soft currents on one hand and the soft charges on the other. 

Indeed, while the higher derivatives arise in the Ward identity as a way of making the soft charge a local quantity, the higher derivatives show up in the two-dimensional models in order to supply the dressing functions that we have used in this appendix to derive the soft theorems. In this sense, as we had alluded to in the beginning of this chapter, the Ward identity approach to soft theorems in a very precise way anticipated the two-dimensional models we have derived, which simply realize the soft charges as currents instead of in terms of presenting an asymptotic symmetry on null infinity. 

We close with some brief comments on how this generalizes to the subsubleading soft graviton theorem. The charges we have discussed have in the literature been derived by simple performing expansions of the field equations at null infinity in inverse powers of $r$, matching coefficients of specific powers and identifying parts of these which are conserved along the null boundary. Deriving the hard charge in this fashion is somewhat involved; partial progress was made in \cite{Campiglia:2016efb,Campiglia:2016jdj} by Campiglia and Laddha but the most general approach taken has been due to Barnich, Mao and Ruzziconi in \cite{Barnich:2019vzx}. The soft sector of the charge however is given by a more familiar structure.

Just as we had to project out the leading order soft theorem to isolate the contribution of the subleading theorem in the last two cases, a similar projection scheme has to be worked out in the case of the subsubleading soft graviton theorem. We choose to follow the convention discussed in \cite{Campiglia:2016efb}. There, by noting that the subsubleading soft theorem is generically proportional to $\omega$, the energy of the soft particle, this contribution was extracted by multiplication with $\omega^{-1}$, and keeping the finite piece. 

Keeping in mind now that the shear $C_{AB}$ is defined on null infinity, the multiplication by $\omega^{-1}$ amounts to performing an integral over the null coordinate $u$, since it is essentially the Fourier conjugate of the energy. Concretely, use is made of the following identity \cite{Campiglia:2016efb} 

\begin{equation}
   \lim_{\omega\rightarrow 0} \frac{1}{\omega}\widetilde{F}(\omega) = -i\int_{-\infty}^{\infty}du\int^{u}_{-\infty}du'F(u'),
\end{equation}
where $\widetilde{F}$ is the Fourier transform of the function $F$. Indeed, this double integral performs the necessary projection onto the subsubleading piece after divergent terms are dropped. The task at hand now is to figure out how the soft piece is constructed, making use of the preceding projection and the shear field $C_{AB}$.

Even without performing any calculations, we can guess that the soft charge must contain fourth order derivatives in the holomorphic and antiholomorphic coordinates, if only to be in agreement with the structure we have already observed. More technically, one would expect this due to considerations of locality. Note that the subsubleading soft theorem in gravity (restricting our attention to the positive helicity) sector, degree $4$ in the parameter $\z$, labelling the direction of the soft particle. To turn this into a local quantity, we have to carry out four derivatives\footnote{This argument is covered in greater detail in the letter \cite{Campiglia:2016jdj} due to Campiglia and Laddha (which also serves as a nice summary of \cite{Campiglia:2016efb}).} using $D_{\z}$ (or $D_{z}$ for the negative helicity case). Consequently, a form $\sim D_{\z}^{4}C_{zz} + \rom{c.c}$ is anticipated as the building block of the soft part of the conserved charge.

This turns out to be the case; the soft charge that leads to the subsubleading soft graviton theorem takes the form

\begin{equation}
    Q^{\rom{soft}}_{X} = \int_{-\infty}^{\infty}du\int^{u}_{-\infty}du'\int d^{2}z X_{zz}D^{4}_{\z}C_{zz}(u,z,\z) + \rom{c.c},
\end{equation}
where the $X_{ab}$ for $ab \in \lbrace{zz,\z\z\rbrace}$ denote the diagonal elements of a second rank symmetric tensor on the sphere, which ensures that the soft charge is well defined on the sphere. 

To derive the soft charge, an appropriate choice of the tensor $X$ needs to be made. It should come as no surprise that the choice (for positive helicity radiation)

\begin{equation}
    X_{zz}(z',\z') = \frac{1}{6}\frac{(\z-\z')^{3}}{z-\z},
\end{equation}
and

\begin{equation}
    X_{\z\z}(z',\z') = 0
\end{equation}
and yields the creation operator for one soft positive helicity graviton in light of the fact that the identity

\begin{equation}
    D^{4}_{\z}\left(\frac{1}{6}\frac{(\z-\z')^{3}}{z-\z}\right) = \pi\dfunction,
\end{equation}
is known to hold. Defining the hard operator however turns out to be a bit of a challenge. Hard operators in the lower order soft theorems in QED and gravity emerge natually alongside the soft operators when field expansions are carried out to a specified order. Accordingly, generic expressions in terms of the stress-energy tensor are typically obtained, as laid out in the preceding discussions. To the extent that the present author is aware, there have not been too many detailed studies of the subsubleading soft graviton theorem in this manner. In \cite{Campiglia:2016efb}, the hard charge was constructed \emph{post facto} by matching to the action on the external states, while in \cite{Barnich:2019vzx} the subsubleading contribution was extracted by performing an expansion in terms of the coefficients due to Newman and Penrose. We omit these details and refer the reader to the relevant papers for more information; we simply state that we require the following condition

\begin{equation}
    \bra{\rom{out}} [Q^{\rom{hard}}_{Y},\mathcal{S}]\ket{\rom{in}} = \left(\sum_{i=1}^{n}\mathcal{J}^{(2)}_{X}(z_{i},\z_{i})\right)\bra{\rom{out}} \mathcal{S}\ket{\rom{in}},
\end{equation}
where the function $\mathcal{J}^{(2)}_{X}$ is expanded as,

\begin{equation}
    \mathcal{J}^{(2)}_{X}(z_{i},\z_{i}) = \eta_{i}\omega\left(\frac{1}{6}D^{2}_{\z}X_{zz}(z_{i},\z_{i}\omega_{i}^{2}\partial^{2}_{\omega_{i}}-\frac{2}{3}D_{z}X_{zz}(z_{i},\z_{i}\omega_{i}\partial_{\omega_{k}}D_{\z_{i}}+X_{zz}(z_{i},\z_{i})D_{\z_{i}}^{2}\right) + \rom{c.c},
\end{equation}
which supplies the positive subsubleading soft graviton theorem when we make the preceding choice for the tensor $X$. The reader will also observe that this matches our choice for the dressing operator $Q^{X}$ defined in terms of the two-dimensional field $X_{2}^{A}$. 

As far as a physical interpretation of the conserved quantity $Q_{X}$ is concerned, it is a little difficult to provide a concrete set of observables whose preservation along the sphere is guaranteed by the corresponding Ward identities. However, it was observed in \cite{Campiglia:2016efb} that the right way of thinking about this question is to think of the conserved charges as due to a class of generalized diffeomorphisms, specifically ones which are parametrized by nonzero vector $X^{A}$ that satisfies

\begin{equation}
    D_{A}X^{A} = 0
\end{equation}
on the sphere. It turns out that these diffeomorphisms are generated by finite charges that correspond to the subsubleading soft graviton theorem. 